\documentclass[
aps,
prd,
reprint,
showpacs,
superscriptaddress,
longbibliography,
floatfix,
nofootinbib
]{revtex4-2}

\usepackage{comment}
\usepackage{amsmath}
\usepackage{amssymb}
\usepackage{multirow}
\usepackage{graphicx}  
\usepackage{dcolumn}   
\usepackage{bm}        
\usepackage{booktabs}
\usepackage{tabularx}
\usepackage{enumitem}
\usepackage[super]{nth}
\usepackage[normalem]{ulem}
\usepackage{wasysym}
\usepackage{microtype}
\usepackage{makecell}

\usepackage[usenames, dvipsnames, svgnames, table]{xcolor}

\usepackage{autofigs9}

\usepackage{braket}
\usepackage{color} 

\usepackage{hyperref}

\usepackage[capitalise]{cleveref}

\graphicspath{{figures/}{figures_new/}}
\newcommand{\LM}[1]{{\color{orange}{\texttt{[LM: #1]}}}}

\newcommand{\conj}[1]{\ensuremath{#1^{*}}}

\newcommand{\red}[1]{\ensuremath{\overline{#1}}}
\newcommand{\cl}[1]{\ensuremath{\check{#1}}}
\newcommand{\qo}[1]{\ensuremath{\hat{#1}}}
\newcommand{\vphi}{\ensuremath{\varphi}}

\newcommand{\VAR}[1]{\ensuremath{\textsc{Var}[#1]}}
\newcommand{\PSD}[1]{\ensuremath{\textsc{Psd}[#1]}}
\newcommand{\CSD}[2][]{\ensuremath{\textsc{Csd}_{#1}[#2]}}

\newcommand{\SNR}[1]{\ensuremath{\textsc{Snr}[#1]}}
\newcommand{\REof}[1]{\ensuremath{\Re\!\{#1\}}}
\newcommand{\IMof}[1]{\ensuremath{\Im\!\{#1\}}}

\newcommand{\FISH}[1]{\ensuremath{\mathcal{I}[#1]}}

\newcommand{\PDF}[1]{\ensuremath{\textsc{Pdf}\left[#1\right]}}

\newcommand{\PCF}[1]{\ensuremath{\textsc{Cf}\left[#1\right]}}

\newcommand{\VNP}[1][ ]{\ensuremath{\cl{\theta}_{#1}}}
\newcommand{\SNP}[1][ ]{\ensuremath{S_{\cl{\theta}#1}}}
\newcommand{\SNPr}[1][ ]{\ensuremath{\red{S}_{\cl{\theta}#1}}}
\newcommand{\VNA}[1][ ]{\ensuremath{\cl{n}_{#1}}}
\newcommand{\SNA}[1][ ]{\ensuremath{S_{\cl{n}#1}}}
\newcommand{\SNAr}[1][ ]{\ensuremath{\red{S}_{\cl{n}#1}}}
\newcommand{\K}[1][ ]{\ensuremath{\kappa}}
\newcommand{\rK}[1][ ]{\ensuremath{\red{\kappa}}}

\newcommand{\HqM}[1][ ]{\ensuremath{\qo{D}}}
\newcommand{\HpM}[1][ ]{\ensuremath{D}}
\newcommand{\Hpm}[1][ ]{\ensuremath{\tilde{d}}}

\newcommand{\PqE}[1][ ]{\ensuremath{\qo{E}}}
\newcommand{\PpE}[1][ ]{\ensuremath{E}}
\newcommand{\Ppe}[1][ ]{\ensuremath{\tilde{e}}}

\newcommand{\qoa}[1][ ]{\ensuremath{\qo{\mathfrak{a}}}}

\newcommand{\gq}{\ensuremath{\mathfrak{g}}}
\newcommand{\PV}{\zeta}
\newcommand{\rmd}{\mathrm{d}}

\newcommand{\armX}{1}
\newcommand{\armY}{2}

\newcommand{\fvar}[1]{\ensuremath{\tilde{#1}}}

\newcommand{\param}{\ensuremath{\alpha}}
\newcommand{\SPEC}{\ensuremath{\Phi}}

\usepackage{lineno}
\usepackage{linenoAMS}
\usepackage[draft]{derivs}

\begin{document}
\title{Single-Photon Signal Sideband Detection for High-Power Michelson Interferometers}

\author{Lee McCuller}
\affiliation{Division of Physics, Mathematics and Astronomy, California Institute of Technology, Pasadena, CA 91125, USA}
\email{mcculler@caltech.edu}

\date{\today}

\begin{abstract}
  The Michelson interferometer is a cornerstone of experimental physics. Its applications range from providing first impressions of wave interference in educational settings to probing spacetime at minuscule precision scales. Interferometer precision provides a unique view of the fundamental medium of matter and energy, enabling tests for new physics as well as searches for the gravitational wave signatures of distant astrophysical events. Optical interferometers are typically operated by continuously measuring the power at their output port. Signal perturbations then create sideband fields, forming a beat-note with the fringe light that modulates that power. When operated at a nearly-dark destructive interference fringe, this readout is a form of homodyne detection, with an imprecision set by a ``standard quantum limit'' attributed to shot noise from quantum vacuum fluctuations. The sideband signal fields carry energy which can, alternatively, be directly observed as photons distinct from the source laser. Without signal energy, vacuum does not form sidebands and cannot spuriously create photon counts or shot noise. Thus, counting can offer improved statistics when searching for weak signals when classical backgrounds are below the standard quantum limit. Here, photon counting statistics are described for optical interferometry, relating the two forms of measurement and showing cases where counting greatly outperforms homodyne readout, even with squeezed state quantum enhancement. The most immediate application for photon counting is improving searches of stochastic signals, such as from quantum gravity or from new particle fields. The advantages of counting may extend to wider applications, such as gravitational wave detectors, and the concept of \emph{Fisher-information representative spectral density} is introduced to motivate further study.
\end{abstract}

\maketitle

\section{Introduction}

Optical interferometry is a powerful technique for experimental physics that combines several key advantages: Short optical wavelengths allow precision sensing even at low power; high quality mirror coatings allow large, kilowatt scale, power levels, boosting the quantum limited precision orders of magnitude; incorporating Fabry-Perot cavities optimizes sensitivity for specific signal frequencies; and high single-photon energy ensures that the probe field is effectively zero temperature, so classical background noises are minimal and introduced primarily by the (excellent) material properties of dielectric mirrors. 

Together, these features have established optical Michelson interferometers as the key experimental implementation for gravitational wave detection\cite{abbottLRR20ProspectsObserving, aggarwalAAPPP20ChallengesOpportunities}, and promising in searches for certain signatures of quantum gravity \cite{zurekPLB22VacuumFluctuations, verlindePLB21ObservationalSignatures, li22InterferometerResponse, chouCQG17HolometerInstrument} as well as dark matter candidates \cite{aielloPRL22ConstraintsScalar, grotePRR19NovelSignatures, liuPRD19SearchingAxion, deroccoPRD18AxionInterferometry, aielloPRL22ConstraintsScalar, vermeulenN21DirectLimits}.

Interferometers are typically operated by reading a timeseries of the power modulations from the interference of a ``DC'' static field and an ``AC'' signal field at an output port. The interference beatnote between the static and signal fields linearly converts modulations of the signal field into a continuous measurement of photopower, implementing a technique known as homodyne readout. The independent arrival of photons in this measurement leads to fluctuations in of the homodyne timeseries, shot noise, from Poisson statistics. The power inside the interferometer determines the optical gain between perturbations-of-interest and the AC signal field. Together, the shot noise and optical gain establish the ``standard quantum limit (SQL)'' for the precision (or rather, imprecision) of interferometers.  

This work derives the physical description of the measurement process for high-power interferometers to directly compare the statistical Fisher information provided by the homodyne timeseries readout against direct photon detection of the signal sidebands. The axion detection community has shown that in regimes where classical thermal noise is sufficiently below than the SQL in resonant microwave detectors, that photon counting accelerates their searches compared to linear amplifiers\cite{lamoreauxPRD13AnalysisSinglephoton}. Relatedly, the ALPS II experiment is designed to incorporate optical-wavelength photon counting for signals from Fabry Perot cavities to search for dark matter \cite{BahreJI13AnyLight, DiazOrtizPotDU22DesignALPS, Dreyling-EschweilerJMO15Characterization1064, GimenoNIaMiPRSAASDaAE22TESDetector}; however, it searches for monochromatic signals coherently generated within the ALPS experiment, so its photon counting application is different than the search applications analyzed here.
The statistics of direct photon detection have also been well known from the Dicke radiometer equation\cite{dickeRSI46MeasurementThermal}, which has been generalized using modern quantum measurement theory towards coherent optical applications as ``quantum spectrum analysis~\cite{ngPRA16SpectrumAnalysis}.''

The conclusion that counting presents potential statistical speedups are explored in the following, finding that optical interferometer experiments can benefit when using photon counting rather than homodyne detection. Interferometers have notable differences from microwave experiments with linear readout amplifiers in that they have highly frequency-dependent and varied classical noise backgrounds as well as a wide variety of physics applications with associated statistical tests. Homodyne detection is a linearization of a nonlinear process, introducing its own subtleties particularly when used with ``signal recycled'' resonant, detuned interferometers. These unique aspects motivate this work: a dedicated, detailed, and direct analysis of photon counting for interferometers.

The analysis of this work recovers several results in chapters VI-VIII of Helstrom\cite{helstrom76QuantumDetection}, applying them specifically to the detection of signals from Michelson interferometers. This provides new insights. It explicitly describes the physical process of homodyne detection and the assembly of statistics for photon counting. This description is fully in the Heisenberg picture, which allows a particularly semi-classical treatment of the optical fields. The derivation develops a temporal-mode basis which relates between spectral analysis, the number of independent measurements, and possible application-specific, optimized basis for waveform discrimination. It also fully incorporates the consequences of a classical noise background, which greatly affects the efficacy of photon counting. 

This work primarily focuses on wideband and narrowband stochastic detection tests, beginning with an application example. These tests are concise to describe and relevant to developing tabletop Michelson interferometer demonstrations of photon counting, optimized to find quantum gravity\cite{li22InterferometerResponse} and new particle fields that interact with light. In particular, the GQuEST experiment\cite{mccullerGQuEST} is optimized for these goals, and other experiments could be adapted\cite{vermeulenCQG21ExperimentObserving}. The applications to waveform feature detection, e.g. for astrophysics, are discussed and motivated in the outlook section, \ref{sec:outlook}.

The constructions of this work contest the conclusion that interferometers using squeezed states saturate the enhancement available from quantum techniques\cite{demkowicz-dobrzanskiPRA13FundamentalQuantum}. Photon counting is generalized for event characterization in \cref{sec:events}, and, for scientific analysis where the classical noise is much smaller than quantum noise, surpasses the benefits of squeezed states. In the outlook and event sections, efficiency bounds are developed to relate timeseries power spectra, which include the measurement penalty of shot noise, against counting, which provides more Fisher information but uses less efficient statistical tests.

The outline of the paper is:
\begin{description}[topsep=2pt,itemsep=2pt,partopsep=2pt, parsep=2pt]
\item[First]  photon counting is applied to show a major acceleration to a stochastic search for signatures of quantum gravity. This motivates rigorous analysis that follows.
\item[Second]  the Michelson detector is described, to establish the optical detection process.
\item[Third]  the signal and noise processes are described and the statistical test for them is defined. The temporal mode basis is then defined and described.
\item[Fourth]  the Homodyne readout method and its statistics is fully explored and related to the power spectral density, typically used as a figure of merit for interferometers.
\item[Fifth]  the photon counting method is analyzed in a manner that relates it, as closely as possible, to timeseries signal analysis.
\item[Sixth]  signal recycled interferometers are investigated as a case analysis between experimental options and physics applications.
\item[Seventh]  the outlook of the technique for broader application is briefly explored, particularly towards future gravitational wave detectors and quantum enhancements.
\item[Finally]  event-based searches are analyzed to characterize the statistical trade-offs between waveform inference and waveform discrimination.
\end{description}

\newcommand{\GQL}{\ensuremath{5~\text{m}}}
\newcommand{\GQP}{\ensuremath{10~\text{kW}}}
\section{Application Example: Stochastic Search for Quantum Gravity}\label{sec:application}

Recent work on Verlinde-Zurek ``geontropic'' entanglement-entropy fluctuations in gravitation\cite{verlindePLB21ObservationalSignatures, zurekPLB22VacuumFluctuations} describes a novel signature visible in interferometers and now includes a calculation of the fluctuation signal spectral density\cite{li22InterferometerResponse}. The power spectral density of the spectrum can be expressed as a phase fluctuation imprinted upon the light or as an equivalent displacement spectral density. This section will apply the photon counting technique to determine the necessary observation time from statistics. Later sections will rigorously calculate the interferometer response, statistical tests, and benefits of photon counting to this application.

\begin{figure}[ht]
  \centering
  \includegraphics[width=.98\linewidth]{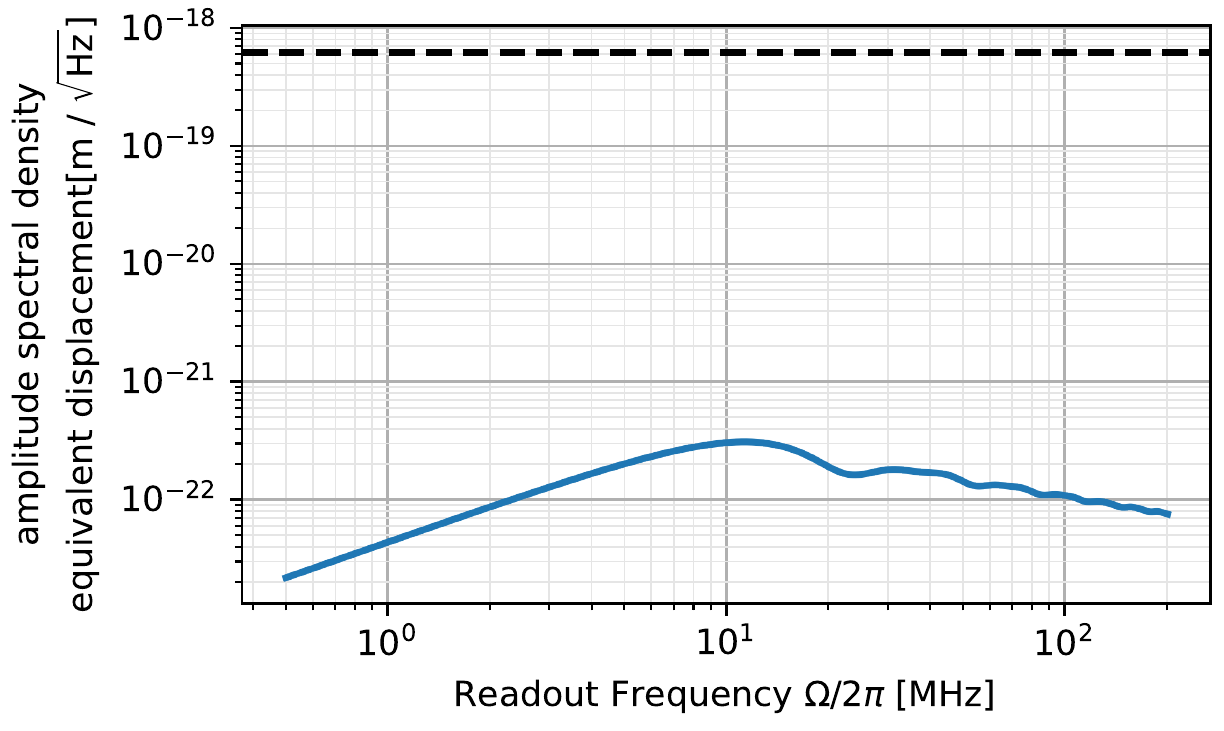}
  \caption{
    \textbf{GQuEST model spectral density} as compared to quantum noise. This is a model of the equivalent displacement spectrum expected from the geontropic signal in a L=5m Michelson interferometer, with 90degree angle between the arm. The dashed line indicates the quantum shot noise contribution, $\sqrt{S_{\HqM}}$, in amplitude spectral density units. Blue is the signal model $\sqrt{\SPEC(\Omega)}$ with $\alpha=1$.
  }
  \label{fig:model_displacement}
\end{figure}

For the purposes of defining a statistical test, the spectrum is decomposed into a scale parameter $\param$ and fiducial displacement spectrum $\SPEC(\Omega)$. The spectrum is given by an extensive analytic construction in \cite{li22InterferometerResponse}, too long to include here but graphed in \cref{fig:model_displacement}.
\begin{align}
  S_{\vphi}(\Omega) &\equiv \param \SPEC(\Omega)
  &&&
      \red{\SPEC} &\equiv \max_{0<\Omega<\infty} \SPEC(\Omega) \equiv \SPEC(\red{\Omega})
\end{align}
The scale parameter $\alpha$ is defined such that the $\alpha=1$ is the expected value established by theory, while $\red{\SPEC}$ is the maximum of the expected density at the peak frequency, $\red{\Omega}$. The peak frequency is useful for reasoning about the scales relevant for the experimental design. \Cref{sec:descriptions}, will revisit this decomposition of the spectrum into a nominal frequency dependence and test scale parameter.

The interferometer design for the GQuEST experiment is to use power recycling and appropriate technologies to circulate $P_{\text{BS}}=\GQP$ of light on the beamsplitter. The laser will have the wavelength $\lambda=1.5\text{um}$, corresponding to optical angular frequency of $\omega=2\pi c/\lambda$ and wavenumber $k=\omega / c$. This results in the interferometer optical gain, $\red{g}$, for a shot noise limited sensitivity $S_{\HqM}$ of 
\begin{align}
  S_{\HqM} &= \frac{1}{2\red{g}^{2}}  \approx \bigg(5{\cdot}10^{-19} \frac{\text{m}}{\sqrt{\text{Hz}}}\bigg)^2
        \bigg( \frac{\GQP}{P_{\text{BS}}} \bigg), &
                                                    \red{g} &= 2k\sqrt{\frac{P_{\text{BS}}}{4\hbar \omega}}
            \label{eq:shot_noise_sensitivity}
\end{align}
This factor is later given by \cref{eq:quantum_PSD}, and, for now, excludes considering classical noise. This relates the notation $S_{\HqM} = \PSD{\text{Qu}}$ for this experiment.
The optical gain is factorized into the mirror-displacement sensitivity $2k$ and two factors-of-two reduction in sensitivity from the beamsplitter, 1: reducing the arm power and 2: the signal power. These two factors represent that the arms/endmirrors are the actual sensor in a Michelson interferometer and that two measurements are needed to perform a differential comparison. \Cref{sec:descriptions} derives this optical gain.

\cref{fig:model_displacement} Shows a representative spectrum that GQuEST is targeting in its search. This spectrum has an analytical description constructed in \cite{li22InterferometerResponse}. The signal is remarkably small, with the scaling
\begin{align}
  \red{\SPEC} &\approx 
                \bigg(3{\cdot}10^{-22} \frac{\text{m}}{\sqrt{\text{Hz}}}\bigg)^2
                \bigg(\frac{L}{\GQL} \bigg)^2
  \label{eq:SPEC_scaling}
\end{align}

The number of independent measurements of the fluctuation of the Michelson fringe light (a form of homodyne readout) required to resolve $\alpha=1$ using a stochastic search with an optimal statistic is later given by \cref{eq:homodyne_chi_sq_var}. The inverse of the variance of the $\param$-parameter estimator corresponds to the square of the (amplitude) signal-to-noise ratio ``$\sigma_{\textsc{S/N}}$'' which expresses the number of standard deviations of significance. This number is
\begin{align}
  N_{\HqM}^{1\sigma} &= \frac{S_{\HqM}^2}{(\alpha \red{\SPEC})^2} \approx 10^{13} & N_{\HqM}^{1\sigma} &= \Delta F_{\HqM} \Delta T_{\HqM}^{1\sigma}
\end{align}
The applicable bandwidth must then be determined. Experimental constraints on data processing can determine it, but fundamentally it is related to the optimally weighted search statistic, built from \cref{eq:Homodyne_statistic_full} and developed in \cref{D:chi_weighted}. 
These signals span an effective bandwidth determined by the signal model. Comparing, \cref{eq:chi_sq_fully_weighted} with \cref{eq:homodyne_chi_sq_var}, The bandwidth is defined for for the optimal statistic as:
\begin{align}
  \Delta F_{\HqM} &= \int_0^{\infty} \left( \frac{\SPEC(\Omega)}{\red{\SPEC}} \right)^2 \frac{\rmd \Omega}{2\pi}
                         \approx \frac{c}{4L}
                         \label{eq:DeltaF_Q}
\end{align}
This assumes constant $S_{\HqM}$ for white measurement noise dominated by the the quantum shot noise of a broadband Michelson. The approximation indicates the calculation, under ideal conditions, for the geontropic signal of \cref{fig:model_displacement}

The number of measurements over a given measurement frequency band and observing time is given by the time-bandwidth product, thus an expected measurement time for a $1\sigma$ measurement is
\begin{align}
  \Delta T_{\HqM}^{1\sigma} &> \frac{N_{\HqM}^{1\sigma}}{\Delta F_{\HqM}} \approx 185\text{ hr} \cdot \bigg(\frac{\GQL}{L}\bigg)^{3} \bigg(\frac{\GQP}{P_{\text{BS}}}\bigg)^2
   \label{eq:homodyne_time_scaling}
\end{align}

This indicates that the experiment is feasible to measure $3\sigma$ level significance on the $\alpha \sim 1$ normalization of the theory, but will struggle to achieve higher a statistical power or equivalently attempt to measure and set limits on $\alpha < \frac{1}{3}$. In practice, the signal is far to small below the noise to properly subtract the shot-noise as a background, thus two interferometers are needed and the data must be cross correlated as derived in \cref{s:correlation}. Using two interferometers at the same power as above will cut the measurement time in half.

\begin{figure}[ht]
\centering
\includegraphics[width=.98\linewidth]{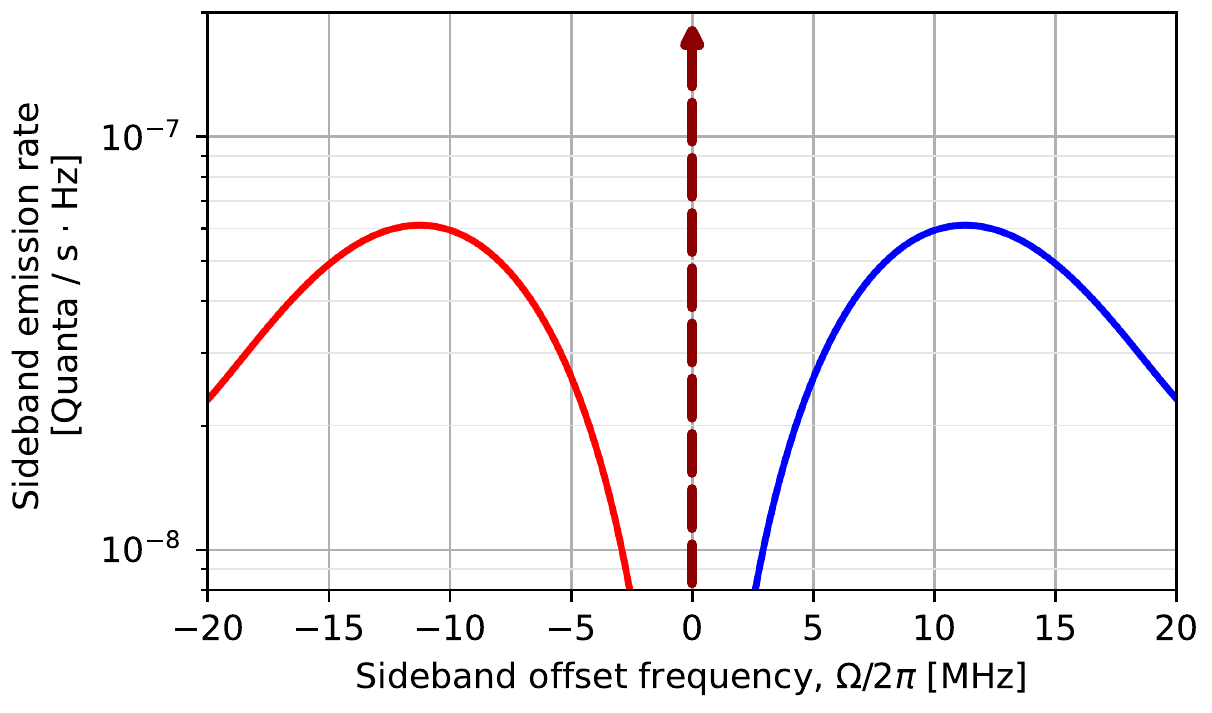}
\caption{
    \textbf{GQuEST signal sideband emission spectrum}. The model of \cref{fig:model_displacement} is converted into photon counts of the signal. The signal modulates into both an upper (blue) and lower (red) frequency sidebands. The dashed arrow in the center indicates the carrier light frequency and the potential for a large photopower from the residual Michelson fringe light.
  }
\label{fig:sb_F}
\end{figure}

\subsection{Photon Counting}

While homodyne readout will be utilized for the GQuEST experiment, a new readout methodology for Michelson interferometers will be developed and demonstrated for this experiment, vastly improving the potential of the Michelson interferometer. The new method is to utilize photon counting as a means of directly measuring the sidebands generated by the geontropic signal. \Cref{fig:sb_F} shows the principle of operation for the geontropic signal and can be related to \cref{fig:model_displacement}. In the first figure, the one-sided spectral density of shot noise has a magnitude of $1/2$ quanta, or equivalently $1/2$ quanta per second-Hz. The vacuum fluctuation has the same 1/2 quanta at all optical frequencies, but it is evenly distributed between the amplitude and phase quadratures, putting 1/4 quanta in each. The upper and lower sidebands, representing the negative and positive frequencies of the two-sided spectrum, then each contribute their 1/4 quanta of phase fluctuation to sum to the 1/2 in the one-sided spectrum of quadrature fluctuations, measured with homodyne readout.

The geontropic signal is in the phase quadrature and the shot noise magnitude then sets its normalization into its sideband flux-spectral-density, $P(\Omega)$.
\begin{align}
  P_{\vphi}(\Omega) &= \frac{ S_{\vphi}(\Omega)}{4 S_{\HqM}(\Omega)}
                              \label{eq:flux_spectral_density}
\end{align}
One can then, in principle, detect that signal photon flux when measuring over a band of frequencies.
\begin{align}
  E_{\vphi} &= \Delta T_E \int_{-\infty}^{\infty} P_{\vphi}(\Omega) \frac{\rmd \Omega}{2\pi}
\end{align}
Thus $E_{\vphi}$ is the accumulated photon number. This can be factorized into a time, bandwidth, and rate prefactor.
\begin{align}
  E_{\vphi} &= \frac{\alpha\red{\SPEC}}{4 S_{\HqM}} \Delta T_E \Delta F_E
            &
  \Delta F_{E} &= \int_{-\infty}^{\infty} \frac{\SPEC(\Omega)}{\red{\SPEC}} \frac{\rmd \Omega}{2\pi}
                         \approx \frac{2c}{L}
\end{align}
Where the approximation assumes the fiducial geontropic spectrum shape. Note that the definition of bandwidth is different for photon counting than for homodyne readout, lacking the square in the integrand compared to \cref{eq:DeltaF_Q}, and including an implicit factor of two by separately measuring the negative optical frequencies.

Still eliding classical noise backgrounds, the $1\sigma$ statistical limit is set by the average time, $\red{T}_E^{1\sigma}$, to observe a single photon, leading to a similar decomposition into a effective number of required measurements:
\begin{align}
  1 = N_E^{1\sigma} &\equiv \Delta \red{T}_E^{1\sigma} \Delta F_E = \frac{4 S_{\HqM}}{\alpha\red{\SPEC}}
                  \label{eq:counting_SNR_example}
\end{align}
Using a readout that only captures $\Delta F_{E}^{\text{read}} < \Delta F_{E}$ of the available bandwidth, centered near the peak of the spectrum $\Omega_{\text{pk}}$, this leads to a measurement time for a 1-sigma significance measurement
\begin{align}
  \Delta \red{T}_E^{1\sigma}
  &\approx 0.1
  \text{ s} \cdot \bigg(\frac{\GQL}{L}\bigg) \bigg(\frac{\GQP}{P_{\text{BS}}}\bigg)
    \bigg(\frac{F_{E}}{\Delta F^{\text{read}}_{E}}\bigg)
  \\
  &\approx 4
  \text{ minutes} \cdot \bigg(\frac{\GQL}{L}\bigg)^2 \bigg(\frac{\GQP}{P_{\text{BS}}}\bigg)
  \bigg(\frac{50\text{kHz}}{\Delta F^{\text{read}}_{E}}\bigg)
  \label{eq:counting_time_scaling}
\end{align}
The time constants show the powerful promise of the photon counting technique, even if the accessible signal bandwidth is used inefficiently. Counting has fundamentally better performance when searching for a stochastic signal when quantum noise is otherwise the dominant experimental background, as it removes quantum noise as a background. The detection rate is still determined by a quantum limit. In doing so, the scalings change from \cref{eq:homodyne_time_scaling} to \cref{eq:counting_time_scaling}, which substantially accelerates the search, accepts lower power interferometers, and presents a more favorable scaling in the instrument length for university-scale tabletop implementation.

Theoretically, photon counting shows extreme promise. Experimentally, however, saturating the statistical potential requires measuring signal photons with a background rate less than $1/\Delta T_E^{1\sigma}$. Considering using an interferometer circulating kilowatts of power, this will necessitate extreme dynamic range isolation between the circulating carrier power and signal sideband field frequencies. Thus, the new readout technique must also drive the experiment design, leading GQuEST to incorporate some unique elements required to use photon counting.

Including classical noise and a background count-rate, $\red{B}$, modifies the scaling
\begin{align}
  \Delta T_E^{1\sigma}
  &=
    \Delta \red{T}_E^{1\sigma}
    \left(1 + \frac{\PSD{\text{Cl}}}{\red{\SPEC}} + \red{B}\Delta T_E^{1\sigma}  \right)
\end{align}

The photon counting section, \ref{sec:counting}, formalizes the arguments above, and includes the classical noise background.
\cref{eq:counting_SNR_example} can be constructed from \cref{eq:photon_counting_G_VAR} or its exact counterpart \cref{eq:Counting_Int}. The derivations relating the statistical expressions and the simplified expressions of this section are given in \cref{D:application_relations}.

Given this considerable acceleration of the search, it is worth more carefully analyzing the relationship between standard time-series stochastic search and photon counting. The following section develops how signals are transduced in an interferometer. Later sections then analyze the quantum measurement and search statistics between homodyne readout and photon counting.

\section{Interferometer Description}\label{sec:descriptions}

This section establishes the quantum theory behind optical interferometers, applicable to either measurement scheme. This establishes the normalizations and conventions used throughout the work.  An input-output formalism in the Heisenberg picture is used. The interferometer late-time output field operators are first described in terms of pre-interaction input operators. Output operators are then assembled into readout observables for analysis. The interferometer output is a function of time (or frequency) and must be reduced into a set of independent quantum states (or density operators). These independent states are aggregated into large-N statistical measurements. The formalism later shown for the homodyne observables is equivalent to the Fourier-domain two-photon formalism\cite{cavesPRA85NewFormalism, schumakerPRA85NewFormalism, corbittPRA05MathematicalFramework, danilishinLRR12QuantumMeasurement} but analyzed in a framework suitable for finite time measurements. That framework is then extended to describe photon counting.


\begin{figure}
  \centering
  \includegraphics[width=0.95\linewidth]{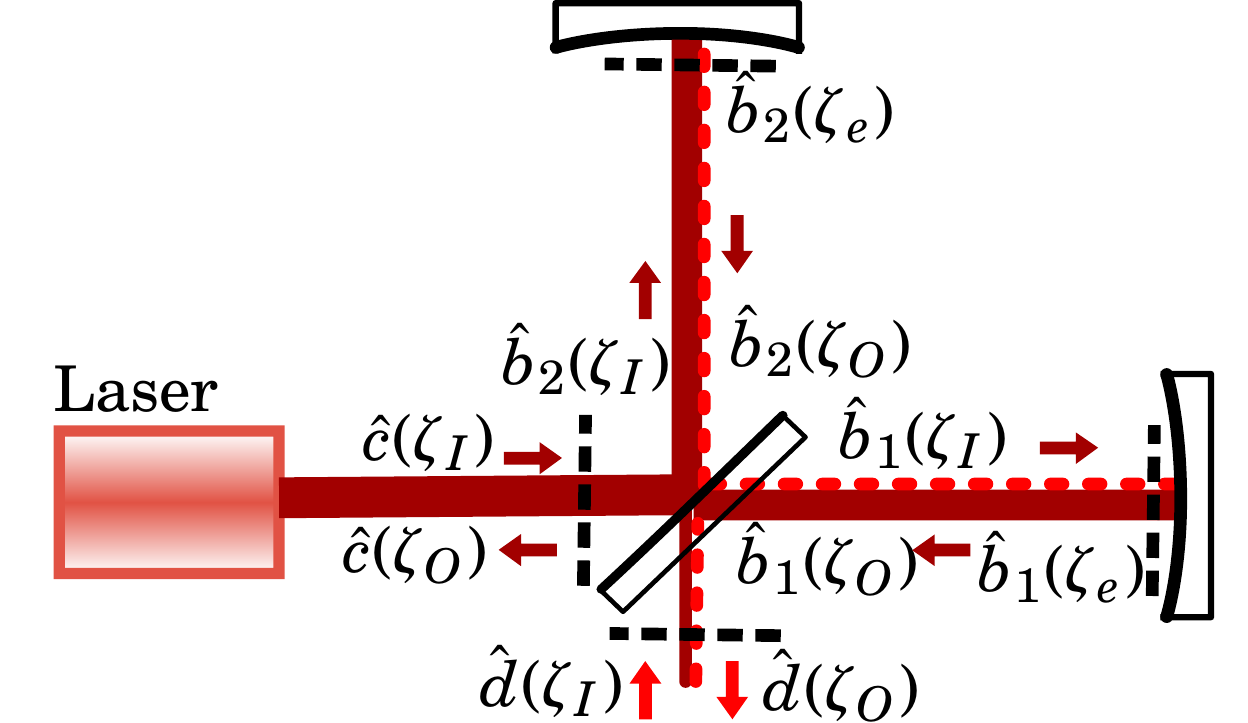}
  \label{fig:IFO_simple},
  \caption{
  Depiction of the fields and variable definitions in a Michelson. The fields are indexed by time (not depicted) as well as the abstract coordinate parameterization $\zeta$. The black dashed lines indicate specific reference planes for $\PV_I$ and $\PV_O$, depending on direction, and for $\PV_e$ at the end mirror of each arm.
 }
\end{figure}

The starting point to describe the interferometer are the input-output relations depicted by \cref{fig:IFO_simple}. The input carrier light at the common-mode port of a Michelson is considered to be from a monochromatic laser source with a static field amplitude of $\red{c}$. Field is expressed in flux units for convenience, so $P_{\text{BS}} = \hbar \omega |\red{c}|^2$ represents the power on the beamsplitter. $\omega$ is the angular frequency of the carrier light. The quantum field operator for the differential port $\qo{d}$, which outputs the signals, is built similarly. It does not include a static field contribution. The lettering for the c and d fields conveniently corresponds to common and differential ports of the interferometer. Along with static field, the inputs into the common and different ports carry quantum fluctuation terms $\qo{c}_I(\tau)$ and $\qo{a}(\tau)$. The interferometer fields are expressed in a rotating frame with respect to their quantum fluctuations and static fields as:
\begin{align}
  \qo{c}(\PV_I, \tau) &= e^{i\omega \tau}\left(\red{c} + \qo{c}_I(\PV_I, \tau)\right)
  \\
  \qo{d}(\PV_I, \tau) &= e^{i\omega \tau}\qo{a}(\PV_I, \tau).
\end{align}
The symbol $\qo{a}(\tau)$ is used for the differential port quantum fluctuation bath term due to its standard usage and relations as a bosonic ladder operator. The quantum fluctuation terms are bosonic bath operators quantization with the canonical commutation relations
\begin{align}
    [\qo{a}(t), \qo{a}^\dagger(t + t')] &= \delta(t')
\end{align}
As such, they do not define quantum states, but rather state density. States can only be defined within suitable integrals, established in the next section.

To analyze interferometer response and noise, the field operators are typically manipulated in the frequency domain. The frequency argument $\Omega$ is expressed as an offset from the carrier frequency $\omega$. This frames it as both the frequency argument for the readout, as well as the frequency offset of sideband fields modulated from the carrier. 
\begin{align}
  \qo{c}(\PV_I, \Omega + \omega) &= \red{c}\delta(\Omega) + \qo{c}_I(\Omega)
  \\
  \qo{a}(\PV_I, \Omega + \omega) &= \qo{a}(\Omega) 
\end{align}
Where $\delta(\Omega)$ is the Dirac delta function.

The input fields at their respective reference plane then interact with the beamsplitter, are converted into the arm fields, propagate to and from the endmirrors as ${\PV_I \rightarrow \PV_e \rightarrow \PV_O}$, then interfere at the beamsplitter. They become the output fields according to:
\begin{align}
  \qo{b}_{\armY}(\tau, \PV_I)
  &= \tfrac{1}{\sqrt{2}}\qo{c}(\tau, \PV_I)
    + \tfrac{1}{\sqrt{2}}\qo{d}(\tau, \PV_I)
  \\
  \qo{b}_{\armX}(\tau, \PV_I)
  &= \tfrac{1}{\sqrt{2}}\qo{c}(\tau, \PV_I)
    - \tfrac{1}{\sqrt{2}}\qo{d}(\tau, \PV_I)
    \\
  \qo{d}(\tau, \PV_O)
  &= \tfrac{-1}{\sqrt{2}}\qo{b}_{\armX}(\tau, \PV_O)
    + \tfrac{1}{\sqrt{2}}\qo{b}_{\armY}(\tau, \PV_O)
  \\
  \qo{c}(\tau, \PV_O)
  &= \tfrac{1}{\sqrt{2}}\qo{b}_{\armX}(\tau, \PV_O)
    + \tfrac{1}{\sqrt{2}}\qo{b}_{\armY}(\tau, \PV_O)
\end{align}
Searches for new physics rely on the field in the arms being perturbed with phase or amplitude modulations from external fields. Here, we assume that new physics will act to modify the dispersion relation - at each point in space as a function of time - for the traveling light. It can also modulate the travel time through the metric. Both kinds of perturbations correspond only to phase modulations. They depend on the values and history of any fields that modulate the optical field frequency $\omega(\tau, \PV) \simeq \omega$, wavenumber ${k(\tau, \PV) \simeq \omega/c \equiv k}$ and metric, $\mathcal{G}$, according to:
\begin{align}
  \theta_{\armX}(\tau) &= \int_{\PV_I}^{{\PV_O}} \left( \omega(\tau, \PV) \frac{\rmd t_{\armX}}{\rmd \PV} - k(\tau, \PV) \left| \frac{\rmd x_{\armX}}{\rmd \PV} \right| \right) \rmd \PV + 2 k \Delta L_{\armX}
                         \label{eq:b_phase}
  \\
  T_{\armX}(\tau) &= \int_{\PV_I}^{{\PV_O}}  \sqrt{\mathcal{G}^{tt}(\tau, \PV)}\frac{\rmd t_1}{\rmd \PV} \rmd \PV
                    \label{eq:b_delay}
\end{align}
via the input-output relation
\begin{align}
  \qo{b}_{\armX}(\PV_O, \tau)
                       &= e^{i\theta_{\armX}(\tau)}\qo{b}_{\armX}(\PV_I, \tau - T_{\armX}(\tau))
                         \label{eq:b_phi_t}
\end{align}
Above, $\Delta L_{\armX}$ expresses the local displacement of an endmirror, factored separately from the parameterized path $x_{\armX}(\PV, \tau)$. New fields, $\varphi(\tau, x, y)$, can drive the phase shift from the dispersion relation as $\omega(\tau) - ck(\tau) \propto \varphi(\tau)$. For gravitational physics, the metric can be viewed as coherently driving a combination of the dispersion relation, delay time, and local displacements, depending on the coordinate parameterization and gauge choice \cite{rakhmanovPRD05ResponseTest}.
The specifics of these expressions are not necessary for the present analysis, though they are essential for specific science-case studies. They are indicated here because their form as time-integrals motivates changing into the frequency domain to simplify using linear relations.

The variations of $T_{\armX}$ and $\theta_{\armX}(\tau)$ combine to create the dynamic phase shift $\tilde{\theta}_{\armX}(\tau)$ and the static time delay $\red{T}_{\armX}$.
\begin{align}
  \tilde{\theta}_{\armX}(\tau) &\equiv \theta_{\armX}(\tau) - \omega\big(T_{\armX}(\tau) - \red{T}_{\armX}\big)
\end{align}
The static delay $\red{T}_{\armX}$, implicitly defined above, does not depend on the time $\tau$ and will be used in following expressions.

Taking the Fourier transform of all components and simplifying in the small signal $\tilde{\theta}_{\armX}(\tau) \ll \pi$ limit then gives:
\begin{align}
  \qo{b}_{\armX}(\omega + \Omega, \PV_O)
  &\approx
    \frac{
    \left(\delta(\Omega) + i\fvar{\theta}_{\armX}(\Omega) \right)
    \label{eq:b_after_fourier}
        \circledast  \qo{b}_{\armX}(\omega + \Omega, \PV_I)}{e^{i (\omega + \Omega)\red{T}_{\armX}}}
\end{align}
Where $\circledast$ indicates convolution of $\Omega$. The $\qo{b}$ field operators are assumed to be sourced by perfectly monochromatic light, delta functions in frequency space, making the convolution trivial for classical fields. Assuming the modulations are small, frequency intermodulations of the quantum fields are ignored.

Expanding these relations for both arms of the Michelson equations, the output of the interferometer then generally takes the form
\begin{align}
  \qo{d}(\Omega) &\equiv \qo{d}(\PV_O, \Omega + \omega)
                   \\
  &= i l \delta(\Omega)  + i g(\Omega) \cl{\vphi}(\Omega)
    + i \VNP(\Omega) + \VNA(\Omega) + \qo{a}(\Omega)
    \label{eq:d_omega_terms}
\end{align}
This form carries five terms each representing different contributions. All of the contributions implicitly scale with the carrier field $\red{c}$, except for the quantum fluctuation terms $\qo{a}$.

The first term, $i l \delta(\Omega)$ represents the light of the Michelson fringe, where the interferometer is operated nearly at a ``dark fringe'' of destructive interference.
\begin{align}
  l &= (e^{i\omega \red{T}_1} - e^{i\omega \red{T}_2}) \frac{\red{c}}{2i} \approx \sin\left(\omega\tfrac{\red{T}_1 - \red{T}_2}{2} \right)\red{c}
      \label{eq:dark_fringe}
\end{align}
The fringe light level is a delta function over frequency, as the input light is assumed to be unvarying and the arm length offset creating the fringe is assumed to be under feedback control. Here, $l/\red{c}$ is the fraction of the carrier field passing from the input to the output due to a small offset in the arm lengths. Operating at nearly-dark fringe, this fraction is assumed to be small.

The second term, $i g(\Omega) \cl{\vphi}(\Omega)$, represents the coupling of external signals, $\cl{\vphi}(\Omega)$, into the phase quadrature of the light in the frequency domain. The down-hat notation for $\cl{\vphi}$ indicates that it is a classical signal with statistical (rather than quantum) fluctuations, with statistics to be later specified. From the mathematical assembly established for the Michelson, these terms can be built:
\begin{align}
  g(\Omega) \cl{\vphi}(\Omega)
  &\equiv
    \left(
    \fvar{\theta}_{\armX}(\Omega) - \fvar{\theta}_{\armY}(\Omega)
    \right) \frac{\red{c}}{2}
\end{align}
The factor $g(\Omega)$ represents the frequency dependence of the signal coupling and the gain of the interferometer.
The Michelson so far depicted is broadband and does not have any frequency dependence. The generalization to $g(\Omega)$ is established to consider additional signal recycling Fabry-Perot cavities to be considered later in \cref{sec:sensitivity_comparison}. Interferometer signals are typically referred to their equivalent differential arm length displacement. Differential arm displacement signals move each arm by $\pm\cl{\vphi}(\Omega)/2$ and a path length change in either arm changes the phase by $2 k$, together, the gain in a broadband interferometer is:
\begin{align}
  g(\Omega) &= \red{g} \equiv k \red{c} = k\sqrt{\frac{P_{\text{BS}}}{\hbar\omega}} = 2k\sqrt{\frac{P_{\text{arm}}}{2\hbar\omega}} 
              \label{eq:g_normalization}
\end{align}
Where the factor $\red{g} $ is defined to specifically refer to the broadband optical gain of a Michelson interferometer as a baseline sensitivity.

The remaining terms in \cref{eq:d_omega_terms} indicate noise processes. The third term, $\VNP(\Omega)$, indicates a coupling of classical phase noises from the carrier into the output. This can include noise processes on the laser itself, passively coupling to the output, or it represents additional processes that modulate the carrier light within the interferometer, such as thermal noises on the end mirrors. Similarly, the fourth term $\VNA(\Omega)$ represents classical amplitude-quadrature noise terms. Homodyne measurements observe only a single quadrature, so the two are expressed as separate factors. In particular, fringe readout does not observe the $\VNA(\Omega)$ term. Photon counting measures both quadratures, so both factors are relevant. As given, these terms implicitly include an optical gain and scale with $\red{c}$. This factorization, implicit in the optical gain, is chosen to simplify expressions. Like $\qo{d}(\tau)$, $\VNP(\tau)$ and $\VNA(\tau)$ are also in units of flux density, so $\VNP(\Omega)$ and $\VNA(\Omega)$ take units of flux field density $[\sqrt{\text{quanta} / \text{s}{\cdot}\text{Hz}}]$, and their squares, e.g. $\VNP^2(\Omega)$ take units of $[\text{quanta}]$ or  $[\text{quanta}/\text{s}{\cdot}\text{Hz}]$. 

Finally, the fourth term represents the reflection of the input quantum noise-density bath operator to the output port. Notably, since the interferometer is operated at the dark fringe, $\qo{c}_I(\Omega)$ does not contribute to $\qo{d}$. Only the operator $\qo{a}(\Omega)$ is present, representing the bath input to differential port reflecting to the output. This is well understood from early analysis of squeezed states in interferometers \cite{cavesPRD81QuantummechanicalNoise}.

The interferometer acts to modulate and affect the input states. For example, the propagation time through the interferometer adds a phase shift $e^{-i \Omega (\red{T}_{\armX} + \red{T}_{\armY})/2}$. For input vacuum states, the output states are also vacuum, unless modulated by physics signals. Thus, to zeroth order, interferometer effects on input vacuum states are negligible and effect-terms are omitted on $\qo{a}(\Omega)$ for simplicity. While this work assumes vacuum state's are (passively) injected, analyzing engineered non-vacuum states requires additional terms in following derived expressions.

\subsection{Signal and Noise Processes}\label{sec:processes}

This work primarily considers time-invariant stochastic sources for both its signal and classical noise processes. By being stochastic, the signal and classical noise carries no coherent value and no well-defined phase. First-moment expectation values are zero. Second moments indicate the noise power, and the signal and noises are assumed to have Gaussian statistics in the frequency domain. Together, these properties are given by
\begin{align}
  \braket{\cl{\vphi}(\Omega)}
  &= 0
  &
    \braket{\cl{\vphi}(\Omega)\conj{\cl{\vphi}}(\Omega + \Omega')}
  &\equiv
    S_{\cl{\vphi}}(\Omega)\pi\delta(\Omega')
    \label{eq:phi_expectations}
  \\
  \braket{\VNP(\Omega)}
  &= 0
  &
    \braket{\VNP(\Omega)\conj{\VNP}(\Omega + \Omega')} &\equiv \SNP(\Omega)\pi\delta(\Omega')
\end{align}
$\cl{\vphi}$ is expected to be a real signal modulating the phase of the light. This entails the additional constraint ${\cl{\vphi}(\Omega) = \conj{\cl{\vphi}}(-\Omega)}$. There is a factor of $\pi$ rather than $2\pi$ on the angular-frequency delta function under the convention that $S_{\cl{\vphi}}(\Omega)$ represents the one-sided power spectral density of the signal.

The one-sided convention is standard in interferometer and signal processing liturature. It reflects that the positive and negative frequencies in the power spectrum are redundant -- not only that their expectation is the same, but also that they are perfectly correlated from the Hermitian symmetry of Fourier transforms of real signals or real observables. Thus, only positive frequencies are needed in signal processing, and the factor of two accounts for the negative frequencies.  This detail will be relevant in the Homodyne derivations and is relevant for photon counting. There, both positive and negative sideband fields are created from the positive and negative components of the power spectrum and can be independently detected.

Classical statistics can be defined using the expectation. In particular, the variance is expressed as,
\begin{align}
  \VAR{\qo{O}} \equiv \Braket{\big(\qo{O} - \braket{\qo{O}} \big)^2} = \braket{\qo{O}^2} - \braket{\qo{O}}^2
\end{align}
This definition can be applied to either quantum $\qo{O}$ or classical, $\cl{O}$ operators. Notably, this definition does not replicate two-point expectations such as \cref{eq:phi_expectations}, as they include two indexing variables $\Omega$ and $\Omega'$. As density operators, the signal and noise stochastic variables have singular correlation functions. Observables constructed from integrals of these density operators can be expected to be Gaussian from the central limit theorem, and, with basis-weighted integrals introduced in the next section, the full distributions and characteristic functions of these classical signals can be defined (cf. \cref{D:Gaussian_Props}) for use in derivations. 

The statistics for the quantum noise arise from the canonical commutation relations and the use of coherent fields. Weak stochastic signals move the large coherent fields circulating in the interferometer into the differential, output port, acting as a quantum ``displacement'' operation on the quantum noise operators. In the Heisenberg formulation here, this is simply the addition of the many classical sources as c-numbers to the quantum term in \cref{eq:d_omega_terms}. The statistics of the quantum terms will impact the calculations in \cref{sec:homodyne} and \cref{sec:counting}. In the Homodyne case, they will appear Gaussian, while in the counting case will appear as a discrete nearly Poisson distribution, with less variance. This is the major result emphasized in this work, indicating a means to accelerate searches for weak, deeply sub-shot-noise signals.
\subsection{Optimal Search}

Searches for new physics are framed here as testing for the existence of a stochastic signal process $\cl{\vphi}$. They can otherwise be framed as establishing posterior probability distributions for parameters that determine that shape and scale for $S_{\cl{\vphi}}(\Omega)$. For simplicity, this work only considers the overall scale as a test parameter. This leads to a decomposition of the spectral density of the signal into the scale, $\param$, and the morphology $\SPEC(\Omega)$
\begin{align}
  S_{\cl{\vphi}}(\Omega) \equiv \param \SPEC(\Omega)
\end{align}
Thus the problem is to design optimal unbiased estimators for the noise power scale parameter, $\param$. The optimal experimental and statistical design will be influenced and affected by the spectral shape of $\SPEC(\Omega)$. New physics searches can generally be considered into wide-band and narrow-band searches for stochastic signals, depending on how the support of $\SPEC(\Omega)$ compares to the physical scales of the interferometer. Those cases are analyzed in \cref{sec:sensitivity_comparison}.

The narrowband or wideband forms for stochastic signals are suitable for a number of searches for new physics. They are specifically amenable to excess power searches using chi-square statistics. These statistics will be shown to naturally occur for Homodyne readouts on stochastic signals with the expectations above, and have a direct counterpart in photon counting. Notably, these forms for signals and this analysis is not immediately appropriate for matched-template searches, such as those for gravitational wave signals from binary inspiral coalescence. Such signals can be considered incoherent from having an unknown arrival time, and can be found by searching for excess noise power in a time-resolved manner; however, their optimal search is not from a chi-square statistic. The relevance of photon counting techniques for matched-template physics and waveform estimation is considered in the outlook and derived in \cref{sec:events}.

The wideband signal search case is assumed for approximations in the following sections. There,  the value $\red{\SPEC}$ is used to indicate the maximum value of the spectrum over a frequency region over which $\SPEC(\Omega)$ is approximately constant. When  $\red{\SPEC}$ is used to indicate a ``typical'' sensitivity over some band, $\red{S}_{\VNP}$ is used to express an average or representative noise level in the same band. 

The reduction of the search problem to a single scale parameter $\alpha$ allows a direct comparison of the techniques by calculating the large-N statistics of homodyne and counting searches. The statistical performance of those searches is then compared to their corresponding classical Fisher information. This comparison indicates that the data analysis is fully efficient in either case, but that the choice of quantum observable to photon counting provides a more sensitive search in certain scenarios.

\subsection{Temporal-Mode Basis Description}

Interferometer response and physical signals are generally best represented in the frequency domain. The two-photon formalism \cite{cavesPRA85NewFormalism, schumakerPRA85NewFormalism, corbittPRA05MathematicalFramework, danilishinLRR12QuantumMeasurement} is developed to describe homodyne quantum observables and describe power spectra. Here, it is re-framed using finite-duration basis-modes rather than the Fourier transform. The basis is later chosen to use Fourier-like functions, relating the general basis description to the typical frequency-domain representation. This navigates several subtleties: first, seemingly non-Hermitian operators from the complex Fourier transform can be expressed as linear combinations of a real temporal-mode basis equivalent to cos and sin transforms. Second, density operators can be converted into explicit states, which are mathematically better behaved while having fewer integrals in expressions. Third, interferometer gravitational-wave signal analysis is often framed in the language of matched templates can lead to a preferred choice of basis for statistical tests to discriminate between waveform candidates. The formulation here expresses a physical counterpart to classical post-detection signal analysis, motivating future developments towards time and frequency resolved quantum interferometry outlined later in \cref{sec:outlook}. 

The classical and quantum noise sources so far have been given as singular density operators. Physically, they can only be employed in a measurement that integrates over time. We can define a single state measurement as
\begin{align}
  \qoa &= \int_{-\infty}^{\infty} \K(t)\qo{a}(t)\rmd t
\end{align}
where the temporal mode $\K(t)$ has the normalization property ${1 = \int_{-\infty}^{\infty} |\K(t)|^2 \rmd t}$. For simplicity, we will also only consider basis modes which have no overlap with DC (static) signals, implying the constraint ${0 = \int_{-\infty}^{\infty} \K(t) \rmd t}$. Quantum operators defined in this manner will then obey the canonical commutation relations based on the density relations of the bath operators
\begin{align}
  [\qoa, \qoa^\dagger] &= 1 &&\leftrightarrow& [\qo{a}(t), \qo{a}^\dagger(t + t')] &= \delta(t')
\end{align}

Now consider a series of measurements from a basis set $\mathcal{K}_N = \{\K_{+i}\}$. The temporal-modes, $\K_{+i}$ within this set are considered to have an orthogonality property
\begin{align}
  \int \K_{i}(t)\K_{j}(t) \rmd t \le \epsilon \approx 0 \text{ for } i \ne j
  \label{eq:orthogonality}
\end{align}
Where $\epsilon$ bounds the overlap terms. It is assumed to be arbitrarily small for simplicity. The basis set leads to a set of independent temporal-mode operators
\begin{align}
  \qoa_{i} &= \int \qo{a}(t)\K_{i}(t) \rmd t
  &
  \qoa_{i} &= \int \qo{a}(\Omega)\K_{i}(-\Omega) \rmd \Omega
\end{align}
To simplify the following discussion and formulas, we will choose a specific basis set $\red{\mathcal{K}}_N = \{\rK_{+i}\}$ to be Fourier-like modes over the frequencies $\Omega_{+i}$.
\begin{align}
  \rK_{\pm i}(t) &= \lim_{\Delta T \rightarrow \infty} \frac{e^{\pm i \Omega_{i} t - t^2/4T^2}}{\Delta T^{1/2} (2\pi)^{1/4}}
  \\
  \rK_{c i}(t) &= \sqrt{2}\lim_{\Delta T \rightarrow \infty} \frac{\cos(\Omega_{i} t )e^{- t^2/4T^2}}{\Delta T^{1/2} (2\pi)^{1/4}}
\end{align}
There is a similar basis waveform, elided, for the sin phasing $\rK_{s i}(t)$.
Note here that positive and negative frequency basis modes are orthogonal, implying independent measurements. These basis waveforms can be seen as taking a Fourier transform of the $\qo{a}(t)$ bath operators. Due to their normalization, these basis functions are more closely related to Fourier series than the Fourier transform, relevant for measuring a finite time interval. The limiting process indicates they are considered arbitrarily narrow in frequency, but represent a finite set of basis temporal modes from finite time and finite observation bandwidth.

The orthonormality of the basis set is necessary to define independent measurements, while the finite duration prevents the appearance of delta functions in expectations. Only sufficiently separated frequencies, $\Omega_i$ can obey the orthogonality condition. For bandwidth, $\Delta F = \max(\Omega_i/ 2\pi) - \min(\Omega_i/ 2\pi)$, over a measurement duration, $\Delta T$, the maximum number of orthogonal basis modes is $N = 2\Delta F \Delta T$, accounting either for the positive and negative complex modes or the independence of sin and cos.

Basis sets such as these enable the following notation to treat quantum and classical noise calculations as discrete states.
\begin{align}
  \qoa_{\pm i} &\equiv \int \qo{a}(t) \rK_{\pm i}(t)\rmd t
  &
               \qoa_{\pm i}^\dagger &\equiv \int \qo{a}^\dagger(t) \conj{\rK}_{\pm i}(t)\rmd t
                                      \\
             \delta_{ij} &= [\qoa_{\pm i}, \qoa_{\pm j}^\dagger] 
                                              \label{eq:a_i_commutator}
  &
    0  &= [\qoa_{\pm i}, \qoa_{\mp j}^\dagger]
  \\
  \delta_{ij} &= [\qoa_{c i}, \qoa_{c j}^\dagger] 
  &
    0  &= [\qoa_{c i}, \qoa_{s j}^\dagger]
\end{align}
Where $\delta_{ij}$ is the discrete delta matrix. Higher moments of these operators are now simpler to define, as they avoid delta-function singularities. The classical noise operators are similarly simplified.
\begin{align}
  \cl{\vphi}_{\pm i} &\equiv \int \cl{\vphi}(t) \rK_{\pm i}(t)\rmd t
  \\
  \braket{\cl{\vphi}_{+i}\conj{\cl{\vphi}}_{+j}} &= \braket{\cl{\vphi}_{+i}{\cl{\vphi}}_{-j}}  \approx S_{\cl{\vphi}}(\Omega_{+i})\delta_{ij}
\end{align}
and 
\begin{align}
  \VAR{\cl{\vphi}_{ci}} &\approx \frac{S_{\cl{\vphi}}(\Omega_{+i})}{2} \approx \VAR{\cl{\vphi}_{si}}
\end{align}

The approximations above arise from the more exact expression 
\begin{align}
  S_{\cl{\vphi}}(\Omega_{+i})
  &\approx
    \Delta T \int_{\Omega_{+i}-\frac{1}{2\Delta T}}^{\Omega_{+i}+\frac{1}{2\Delta T}}
    S_{\cl{\vphi}}(\Omega)\frac{\rmd \Omega}{2\pi}
    \label{eq:S_idx_expr}
\end{align}
With the integral arising as an approximation to the frequency-domain expression of $|\rK_{\pm i}|^2$.
The integral form is necessary when the spectral density itself includes narrow-frequency delta-like features. When a basis set is used that saturates the time-bandwidth product, sums can be converted into integrals:
\begin{align}
  \sum_{i=1}^{N}
  S_{\cl{\vphi}}(\Omega_{+i})
  &\approx
    \Delta T \int_{\Omega_{1}}^{\Omega_{+N}}
    S_{\cl{\vphi}}(\Omega)\frac{\rmd \Omega}{2\pi}
    \label{eq:S_idx_expr_full}
\end{align}
Which can be used simplify later basis-mode expressions to common integral expressions arising in signal analysis.

A noteworthy subtlety in the relationship of this basis to Fourier transforms is the act of conjugation. By this definition, the conjugation of a basis-mode operator also flips the sign of the frequency, such that $\qoa_{+i} \propto \qo{a}(\Omega_{+i})$ while $\qoa_{+i}^\dagger \propto \qo{a}^\dagger(-\Omega_{+i})$. This subtlety is well established in the two-photon formalism, and will appear when calculating Homodyne-based observables.

\section{Homodyne Readout}\label{sec:homodyne}
Consider first a continuous measurement of the optical power, which appears with the form of a number operator ~\footnote{This form functions for this analysis, but is given a better physical definition in \cref{D:band_limit}}.
\begin{align}
  \qo{P}_{d}(t) = \qo{d}^\dagger(t)\qo{d}(t)
  \label{eq:P_d_form}
\end{align}
Consider now the power operator in the frequency domain for the output port of the interferometer.
It takes the concise form of a number operator using a convolution, mixing frequencies
\begin{align}
  \qo{P}_d(\Omega) = \qo{d}(\Omega) \circledast \qo{d}^\dagger(\Omega)
\end{align}
applying the form assumed for $\qo{d}$ from \cref{eq:d_omega_terms} gives:
\begin{align}
  \qo{P}_d(\Omega)
  &=
    |l|^2\delta(\Omega) + \qo{\mathcal{P}}_1(\Omega) + \qo{\mathcal{P}}_2(\Omega)
    \label{eq:power_factorized}
\end{align}
Composed of a static fringe-field term and two fluctuation terms:
\begin{align}
  \qo{\mathcal{P}_1}(\Omega)
 &\equiv (g(\Omega)\cl{\vphi}(\Omega)\conj{l} + \conj{g}(-\Omega)\conj{\cl{\vphi}}(-\Omega)l)
   \label{eq:power_factor_1}
\\ \nonumber
  &(\VNP(\Omega)\conj{l} + \conj{\VNP}(-\Omega)l) + (i\qo{a}(\Omega)\conj{l} - i\qo{a}^\dagger(-\Omega)l)
  \\ \nonumber
  & + (i\VNA(\Omega)\conj{l} - i\conj{\VNA}(-\Omega)l)
\end{align}
and
\begin{align}
  \qo{\mathcal{P}_2}(\Omega)
  &\equiv
    \int_{-\infty}^{\infty}\big(
    \qo{a}(\Omega) + g(\Omega)\cl{\vphi}(\Omega) + \VNP(\Omega) + \VNA(\Omega)
    \big)^\dagger
    \label{eq:power_factor_2}
    \\&\hspace{2em}\cdot \nonumber
    {\big(
  \qo{a}(\Omega') + g(\Omega')\cl{\vphi}(\Omega') + \VNP(\Omega') + \VNA(\Omega')
    \big)}\rmd \Omega'
\end{align}

These two terms represent the linear and quadratic contributions to the photopower at the detector with respect to the signal and noise terms. The typical operation of the Michelson is to choose $l$ sufficiently large that quantum and statistical fluctuations from the $\mathcal{P}_2$ term are strongly subdominant to those from $\mathcal{P}_1$.  The remainder of this section uses $\mathcal{P}_2(\Omega) \ll \mathcal{P}_1(\Omega)$ to implement this limit, while the following section for photon counting uses the opposite limit . That assumption is the basis of the using the Michelson fringe as a homodyne readout. For general homodyne readouts, the local oscillator is determined by a complex $l$, where the phase determines the readout quadrature. The form of conjugations and negative frequencies in $\mathcal{P}_1$ is the foundation of the two-photon formalism. Michelson interferometers, using fringe light as a homodyne readout, are constrained to have real $l$ when operating near the dark fringe, per \cref{eq:dark_fringe}. This enforces phase-quadrature readout that is sensitive to arm length variations. The expression $l \rightarrow \red{l}$ is used where $l$ is assumed to be purely real.

To simplify expressions, we can assume an interferometer with no cavities, or at least with cavities that are on resonance. For this case, the optical gain is expressed as $g(\Omega) \rightarrow \red{g}(\Omega)$, which assumes the constraint $\red{g}(\Omega) = \conj{\red{g}}(-\Omega)$. This will allow future expressions to use the concise factor $|2\red{g}(\Omega)|$. This factor can be expanded to return to the general case
\begin{align}
  |2\red{g}(\Omega)|^2
  &\rightarrow \left|g'(\Omega) + \conj{g'}(-\Omega)\right|^2
\end{align}

With this particularly simple form of Fourier-like basis and the above assumptions, the measurements can be concisely expressed as
\begin{align}
  \HqM_{+i} &= \int \qo{P}_d(\Omega)\conj{\rK}_{+i}(\Omega) \rmd \Omega
              &
  \HqM_{ci} &= \int \qo{P}_d(\Omega)\conj{\rK}_{ci}(\Omega) \rmd \Omega
\end{align}
These expressions expand for the real and complex types of basis-sets:
\begin{align}
  \HqM_{+i}
  &= \big(2\red{g}(\Omega_{i})\cl{\vphi}_{+i} + 2\VNP[+i] + i\qoa_{+i} - i\qoa_{-i}^\dagger\big)\red{l}
    \label{eq:M_fourier}
    \\
    \HqM_{ci}
  &= \big(2\sqrt{2}\REof{\red{g}(\Omega_{i})\cl{\vphi}_{+i} + \VNP[+i] } + i\qoa_{ci} - i\qoa_{ci}^\dagger\big)\red{l}
    \label{eq:M_sincos}
\end{align}
With analogous definitions for $\HqM_{-i}$ and  $\HqM_{s i}$. Note that this reduced expression uses the simplification $\VNP[+i] = \VNP[-i]^\dagger$, which assumes that $\VNP(t)$ is a real signal representing phase quadrature noise. This simplification cannot apply to $\qoa_{-i}^\dagger$, so the complex temporal modes contain two quantum operators, unlike the real basis modes.

The relationship between the sin/cos and complex temporal modes entails the relations
\begin{align}
  \HqM_{\pm i} &= \frac{\HqM_{c i} \pm i\HqM_{s i}}{ \sqrt{2}} & \HqM_{\pm i}^\dagger &= \HqM_{\mp i}
\end{align}
So far, our interferometer computations are strictly using a Heisenberg picture of quantum mechanics, transforming asymptotically early operators, $\hat{a}$ and $\hat{c}_I$ into the measurable, post-interaction and asymptotically late operator $\qo{d}$. We have not explicitly computed the state at the output of the interferometer, and we can only take expectation values for the output state using our output operators. Nevertheless, we construct the probability distribution, $\PDF{\HpM, \HqM_{c i}}$, of single Hermitian measurements such as $\HqM_{c i}$ by using the expression of $\qo{d}$ via $\qo{a}$, through the formal definition
\begin{align}
  \PDF{\HpM, \HqM_{c i}}
  &= \braket{\delta(\HpM - \HqM_{c i})}_{q}
         \\
  &= \frac{1}{2\pi}\int_{-\infty}^{\infty} e^{i \HpM \Hpm} \PCF{\Hpm, \HqM_{c i}}\rmd \Hpm
\end{align}
The term, $\PCF{\Hpm, \HqM_{c i}}$, is the characteristic function of the probability. The subscript q on the expectation is that only the quantum expectation (equivalently, trace) is being taken over $\qo{a}$, and the classical variables are maintained.
This expression can be computed (cf. \cref{D:PCF_Gaussian}) and expanded
\begin{align}
  \label{eq:PCF_Gaussian}
  \PCF{\Hpm, \HqM_{c i}}
  &\equiv
    \Braket{e^{-i \HqM \Hpm}}_q
    \\
  &=
    e^{i2\sqrt{2}\REof{\red{g}(\Omega_{c i})\cl{\vphi}_{c i} + \VNP[c i]}\red{l}\Hpm - \frac{\Hpm^2}{2}\red{l}^2}
\end{align}
The computation uses the Baker–Campbell–Hausdorff formula to normally order the exponential function into independent left and right exponentials of $\qoa_{c i}$ and  $\qoa_{c i}^{\dagger}$. The derivation ultimately shows that quantum noise in homodyne readout has the Gaussian distribution for either cos or sin modes:
\begin{align}
  \PDF{\HpM, \HqM_{c i}}
  & =
    \frac{1}{\red{l}\sqrt{2 \pi}}e^{-\frac{1}{2\red{l}^2}(2\sqrt{2}\REof{\red{g}(\Omega_{i})\cl{\vphi}_{+i} + \VNP[+i] }\red{l} - \HpM)^2}
\end{align}
The expectation for $\cl{\vphi}_{c i}$ and $\VNP[c i]$ was not taken, so these random variables show to first order. They shift the mean of the distribution in either direction, and are difficult to distinguish from the randomness inherent to the quantum noise distribution.

The complex basis of the $\HqM_{+i}$ require more care to construct the distribution, as they are not Hermitian. They are instead benignly composed of two Hermitian observables, the sin and cos modes. The complex distribution must then be composed separately of the real and imaginary components of the temporal mode observation.
\begin{align}
  \PDF{\HpM, \HqM_{\pm i}}
  &= \braket{\delta(\REof{\HpM - \HqM_{\pm i}})\delta(\IMof{\HpM - \HqM_{\pm i}})}_{q}
  \\
  &= 2 \PDF{\REof{\HpM}, \frac{\HqM_{c i}}{\sqrt{2}}}\PDF{\IMof{\HpM}, \frac{\HqM_{s i}}{\sqrt{2}}}
  \\
  & = \frac{1}{\pi\red{l}^2}e^{-\frac{1}{{\red{l}^2}}|2({\red{g}(\Omega_{i})\cl{\vphi}_{\pm i} + \VNP[\pm i] })\red{l} - \HpM|^2}
\end{align}
This Gaussian distribution on a complex variable takes on a different standard deviation than for the purely real basis mode. This can be expected from the scaling of the relationship between \cref{eq:M_fourier} and \cref{eq:M_sincos}. Notably however, the scaling on the $\cl{\vphi}$ term also changes by the same amount. This shows the curious fact that the positive and negative complex basis modes each define independent measurements, leading to the commutator \cref{eq:a_i_commutator}, but measuring the complex Fourier-like mode on the homodyne signal is actually measuring both the sin and cos modes at once. This is due to \cref{eq:M_fourier} having both the positive and negative frequency quantum operators in its expression.

It's worth reviewing the multiple elements of this observable and its distribution. First, it is taken in the limit where the local oscillator field is in a specific phase such that $\red{l}$ is real and where the field strength, $\red{l}$, is larger than any arising from signal or noise terms. This causes the Michelson fringe light to implement a \emph{homodyne} observable. A homodyne measures a single quadrature of signal field with a Gaussian distribution of quantum noise arising from the distribution of vacuum's wavefunction along the quadrature $i\qoa_{c i}^{\dagger} - i\qoa_{c i}^{\dagger}$. Second, it is implementing a single measurement in a specific temporal mode, defined from Fourier-like temporal mode $\rK_{c i}$. The ``raw'' power observable, $\hat{P}_d$ represents a time-series and is defined using density operators. Integrating it with appropriately normalized basis mode implements observations of well defined and independent quantum states out of the channel $\qo{d}(t)$. Third, the distribution of $\HqM_{c i}$ shows that the mean value is directly measuring the optical field of the signal and noise $\red{g}(\Omega_{c i})\cl{\vphi}_{c i} + \VNP[c i]$.

Finally, the distribution above is scaled by the local oscillator field $\red{l}$, but so is the coefficient of $\cl{\vphi}_{c i}$. That scaling can be removed to show that the magnitude $\red{g}(\Omega_{c i})\cl{\vphi}_{c i}$ determines the size of the signal relative to quantum noise. The local oscillator magnitude, from the Michelson fringe offset, does not affect signal to noise as long as it is sufficiently large to operate in the homodyne readout regime.

\subsection{Signal Power Measurements}\label{ssec:signal_power}
Since the signals are stochastic in nature, with the expectations \cref{eq:phi_expectations}, we find that $\braket{\HqM_{c i}} = 0$ thus only the variance or ``signal power'' of each measurement can be estimated. From above, we know that the measurement can be rescaled, so the power measurement on a real basis mode is defined as:
$\qo{V}_{c i} \equiv (\HqM_{c i}/\red{l})^2$.
For the complex Fourier-like basis, the power measurement is defined as 
\begin{align}
  \qo{V}_{+i} = \qo{V}_{-i} &\equiv \frac{|\HqM_{\pm i}|^2}{|\red{l}|^2} = \tfrac{1}{2}(\qo{V}_{c i} + \qo{V}_{s i})
\end{align}
Its mean and variance contains information about the signal process, $\cl{\vphi}$.
\begin{align}
  \braket{\qo{V}_{c i}} 
  = \braket{\qo{V}_{+i}}
  &=
    2|\red{g}(\Omega)|^2S_{\cl{\vphi}}(\Omega_{c i}) +
    2\SNP(\Omega_{c i}) + 1
\end{align}
The factor $1 = \braket{|i\qoa_{c i}-i\qoa_{c i}^\dagger|^2}$ arises from quantum terms. For the real temporal mode case (ci or si subscripts), the factor of $\sqrt{2}$ is missing due to the real part carrying only half of the signal power.

These observables are constructed to measure the variance of the Gaussian observables $\HqM$, including the contribution from the signal and noise processes.
From the Gaussian distribution of $\PDF{\HpM, \HqM_{c i}}$, we can infer that the observable $\qo{V}_{c i}$ has a chi-square distributed, $\chi^2_1$, over one degree of freedom. $\qo{V}_{+ i}$, by measuring variance of both sin and cos, has a $\chi^2_2$ distribution.

For this paper, we will assume that the signal and background are small compared to the quantum noise term $2|\red{g}(\Omega)|^2S_{\cl{\vphi}}(\Omega_{c i}) + 2\SNP(\Omega_{c i}) \ll 1$.
The variance of the power observables can be calculated.
\begin{align}
  \VAR{\qo{V}_{c i}} &= 2\braket{\qo{V}_{c i}}^2 \approx 2, & \VAR{\qo{V}_{+ i}} &= \braket{\qo{V}_{+ i}}^2 \approx 1
\end{align}
The approximation symbols are used to indicate calculations in the weak signal limit.
The coefficients indicate the expected variance in relation to the mean for $\chi^2_1$ and  $\chi^2_2$ processes.

\subsection{Aggregating Measurements}
With all of the above distributions, means and variances calculated, we can then produce a statistic to combine many independent measurements. Combining measurements improves the sensitivity through the central limit theorem, allowing us to constrain the signal power $S_{\varphi}$ when it is much smaller than the quantum and background noise power.

The statistic is defined as the sum of measurements of some finite set of measurement basis modes, typically associated to a set of frequencies and time intervals for a Welch-method power spectrum estimate \cite{harrisPI78UseWindows, heinzel02SpectrumSpectral, welchITAE67UseFast} or similar technique. The expectation value of the power measurement, when the signal is not present, is subtracted. This produces an unbiased statistic, though is practically difficult. The ``correlation'' subsection below indicates a practical scheme to create unbiased statistics.

The unbiased estimator is defined as:
\begin{align}
  \qo{\chi}^2_{2N}
  &\equiv
    \frac{1}{(\sum_{i=0}^{N_{\HqM}} W_i) } \sum_{i=0}^{N}
    \frac{W_{i}\left(\qo{V}_{+i} - \braket{\qo{V}_{+i}}\big|_{S_{\varphi} =0}\right)}{2|\red{g}(\Omega_i)|^2\SPEC(\Omega_i)}
    \label{eq:Homodyne_statistic_full}
  \\
  &\approx
    \frac{1}{N_{\HqM}} \sum_{i=0}^{N_{\HqM}}
    \frac{1}{2|\red{g}|^2\red{\SPEC}}\left(\qo{V}_{+i} - \braket{\qo{V}_{+i}}\big|_{S_{\varphi} =0}  \right)
    \label{eq:Homodyne_statistic_red}
\end{align}
Where $W_{i}$ are weight factors for each measurement. They optimize the statistic when the signal power varies as a function of the chosen basis-mode or frequency. The choice of cutoff frequencies for the minimum and maximum of $\Omega_{i}$ can act with a similar role as weights. Conversely, the variation in the weights changes the equivalent bandwidth. In the approximation of \cref{eq:Homodyne_statistic_red} and following expressions, the signal, optical gain and noise is assumed to be constant, and the weights are taken to be $W_i = 1$ for simplicity. Full expressions for optimal weights are in \cref{D:chi_weighted}.

This statistic can be equivalently defined in terms of the sin and cos basis operators. By the central limit theorem, over many measurements the aggregate statistic will approach a Gaussian distribution with mean and variance
\begin{align}
  \braket{\qo{\chi}^2_{2N_{\HqM}}} &= \param
                              &
    \VAR{\qo{\chi}^2_{2N_{\HqM}}}
  &\approx \frac{1}{N_{\HqM}} \left( \frac{1}{2|\red{g}|^2\red{\SPEC}} \right)^2\left(1 + 2\red{S}_{\VNP}\right)
    \label{eq:homodyne_chi_sq_var}
\end{align}
Where the variance decreases with the number of measurements due to the normalization by $1/N$ in the definition. This normalization arises naturally from constructing an unbiased estimator.

The SNR can then be used to determine the $1\sigma$ significance bound for signal exclusions from our statistic combining many measurements.
\begin{align}
  \alpha\red{\SPEC} \lesssim \sqrt{\VAR{\qo{\chi}^2_{2N_{\HqM}}}} = \frac{1}{\sqrt{N_{\HqM}}2|\red{g}|^2} \approx \frac{1}{\sqrt{\Delta T \Delta F}}\frac{1}{2|\red{g}|^2}
\end{align}
Where the factor $\Delta T \Delta F$ is the time-bandwidth product expressing the maximum size of the utilized basis set.

This expression revisits the standard quantum limit as it applies to broadband stochastic searches. The factor of $1/2$ is the single-sided power spectrum of the vacuum fluctuation. The vacuum has $1/2$ quanta of noise power in every temporal mode of the Michelson homodyne measurement, even though it does not carry physical energy. The factor $|\red{g}|^2$ is the optical gain, in power units. For a Michelson without any recycling, it is given by \cref{eq:g_normalization}

\subsection{Fisher Information}
After computing the mean and variance of the power operators, we also know now the variance of the original homodyne observable $\HqM_{c i}$, as $\VAR{\HqM_{c i}} = \braket{\qo{V}_{c i}}\red{l}^2$. If we also include the Gaussian distributions of $\cl{\vphi}$ and $\VNP$, their distributions convolve with the quantum noise distribution, resulting in a distribution with the variance computed above:
\begin{align}
  \PDF{\HpM, \HqM_{c i}}
  & =
    \frac{1}{\red{l}\sqrt{2 \pi \braket{\qo{V}_{c i}}}}e^{-\frac{\HpM^2}{2\red{l}^2}
    \braket{\qo{V}_{c i}}^{-1}
    }
\end{align}

Given our parameter of interest, $\param$, the Fisher information, $\FISH{}$, of the above Gaussian distribution with respect to that parameter is 
\begin{align}
  \FISH{{\HqM_{c i}} | \param}
  &\equiv
    \Braket{\left(
    \frac{\partial}{\partial \param}
    \log\left(\PDF{\HqM_{c i}, \HqM_{c i}}\right)
    \right)^2}
    \label{eq:fisher_def}
  \\
  &= \frac{1}{2}\left( \frac{2|\red{g}(\Omega)|^2\SPEC(\Omega)}{\braket{\qo{V}_{c i}}}  \right)^2
    \approx 2|\red{g}(\Omega)|^4 \SPEC^2(\Omega)
    \label{eq:Homodyne_fisher}
\end{align}
Where the approximation indicates the low signal and low background limit, with respect to the quantum noise.

The complex Fourier-like basis-modes measure both the cos and sin quadratures of the signal simultaneously, giving:
\begin{align}
  \FISH{{\HqM_{\pm i}} | \param}
  &= \FISH{{\HqM_{c i}} | \param} +  \FISH{{\HqM_{s i}} | \param}
    =
    \frac{\left(2|\red{g}(\Omega)|^2\SPEC(\Omega)\right)^2}{\braket{\qo{V}_{\pm i}}^2} 
\end{align}

The classical Cramer-Rao bound establishes that this Fisher information is the upper limit for the variance of any unbiased statistic established from the classical (Hermitian) observables:
\begin{align}
  \frac{1}{N\cdot\VAR{\qo{O}_N(\{\HqM_i\})}} \le \FISH{{\HqM_i} | \beta} \text{ where } \braket{\qo{O}_N - \beta} = 0
\end{align}

Accounting for the factor of two that complex basis-modes incorporate two Gaussian measurements, the Fisher information calculated above indicates that the defined chi-square statistic, $\qo{\chi}^2_{2N}$ from the sum of squares of Fourier-like basis will saturate the classical Cramer-Rao bound for the Gaussian homodyne observable. Thus, to further searches for incoherent signal power, an alternative \emph{quantum observable} must be employed than the homodyne measurement.

\subsection{Relation To Power Spectrum Estimation}\label{sec:psd_estimation}
The above derivation for an unbiased estimator of the signal power is closely related to power spectrum estimation, which simply does not strive to remove the biasing terms from the power observables:
\begin{align}
  \PSD{S_{\cl{\vphi}}(\Omega) | \{\HqM_i\}}
  &\equiv
    \frac{1}{N} \sum_{i=0}^{N} \frac{\braket{\qo{V}_{+i}}}{2|\red{g}(\Omega_i)|^2}
\end{align}
Here, the statistic is computed using a set of Fourier-like basis modes with $\Omega_i \approx \Omega$, such as from Welch-method apodized (windowed) finite Fourier transforms. The PSD then allows one to estimate the variance through the relation.
\begin{align}
  \sqrt{\VAR{\qo{\chi}^2_{2N}}}
  &=
    \frac{\PSD{S_{\cl{\vphi}}(\Omega) | \{\HqM_i\} }}{\sqrt{N}\SPEC(\Omega)}
\end{align}

Interferometer sensitivities are often given as power spectrum budgets, separating the noise contributions into various physical effects. This work distinguishes three such effects: signal, $S_{\vphi}$; classical noise, $\SNP$; and quantum noise, $\qo{a}$. Given an interferometer noise budget, the goal is to determine the classical noise level in the units used in this work. The quantum noise spectrum is primarily a representation of the optical gain, as the noise is $1$ in the chosen units.
\begin{align}
  {\PSD{\text{Quantum}(\Omega) | \{\HqM_i\}}}
  &= \frac{1}{2|\red{g}(\Omega_i)|^2}
    \label{eq:quantum_PSD}
\end{align}
As described in \cref{sec:processes}, the classical noise is factorized to scale with the optical gain. This scaling gets divided away for classical noise in units of arm displacement:
\begin{align}
  {\PSD{\text{Classical}(\Omega) | \{\HqM_i\}}}
  &=
    \frac{2\SNP}{2|\red{g}(\Omega_i)|^2}
    \label{eq:classical_PSD}
\end{align}
The above two relations allow one to use a budgeted noise curve to estimate the underlying processes in the conventions of this paper:
\begin{align}
  \SNP(\Omega)
  &=
    \frac{
    {\PSD{\text{Classical}(\Omega) | \{\HqM_i\}}}
    }{
    2 {\PSD{\text{Quantum}(\Omega) | \{\HqM_i\}}}
    }
    \label{eq:classical_estimation}
\end{align}
This relation indicates that for classical noise that is equal in spectral density to quantum noise, the classical noise is emitting half a photon per Hz per second individually into the upper and lower sidebands. This will be relevant in estimating the performance of photon counting, which is limited by classical photon emission from both phase and amplitude quadratures. Since fringe Homodyne readout does not measure the amplitude quadrature, $\SNA(\Omega)$ cannot be directly estimated from a noise budget. 

\subsection{Two-Interferometer Correlation Measurements}
\label{s:correlation}


Before delving into alternative quantum measurements, it is worth a short diversion to apply the above techniques to cross correlation measurements. The aggregate statistic \cref{eq:Homodyne_statistic_full} required a precise subtraction of the expectation value of the noise power, less the signal power. For deep searches of new physics, the signal may be 1\% or less of the quantum noise. This subtraction can be limited by the precision in calibrating an experiment.

In the event, that a single signal $\cl{\vphi}$ will coherently appear in two instruments A and B, this practical limitation can be removed. The two instruments are assumed to have independent background noises $\VNP[A]$, $\VNP[B]$ and quantum noise baths  $\qoa_A$, $\qoa_B$, leading to respective output field operators $\qo{d}_A(t)$ and  $\qo{d}_B(t)$, sharing only the signal term $\cl{\vphi}(t)$. Respectively, temporal mode measurements can be created for the timeseries from each instrument. Finally, a correlated noise power observable may be formed for the sin and cos components 
\begin{align}
  \qo{V}_{AB,ci} &\equiv \HqM_{A, ci} \HqM_{B, ci} \label{eq:V_cross_ci}
                   &
                   \qo{V}_{AB,si} &\equiv \HqM_{A, si} \HqM_{B, si}
\end{align}
The complex mode observables require some care, as they create a power operator that is complex, unlike the single interferometer case.
\begin{align}
  \qo{V}_{AB+i} &\equiv \HqM_{A, +i} \HqM_{B, +i}^\dagger
                  \\ &=
                  \frac{1}{2}(\qo{V}_{AB,ci} + \qo{V}_{AB,si} )
                  +
                  i(\HqM_{A,ci}\HqM_{B,si})
\end{align}

These have considerable experimental advantage that the mean does not include the quantum noise or background processes, and so no subtraction is required to test for excess noise.
\begin{align}
  \braket{\qo{V}_{AB, c i}} 
  = \braket{\qo{V}_{AB,+i}}
  &=
    2|\red{g}(\Omega)|^2S_{\cl{\vphi}}(\Omega_{c i})
\end{align}
Note that the expectation is purely real even for the complex basis modes. For this reason, only the variance on the real part of the statistic is meaningful. The imaginary part carries no information and can be ignored\footnote{This assumes the phasing between the two instruments is well established. The imaginary term may be necessary to account for experimental uncertainties.}.

By using the fact that the A and B noise and quantum operators commute, we then can calculate the variances:
\begin{align}
  \VAR{\qo{V}_{AB,c i}}
  &= \braket{\qo{V}_{A,c i}} \braket{\qo{V}_{B,c i}} + \braket{(2\sqrt{2}\REof{\red{g}(\Omega_{i})\cl{\vphi}_{+i}})^2}
     \\
     &\approx 1
\end{align}
The notable aspect of this expression is that although the cross spectrum avoids the quantum noise power in the mean, the quantum (and background) noise powers still contribute to the variance. For this reason, cross spectrum observations are not statistically more powerful than single instrument excess-power measurements. They simply have the advantage of simplified experimental systematics, as no precise subtraction is required.

Two further aspects worth noting are, firstly, that the variance of the cross-power is half of either instrument's power observable. This shows a minor statistical advantage over a single instrument, but is not an advantage over combining the independent measurements of two identical instruments. Secondly, the distribution of $\qo{V}_{AB, c i}$ is not a $\chi^2_1$, but rather is Gaussian distributed. This is due to being composed of a product of \emph{independent} Gaussian measurements in \cref{eq:V_cross_ci}. The contribution of $\cl{\vphi}$ to $\qo{V}_{AB, c i}$ is $\chi^2_1$, so where the correlated signal dominates, the distribution becomes $\chi^2$.

We can also consider the complex temporal modes, where the signal appears only in the real phase of the cross correlation.
\begin{align}
  \VAR{\REof{\qo{V}_{AB,+i}}}
  &=
    \frac{1}{4}\left(\VAR{\qo{V}_{AB,c i}} + \VAR{\qo{V}_{AB,s i}} \right)
  \approx \frac{1}{2}
\end{align}
Where the approximation assumes the weak signal limit. This expression is notable, as the variance is half of the single real mode case. This is again from the observable measuring in two basis-modes at once; however, instead of making a $\chi^2_2$, it is making a Gaussian distribution. Like the single temporal-mode case, as the signal dominates the quantum and background noise, the distribution becomes $\chi^2_2$.

The definitions above allow us to define the correlation as an unbiased estimator, $\qo{\mathcal{N}}^{AB}_{2N}$, and relate it to a cross spectrum
\begin{align}
  \CSD[AB]{\cl{\vphi}(\Omega_i) | \HqM_{+i}}
  &\equiv
    \frac{1}{N} \sum_{i=0}^{N} \frac{\qo{V}_{AB+i}}{2|\red{g}(\Omega_i)|^2}
\end{align}
Only the real part of the cross spectrum contains signal, giving:
\begin{align}
  \qo{\mathcal{N}}^{AB}_{2N} &\equiv \frac{\REof{\CSD[AB]{\cl{\vphi}(\Omega_i) | \HqM_{+i}}}}{\SPEC(\Omega)}
\end{align}
Which has the expectations
\begin{align}
  \braket{\qo{\mathcal{N}}^{AB}_{2N}}
  &= \param
  &
    \VAR{\qo{\mathcal{N}}^{AB}_{2N}}
  &\approx \frac{1}{2N} \left( \frac{1}{2|\red{g}|^2\red{\SPEC}} \right)^2
\end{align}
Which again shows the factor of 2 improvement from using two instruments.

The above formulas can be worked more fully to show the usual relationship between the variance of the cross spectrum and the power spectra. 

To be concise, let $\cl{C} = \CSD[AB]{\cl{\vphi}(\Omega_i) | \HqM_{+i}}$, then
\begin{align}
  \overline{\VAR{\cl{C}}}
  &\equiv
    \VAR{\REof{\cl{C}}} + \VAR{\IMof{\cl{C}}}
  \equiv
    \braket{\cl{C}\conj{\cl{C}}} - \braket{\cl{C}}\braket{\conj{\cl{C}}}
    \\ &=
         \frac{
         \PSD{\cl{\vphi}(\Omega_i) | \HqM^A_{+i}}
         \PSD{\cl{\vphi}(\Omega_i) | \HqM^B_{+i}}
         }{N}
         \\&=
  2\VAR{\qo{\mathcal{N}}^{AB}_{2N}}\ \SPEC(\Omega)
\end{align}

Altogether, the cross variables are shown to have the same statistical power as the combination of two independent observations from similar or identical instruments. Their advantage is to avoid the quantum noise power from appearing in the observable, which in turn prevents the need for a precise subtraction to search for excess signal power. This is a powerful experimental technique, but not a fundamental statistical advantage.

There is also an alternative derivation of the cross spectrum, where the homodyne timeseries measurements of two interferometer are added and subtracted into sum (common) and difference channels. After the linear combinations, then basis modes are applied as a discrete Fourier transforms, followed by power observables for spectrum estimation. This avoids the need for cross spectra. Instead one can show that the sum channel includes the signal spectrum along with all of the background noises, while the difference channel includes no signal, but identical noises. Thus, the power spectrum of the difference channel creates the estimate of the background noise. Subtracting the spectra of the sum and difference channels creates the spectrum of the signal. This method is statistically equivalent to producing a cross spectrum and the two methods can be related using basis changes, see sec. 6 of \cite{chouCQG17HolometerInstrument}.

This alternative method is mentioned as there is an analogous technique to cross correlation for counting statistics, which is more closely related to the sum/difference description than the cross-spectrum description of this bias-removal technique.

\section{Photon Counting}\label{sec:counting}

This section follows a similar pattern as the previous section. The quantum observation operator is defined from the definitions and experimental setup established in \cref{sec:descriptions}. For photon counting, the observation naturally sums observations in many temporal modes, naturally creating an aggregate statistic that must be simply normalized to form an unbiased estimator. 

First, the operating regime of the experimental setup must be defined. The fundamental change from homodyne readout is that local oscillator light is minimized, thus only $\mathcal{P}_2$ (\cref{eq:power_factor_2}) contributes to the power read from \cref{eq:power_factorized}.
Photon counts from frequencies that contain noise but no signal will add background, and in particular, any residual Michelson fringe light can add considerable background counts. In practice, this requires sharp optical filters between the interferometer and the readout to suppress light near the carrier frequency $\Omega \sim 0$. The frequency response of the readout filters is expressed as $\mathfrak{q}(\Omega)$. The filters are assumed to be passive, implying $|\mathfrak{q}(\Omega)| < 1$. To be concise, the factor $\gq(\Omega) = \mathfrak{q}(\Omega)g(\Omega)$ is used to indicate the optical gain for the signal, while factors of $\mathfrak{q}(\Omega)$ are included with the output noise.

With the filters applied, the field at the output port of the interferometer is:
\begin{align}
  \qo{d}(\Omega) 
  &= i \gq(\Omega) \phi(\Omega) + \mathfrak{q}(\Omega)\big(i\VNP(\Omega) + \VNA(\Omega)\big) + \qo{a}(\Omega)
\end{align}
Consider a measurement of the photon flux integrated over an interval from time $0$ to $\Delta T$. This gives the total observed energy as a number of photon counts: 
\begin{align}
  \qo{E} &= \int_{0}^{\Delta T}\qo{P}_{d}(t)\rmd t
\end{align}
On this interval, a Fourier series can be established using a basis set of temporal waveforms $\rK_{\pm j}(t) = e^{i t \Omega_{\pm j}} / \sqrt{T}$, with $\Omega_{\pm j} = \pm j 2\pi / \Delta T$. From this basis set, the field can be decomposed over the measurement interval:
\begin{align}
  \qo{d}(t)
  &\approx \sum_{i=1}^{\infty} \rK_{-i}(t)(i \gq(\Omega_{+i}) \phi_{+i} + i \mathfrak{q}(\Omega_{+i})\VNP[+i] + \qoa_{+i})
    \\ & \nonumber
         \hspace{3em} + \rK_{+i}(t)(i \gq(\Omega_{-i}) \phi_{-i} + i \mathfrak{q}(\Omega_{-i})\VNP[-i] + \qoa_{-i})
\end{align}
The observable can then be expressed as the sum of the basis operators.
\begin{align}
  \PqE
  &=
    \int_0^{\Delta T} \qo{d}^\dagger(t)\qo{d}(t) \rmd t
    \label{eq:E_power_observable}
  \approx \sum_{i=1}^{\infty} \PqE_{+i} +  \PqE_{-i}
\end{align}
Where each individual basis has a corresponding number operator, $\PqE_{\pm i}$, for the given temporal mode
\begin{align}
  \PqE_{\pm i}
  &\equiv \big(\gq(\Omega_{\pm i}) \cl{\vphi}_{\pm i} + \mathfrak{q}(\Omega_{\pm i})\VNP[\pm i] - i\qoa_{\pm i}\big)^\dagger
    \label{eq:E_template_observable}
  \\ & \nonumber
       \hspace{3em} \cdot \big(\gq(\Omega_{\pm i}) \cl{\vphi}_{\pm i} + \mathfrak{q}(\Omega_{\pm i})\VNP[\pm i] - i\qoa_{\pm i}\big)
\end{align}
It can be explicitly given by
\begin{align}
  \qo{d}_{\pm i}
  &\equiv \int_{-\infty}^{\infty}\qo{d}(t)\K_{\pm i}(t) \rmd t \equiv \int_{-\infty}^{\infty}\qo{d}(\Omega)\K_{\pm i}(-\Omega) \rmd \Omega
\end{align}
With
\begin{align}
  \PqE_{\pm i}
  &\equiv
    \qo{d}_{\pm i}^\dagger 
    \qo{d}_{\pm i}
    \label{eq:E_power_d}
\end{align}
These integrals and decomposition of $\qo{E}$ into $\qo{E}_{\pm i}$ is a formal technique here to analyze stochastic signals, where ``all'' templates are measured and the filter $\mathfrak{q}(\Omega)$ weights or selects a limited set. With quantum memory technology, individual templates can, in principle, be individually measured and addressed.

Expectation values of the $\qo{E}_{\pm i}$ operators can be given in terms of the constituent signal and noise power:
\begin{align}
  \red{S}_{+i}
  &\equiv
    |\gq(\Omega_{+i})|^2  \frac{S_{\cl{\vphi}}(\Omega_{+i})}{2}
    + |\mathfrak{q}(\Omega_{+i})|^2\frac{\SNP(\Omega_{+i}) + \SNA(\Omega_{+i})}{2}
    \label{eq:constituent_S}
\end{align}

Which allows the concise expressions for the mean and variance of the energy detection operators:
\begin{align}
  \braket{\PqE_{\pm i}}
  &=
    \braket{|\gq(\Omega_{\pm i}) \cl{\vphi}_{\pm i} + \mathfrak{q}(\Omega_{\pm i})\VNP[\pm i]|^2}
    = \red{S}_{\pm i}
    \\
  \VAR{\PqE_{\pm i}}
  &=
    \braket{\qoa_{\pm i}|\gq(\Omega_{\pm i}) \cl{\vphi}_{\pm i} + \mathfrak{q}(\Omega_{\pm i})\VNP[\pm i]|^2\qoa_{\pm i}^\dagger}
       = \red{S}_{\pm i}
    \label{eq:var_Energy_sm}
\end{align}

This variance hints at the information content of a single temporal mode. The signal emission rate, indicated within the mean, is determined by the spectral density and average optical gain of the signal over the frequency and duration of the single temporal mode. For weak signals, the mean is expected to be much less than 1 photon. Similarly, classical noises also cause photon emission, adding to the variance. A single measurement of a single $E_{\pm i}$ is most likely to measure zero photons, but when it measures even a single one, it will be highly significant when the classical noise affecting $\VAR{E_{\pm i}}$ is also small. Many such measurements over a set ${\pm} i$ of temporal modes will be necessary to properly estimate the signal rate, given the small probability in any single detection. The high significance of each detection contributes to a high Fisher information of aggregate measurements. Aggregation in the case of stochastic signals is generally over some time-bandwidth product. In the case of waveform discrimination, it may require multiple instances of candidate events to be observed, while using a common set of temporal-mode envelopes that are orthogonal by being offset in time for each measurement.

\subsection{Aggregate Measurements}

Similarly to the homodyne readout and its associated power observable, we can define an unbiased statistic from the individual power observations over a basis set of temporal modes, building from the definition \cref{eq:E_power_observable}:
\begin{align}
  \qo{G}_{\param}
  &=
    \frac{1}{\red{G}}\left(\PqE + \cl{B} - \braket{\PqE + \cl{B}}\bigg|_{S_{\vphi} = 0} \right)
           \\
  \red{G} &\equiv \frac{1}{2}\sum_{i=1}^{\infty} \left( |\gq(\Omega_{+i})|^2\SPEC(\Omega_{+i}) +  |\gq(\Omega_{-i})|^2\SPEC(\Omega_{-i}) \right)
            \label{eq:redG_norm_factor}
\end{align}
The additional random variable $\cl{B}$ represents a background count process that introduces additional variance. An example are the background counts rate of single-photon detectors. Those should be expected have Poisson statistics with a mean photon count $\braket{\cl{B}} = \red{B}\Delta T$ and variance $\VAR{\cl{B}} = \braket{\cl{B}}$.

The mean and variance of the statistic is computed to be
\begin{align}
  \braket{\qo{G}_{\param}}
  &= \param
  &
    \VAR{\qo{G}} &= \frac{1}{\red{G}^2}\left( \sum_{i=1}^{\infty} (\red{S}_{+i} + \red{S}_{-i}) + \VAR{\cl{B}} \right)
\end{align}

Now, the measurement problem should be simplified to relate this statistic to the Fisher Information on a single temporal mode. Assume that the filter function ${\mathfrak{q}(\Omega_0 + \Omega_1)}\approx 1$ is a steep bandpass filter which passes a (frequency) bandwidth ${|\Omega_1| < \pi\Delta F_{\qo{E}}}$, centered at $\Omega_0$, and is zero outside of that band. Furthermore, assume that the signal power spectrum and the interferometer gain is nearly constant within the band pass ${|\gq(\Omega_0 + \Omega_1)|^2\SPEC(\Omega_0 + \Omega_1) \approx |\red{g}|^2\red{\SPEC}}$ and zero outside of the bandpass. Similarly, assume that the background noise is constant within that band $\SNP(\Omega_0 + \Omega_1) \approx \red{\SNP}$.

The number of independent basis temporal-modes spanned depends on the measurement time, $\Delta T$ and is $N_{\qo{E}} = \Delta F_{\qo{E}} \Delta T$. Under these assumptions
\begin{align}
  \red{G}
  &\approx
    \frac{N_{\qo{E}}}{2}|\red{g}|^2\red{\SPEC}
  &
    \braket{\PqE}
  &=
    \frac{N_{\qo{E}}}{2}\left(|\red{g}|^2\param \red{\SPEC} + \red{S}_{\VNP} + \red{S}_{\VNA}\right)
\end{align}
Applying these to the definitions gives:
\begin{align}
      \VAR{\qo{G}_\param}
  &\approx
    \frac{2}{N_{\qo{E}}}\left(\frac{1}{|\red{g}|^2\red{\SPEC}}\right)^2\left(\red{S}_{\VNP} + \red{S}_{\VNA} + \frac{2\red{B}\Delta T}{N_{\qo{E}}} \right)
    \label{eq:photon_counting_G_VAR}
\end{align}
The term including the noise sources also has a factor of $|\red{g}|^2\param \red{\SPEC}$, but it is elided for simplicity in the weak-signal limit. Additionally, the small noise limit is applied, and square terms $\SNP^2$ are omitted. \Cref{D:PCF_full_variance} shows the complete variance expressions that are computed using the exact expression for the probability mass distribution.

The expectations were derived using a basis of temporal-mode waveforms and several assumptions for constant signals and noise. The expectations can be expanded to give the integral forms where no specific basis is assumed, giving:
\begin{align}
  \red{G}
  &=
    \frac{\Delta T}{2}\int_{-\infty}^{\infty} |\gq(\Omega)|^2\SPEC(\Omega) \frac{\rmd \Omega}{2\pi}
            \label{eq:Counting_Norm_Int}
  \\
  \VAR{\qo{G}_\param}
  &=
    \frac{\Delta T}{\red{G}^2} \int_{-\infty}^{\infty}\hspace{-1em}|\mathfrak{q}(\Omega)|^2\frac{\SNP(\Omega)+ \SNA(\Omega)}{2} \frac{\rmd \Omega}{2\pi}
    + \frac{\VAR{\cl{B}}}{\red{G}^2} 
    \label{eq:Counting_Int}
\end{align}
Note that without the assumptions of constant signal and noise, the Fisher information, integrated over all frequencies, is not necessarily saturated. This is because only passive optical filters in the interferometer to shape $g(\Omega)$ passive readout filters to shape $\mathfrak{q}(\Omega)$, are assumed. The aggregate statistic for the Homodyne variable includes weight factors that can be included during classical computations. Such weight factors can, in principle, be used with photon counting. Present technology uses cavities to define the basis set, but cannot individually observe each $\qo{E}_{\pm i}$, preventing weight factors from being included.
  
\subsection{Comparison to Homodyne Readout}

Both homodyne readout and photon counting readout have statistics which saturate the Fisher information of their respective classical observables, to optimally obtain information about the signal process. There are some notable differences in their implementation which must be addressed to directly compare the two approaches.

First, the effective number of temporal modes for the homodyne observable is quoted to be $2\Delta F_{\HqM} \Delta T$, accounting for the positive and negative frequency basis waveforms that are simultaneously observed by the power operators $\qo{V}_{+ i}$ that compose $\qo{\chi}^2_{2N_{\HqM}}$. This leads to $N_{\HqM}\approx\Delta F_{\HqM} \Delta T$ complex measurements, which use $2N_{\HqM}$ independent basis modes.

For photon counting, the optical bandwidth is twice as large as the classical bandwidth, as the signal spectrum $S_{\cl{\vphi}}(\Omega)$ is imprinted into both positive and negative sidebands.
Thus, double the bandwidth is available in the photon counting case, $\Delta F_{\qo{E}} \approx 2\Delta F_{\HqM}$ for double the number of complex measurements $N_{\qo{E}} \approx 2 N_{\HqM}$. However, the same number of basis waveforms are observed in both cases. With this accounting established, the relative sensitivity of the two techniques can be compared:
\begin{align}
  \frac{\VAR{\qo{G}_{N_{\qo{E}}}}}{\VAR{\qo{\chi}^2_{2N_{\HqM}}}}
  &\approx
    4\left(\SNP + \SNA + \frac{2\red{B}}{\Delta F_{\qo{E}}}  \right) \left( \frac{2\Delta F_{\HqM}}{\Delta F_{\qo{E}}} \right)
    \label{eq:readout_comparison}
\end{align}
This has the simple interpretation that photon counting has an advantages for stochastic noise searches as long as the background noise power spectrum is below $1/4$ of the quantum noise. This comparison does not include squeezing, but heuristics are extended in \cref{sec:outlook} to include it.

A second notable point to compare is ease of implementation. Specifically, the homodyne readout combines all of the independent measurements across the waveform basis using classical computation, enabling the weighting parameters $W_i$ to be chosen after the measurement is taken. Photon counting can not easily apply such weightings, except by changing the filter $\mathfrak{q}(\Omega)$. Thus, optimizing it for SNR is limited, as sharp or narrow filtering using all-optical cavity filters are practically challenging to construct and operate. In particular, thermal acoustic resonances of interferometer mirrors can form large, narrow and regular noise features that readout cavity filters must be designed to avoid. Practically, this can cause some inevitable background counts, additional optical losses $q(\Omega) < 1$ from high-finesse cavities, and substantially less bandwidth $\Delta F_{\qo{E}} \ll \Delta F_{\HqM}$. Many of these issues can be compensated using signal recycling, analyzed in \cref{sec:sensitivity_comparison}.

\subsection{Fisher Information}

Higher moments can be considered from the probability mass function of the operators $\PqE_i$.
\begin{align}
  \PCF{\Ppe, \PqE_{+ i}}
  &\equiv
    \Braket{e^{-i \PqE \Ppe}}
    =
    \frac{1}{\red{S}_{+i}(e^{i\Ppe }-1)+1}
    \label{eq:CF_Energy}
  \\
  \label{eq:PMF_Energy}
  \PDF{\PpE, \PqE_{+ i}}
  &\equiv \braket{\delta(\PpE - \PqE_{+ i})}
  \\
  &= \frac{1}{2\pi}\int_0^{2\pi} e^{i \PpE \Ppe} \PCF{\Ppe, \PqE_{i}} \rmd e
  \\
  &=
    \frac{1}{\red{S}_{+i}}
    \left(1 + \frac{1}{\red{S}_{+i}}\right)^{-(1 + \PpE)}
\end{align}
Which are derived in \cref{D:CF_Energy}, and \cref{D:PMF_Energy}.
The resulting characteristic function and probability mass function is known as the geometric distribution. Perhaps surprisingly, it is not a Poisson distribution expected given the coherent laser light powering the interferometer, but rather equivalent to a Bose distribution on a single oscillator mode.
This discussion is related to equations 4.13 to 4.14 of Helstrom\cite{helstrom76QuantumDetection}.

The Fisher information (defined in \cref{eq:fisher_def}) for determining the signal power, $S_{\vphi}$, is then
\begin{align}
  \FISH{{\PqE_{+i}} | \param}
  &=
    \frac{|\gq(\Omega_{+i})|^4\SPEC^2(\Omega_{+i})}{4 \red{S}_{+i}(1 + \red{S}_{+i})}
    \approx
    \frac{|g(\Omega_{+i})|^4\SPEC^2(\Omega_{+i})}{2 \SNP(\Omega_{+i})}
\end{align}
Where the approximation is taken in the small signal limit. In that limit, the Fisher information is limited only by the background count rate of the instrument. The fundamental advantage for the photon counting method is thus demonstrated: the limit is determined entirely by instrumental processes and not quantum noise of the signal process; however, the optical gain does establish the quantum limit of the information rate.

Thus, the estimator of \cref{eq:photon_counting_G_VAR} saturates the classical Cramer-Rao bound for the Geometric/Bose distribution when it is in the small signal limit. The effective number of measurements is determined by the total optical bandwidth and the measurement time.

\subsection{two-interferometer correlated signals}

Similarly to homodyne readout, photon counting can only create an unbiased estimator by subtracting an estimate of the background count rates from classical noise sources in the interferometer and detector counts. Practically, these background count estimates are challenging to determine, and have their own statistical uncertainties. However, photon counting can employ an analogous two-interferometer measurement technique to \cref{s:correlation}. This allows one to simultaneously determine background counts while search for signal. This prevents the need for configurations that modulate the signal level, as such reconfiguration can also modify the background count rate, leading to systematic biases in the background count rate.

\begin{figure}
  \centering
  \includegraphics[width=0.95\linewidth]{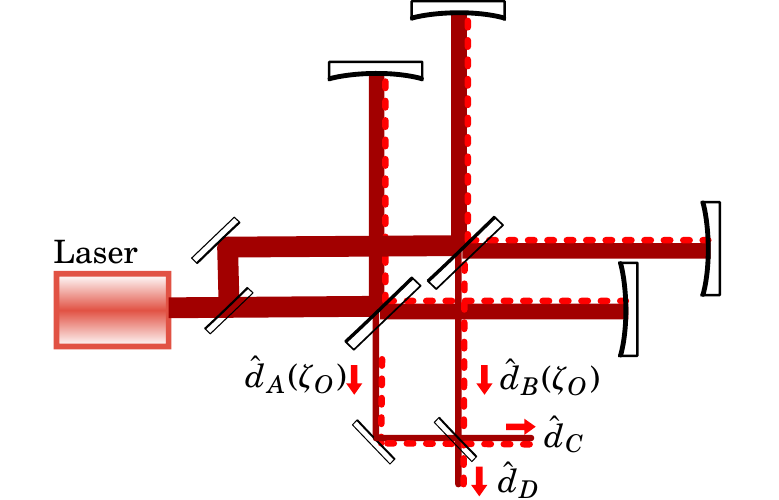}
  \label{fig:IFO_simple_corr},
  \caption{
    Depiction of the cross correlation technique using two instruments with photon counting. The two interferometers must be coherently fed by the same laser or locked lasers, and their outputs must be fed to a beamsplitter.
  }
\end{figure}

The method is depicted in \cref{fig:IFO_simple_corr}. The basis of the method is to use the coherence of the signal in two detectors to direct the signal to one of two photon counting readout ports, with fields $\qo{d}_C$ and  $\qo{d}_D$. In doing so, both readout ports gather counts from backgrounds and incoherent noise sources, but only one gathers signal. This allows the second output to estimate the backgrounds of the first.

Practically, the coherent combination of two interferometers is implemented using a beamsplitter and the choice of which port sees the signal is determined by the path lengths to that beamsplitter. This allows the signal to be swapped between the two output ports. By swapping the ports, only one low-noise readout system is required, since the signal can be enabled and disabled. Even with a readout system on both output ports, swapping is still essential as a calibration mechanism, as both readout systems can not be expected to be identical.


Mathematically, the fields output at each interferometer can be described as
\begin{align}
  \qo{d}_A(\Omega)
  &= i g_A(\Omega) \phi(\Omega) + i \VNP[A](\Omega) + \VNA[A](\Omega) + \qoa_A(\Omega)
    \\
  \qo{d}_B(\Omega) 
  &= i g_B(\Omega) \phi(\Omega) + i \VNP[B](\Omega) + \VNA[B](\Omega) + \qoa_B(\Omega)
\end{align}
Using subscripts to distinguish the noise sources and optical gain of each interferometer.
The beamsplitter operation that creates the output port fields from the interferometer fields is:
\begin{align}
  \qo{d}_C(\tau)
  &= \tfrac{1}{\sqrt{2}}\qo{d}_{A}(\tau, \PV_O)
    + \tfrac{1}{\sqrt{2}}\qo{d}_{B}(\tau, \PV_O)
  \\
  \qo{d}_D(\tau)
  &= \tfrac{1}{\sqrt{2}}\qo{d}_{A}(\tau, \PV_O)
    - \tfrac{1}{\sqrt{2}}\qo{d}_{B}(\tau, \PV_O)
\end{align}
The readout filters $\mathfrak{q}(\Omega)$ are then applied after the beamsplitter. For simplicity, we now assume that $g_A(\Omega) = g_B(\Omega) = g(\Omega)$. This gives for the fields of each output port.
\begin{align}
  \qo{d}_C(\Omega)
  &= i \sqrt{2} \gq(\Omega) \phi(\Omega) + \qoa_C(\Omega)
  \\\nonumber
  & + \frac{\mathfrak{q}(\Omega)}{\sqrt{2}}(i\VNP[A](\Omega) + i\VNP[B](\Omega) + \VNA[A](\Omega) + \VNA[B](\Omega))
  \\
  \qo{d}_D(\Omega) 
  &= \qoa_D(\Omega)
  \\\nonumber
  & + \frac{\mathfrak{q}(\Omega)}{\sqrt{2}}(i\VNP[A](\Omega) - i\VNP[B](\Omega) + \VNA[A](\Omega) - \VNA[B](\Omega))
\end{align}
Here, the quantum fluctuation operators are given indexed by the output port. Physically, these operators are linear combinations of the quantum fluctuation operators from each interferometer. When no quantum state engineering is performed, there is no need to expand the operators in terms of inputs, as they have all the same relations as vacuum bosonic ladder operators. 

There then exists photon counting operators $\PqE_C$ and $\PqE_D$ associated with each output field. These are composed of individual temporal-mode $\PqE_{C, \pm i}$ and $\PqE_{D,\pm i}$ states as for \cref{eq:E_power_observable} and \cref{eq:E_template_observable}.
\begin{align}
  \braket{\PqE_{D, +i}}
  &= 
    \frac{|\mathfrak{q}(\Omega_{+i})|^2}{2}
    \Big(
    \SNP[A](\Omega_{+i}) + \SNP[B](\Omega_{+i})
    \\ &\hspace{6em} \nonumber
    + \SNA[A](\Omega_{+i}) + \SNA[B](\Omega_{+i})
    \Big)
  \\
  \braket{\PqE_{C, +i}} &= 
                      |\gq(\Omega_{+i})|^2 S_{\cl{\vphi}}(\Omega_{+i}) +
                      \braket{\PqE_{D, +i}}
\end{align}

The same factor $\red{G}$ from \cref{eq:redG_norm_factor} can be used to normalize the estimator for the parameter. This then produces equivalent simplified expressions as \cref{eq:photon_counting_G_VAR}.
\begin{align}
\qo{G}_{\param}^{AB}
&=
                  \frac{1}{2\red{G}}\left(\PqE_C - \PqE_D + \cl{B}_C + \cl{B}_D - \braket{\cl{B}_C + \cl{B}_D} \right)
  \\
  \red{S}_{AB} &\equiv \red{\SNP[A]} + \red{\SNP[B]} + \red{\SNA[A]} + \red{\SNA[B]}
  \\
  \VAR{\qo{G}_{\param}^{AB}}
  &\approx
    \frac{1}{N_{\qo{E}}}\left(\frac{1}{|\red{g}|^2\red{\SPEC}}\right)^2\left(\red{S}_{AB} + 2\red{B}\Delta T \right)
\end{align}
As in the homodyne case, these expressions show a factor of 2 improvement in the variance, expected from using two instruments to search for signals. If only a single readout is used, swapping between signal and quiet modes, the factor of two improvement is lost from the duty cycle of using a single readout. Comparing the derivation here to the homodyne case appears that the two techniques are fundamentally different, as the two-interferometer cross spectrum is computed differently than a power spectrum, while this technique only measure power, but linearly combines the signals first.

Altogether, the ability to use a cross correlation technique shows that photon counting has no fundamental statistical disadvantages over homodyne readout. Many of the practical challenges must be explored and tested to fully develop it as a tool to search for new physics using Michelson interferometers.

\section{Sensitivity Comparisons, including Signal Recycling}\label{sec:sensitivity_comparison}


\begin{figure}
  \centering
  \includegraphics[width=0.95\linewidth]{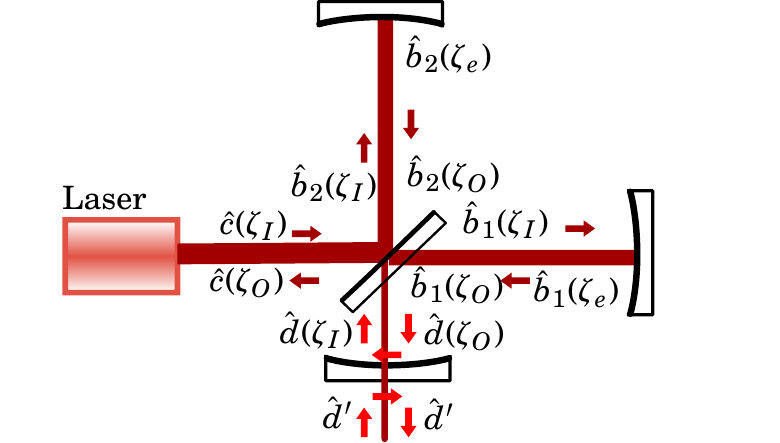}
  \label{fig:IFO_simple_recycling},
  \caption{
    Depiction of the fields and variable definitions in a Michelson
  }
\end{figure}

\newcommand{\TSRM}{\ensuremath{T_{\text{srm}}}}
Adding a appropriate mirrors to the interferometer modifies the optical gain, $g(\Omega)$. Physically, this can be viewed as changing the density of states within the interferometer, such that the total number of states, averaged over frequency, remains constant. For a single mirror with transmissivity of $\TSRM$, added at the output port of the interferometer, the optical gain is transformed from $g(\Omega) = \red{g}$ without the mirror, to $g'(\Omega)$, as:
\begin{align}
  g'(\Omega) &=  \frac{\sqrt{\TSRM}}{1 - \sqrt{1 - \TSRM}\sqrt{\eta}e^{-i(\Omega - \Omega_g)2L/c}}\red{g}
\end{align}
Where $L$ is the length of the arm and $\eta$ is the round-trip optical efficiency. $\Omega_g$ is the center frequency of the resonance, which is determined by the exact path length to the recycling mirror.

When $T_{\text{srm}}$ is small, the signal recycling is in the high finesse limit, and the expression can be simplified.
\begin{align}
  g'(\Omega) &=  \sqrt{\frac{\gamma_{g} c}{L}}\frac{\red{g}}{\gamma_{g} + \gamma_{\eta} - i(\Omega - \Omega_g)}
\end{align}
Using the bandwidth parameters
\begin{align}
  \gamma_{g} &= \frac{c \TSRM}{4 L}
  &&&
    \gamma_{\eta} &= \frac{c }{4 L}(1 - \eta) &
\end{align}
The first of these, $\gamma_{g}$, is adjustable using the recycling mirror transmission, while the second is determined by the round-trip optical efficiency of the interferometer.

With these parameters, the optical gain takes the maximum value ${|g'(\Omega_g)|^2 = 4/(T + 1-\eta)}$ with a HWHM of ${\gamma_{g} + \gamma_{\eta}}$. Note also that 
\begin{align}
  \int_{-\infty}^{\infty} |g'(\Omega)|^2 \frac{\rmd \Omega}{2\pi} &= |\red{g}|^2\frac{c}{2L}\eta_g
  &
  \eta_g &= \frac{\gamma_{g}}{\gamma_{g} + \gamma_{\eta}}
\end{align}

Thus, the integrated optical gain is constant and does not depend on the chosen bandwidth $\gamma_{g}$, except when limited by the optical efficiency (losses). This is the sense in which interferometer gain manipulates the local density of states while conserving the total number.

The general motivation of signal recycling during homodyne readout is to match the interferometer response $g'(\Omega)$ to the signal spectrum $\SPEC(\Omega)$. Exactly matching the two optimizes the total fisher information integrated over the spectrum \cref{eq:Homodyne_fisher}. It can be considered as implementing a physical counterpart to the weighting factors of \cref{eq:Homodyne_statistic_full} (analyzed in \cref{D:chi_weighted}). 

For the purposes of comparing the two readout schemes, it is worth considering a simple Lorentzian signal model with a resonant peak
\begin{align}
  \SPEC(\Omega)
  &=
    \frac{2\gamma_{\vphi}}{\gamma_{\vphi}^2 + (\Omega - \Omega_{\vphi})^2} + \frac{2\gamma_{\vphi}}{\gamma_{\vphi}^2 + (\Omega + \Omega_{\vphi})^2}
    \label{eq:SPEC_narrow_ansatz}
\end{align}

Which is normalized to unit power ${\int_0^{\infty} \SPEC(\Omega) \rmd \Omega / 2\pi = 1}$. Under this normalization, the fit parameter $\param$ takes on units, typically square-length or square-phase, depending on the chosen normalization of $g(\Omega)$, for the sensitivity to displacement signals or to phase shifts.

In the limit where $\Omega_{\vphi} > \gamma_{\vphi}$, the Lorentzian signal has the following related integrals: ${\int_0^{\infty} \SPEC(\Omega)^2 \rmd \Omega / 2\pi = \gamma_{\vphi}}^{-1}$ and  ${\int_0^{\infty} \SPEC(\Omega)^4 \rmd \Omega / 2\pi = \gamma_{\vphi}^{-3} 5/2}\approx$, used in the following.

This form of spectrum can also be parameterized using the peak spectral density $\red{\SPEC}$
\begin{align}
  \SPEC(\Omega)
  &=
    \frac{\red{\SPEC}\gamma^2_{\vphi}}{\gamma_{\vphi}^2 + (\Omega - \Omega_{\vphi})^2} + \frac{\red{\SPEC}\gamma^2_{\vphi}}{\gamma_{\vphi}^2 + (\Omega + \Omega_{\vphi})^2}
    \label{eq:SPEC_narrow_ansatz}
\end{align}
It is parameterized in both ways, as the peak density is a more applicable parameterization to wideband search calculations while the unit-power normalization is more appropriate for narrowband searches.

\subsection{Homodyne Readout}

Calculating the optimal search statistic using the weight factors of \cref{D:chi_weighted} gives a variance of:
\begin{align}
  \frac{1}{
  \VAR{\qo{\chi}^2_{2N}}
  }
  &=
\frac{
  \Delta T
}{2\pi}
\int_0^{\infty}
  \left(
    2|\red{g}(\Omega)|^2\SPEC(\Omega)
  \right)^2
\rmd \Omega
\end{align}
This can be compared to signal recycling. Here, some of the simplifications taken during the derivations must be reconsidered, namely
\begin{align}
  |2\red{g}(\Omega)|^2
  &\rightarrow \frac{1}{2}|g'(\Omega) + \conj{g'}(-\Omega)|^2
  \\
  &\approx
    |\red{g}|^2\frac{c}{2L}
    \left(
    \frac{\gamma_{g}}{(\gamma_{g} + \gamma_{\eta})^2 + (\Omega - \Omega_g)^2}
    +
    \ldots
    \right)
\end{align}
where the approximation is in the limit $\Omega_g > \gamma_{g}$ and the ellipses indicate terms with $(\Omega + \Omega_g)$ factors.

From \cref{eq:g_normalization}, a broadband Michelson with a flux of $F_{\text{BS}} = P_{\text{BS}} / \hbar \omega$ on the beamsplitter has a sensitivity of 
\begin{align}
  \red{g} &= \frac{\omega}{c} \sqrt{F_{\text{BS}}}
\end{align}

Various choices signal recycling bandwidths $\gamma_{g}$ then give
\begin{align}
  \frac{1}{
  \VAR{\qo{\chi}'^2_{2N}}_{\text{SR}}
  }
  &\approx
    \left( \frac{\omega^4}{c^4}\frac{F_{\text{BS}}^2}{\gamma_{\vphi}^2} \right)
    \left(\frac{ \eta_g c/2L}{\gamma_{\vphi} + \gamma_{g} + \gamma_{\eta}}\right)
    \left( \Delta T \frac{c}{2L}\right)
  \\
  \frac{1}{
  \VAR{\qo{\chi}'^2_{2N}}_{\text{BB}}
  }
  &\approx
    \left( \frac{\omega^4}{c^4}\frac{F_{\text{BS}}^2}{\gamma_{\vphi}^2} \right)4\left( \Delta T\gamma_{\vphi}\right)
    \text{ no recycling} 
\end{align}

These can also be expressed in terms of the peak spectral density $\red{\SPEC}$.
\begin{align}
  \frac{1}{
  \VAR{\qo{\chi}'^2_{2N}}
  }
    &\approx
      \left( \frac{\omega^4}{c^4}F_{\text{BS}}^2 \red{\SPEC}^2 \right)
      \left(\frac{\eta_g c/8L}{\gamma_{\vphi} + \gamma_{g} + \gamma_{\eta}}\right)
      \left( \Delta T \frac{c}{2L}\right)
  \\
  \frac{1}{
  \VAR{\qo{\chi}'^2_{2N}}_{\text{BB}}
  }
    &\approx
      \left( \frac{\omega^4}{c^4}F_{\text{BS}}^2 \red{\SPEC}^2\right)\left( \Delta T \gamma_{\vphi}\right)
      \text{ no recycling} 
\end{align}

Where the approximations are in the limit $\Omega_{\vphi} > \gamma_{\vphi}$ that the peak signal frequency is larger than the signal bandwidth.

Thus, resonant signal recycling can be an advantage for Homodyne readout experiments where:
\begin{align}
  \frac{c}{8\gamma_{\vphi}} < L
\end{align}

For narrow band searches of signals like dark matter, this constraint can be easily met. Due to losses in the interferometer, $\TSRM$ should generally not be substantially below 1000ppm, limiting the ability to saturate signal recycling enhancements for short interferometers.

The advantages of signal recycling are different for photon counting. First, existing technology in photon counting cannot use the weight factors $w_i$ to create statistics that saturate the Fisher information over wide frequency bands.
The only ability to do so is provided using the filter $\mathfrak{q}(\Omega)$.

Namely, when $\SNP(\Omega)$ varies significantly, frequencies where it is large can overwhelm the advantages gained where it is small. An example of this is the thermal motion of bulk acoustic modes of interferometer mirrors. Each acoustic mode has spectral peaks with
\begin{align}
  \SNP[\text{,bulk}](\Omega) &= |2\red{g}(\Omega)|^2\sum_i^\infty \frac{k_b \mathcal{T}}{m\Omega_{\text{blk}, i}^2} 2\pi\delta(\Omega - \Omega_{\text{blk}, i})
\end{align}
given the Boltzmann constant $k_b$, temperature $\mathcal{T}$, mirror mass $m$, and acoustic longitudinal bulk resonance frequencies $\Omega_{\text{blk}, i}$. This motion is delta-function like for mirrors of high mechanical quality. The delta-function nature of these resonances indicates that the optimal strategy uses $\mathfrak{q}(\Omega_{\text{blk}, i}) = 0$ for all of the resonances. However, practically such a zero can only be approximated. Using optical cavities to implement $q(\Omega_{\text{blk}, i})$ requires that the center frequency and passband of the cavities be sufficiently separated to reduce background photon counts coming from the thermal resonance frequencies. 

The technical constraints and thermal background practically constrain the filter $\mathfrak{q}(\Omega)$ to be formed as a series of $N_q$ (typically) identical cavities, giving the form:
\begin{align}
  \mathfrak{q}(\Omega)
  &=
    \left( \frac{\gamma_{q}}{\gamma_{q} - i(\Omega - \Omega_q)} \right)^{N_q}
\end{align}
This series suppresses background noise from each resonance by a factor of $|\gamma_{q} / (\Omega_{\text{res}, i} - \Omega_q)|^{N_q}$.

This narrow filter constraint entails is an additional frequency scale from $\gamma_{q}$ to consider in a sensitivity analysis for signal recycling. There are two regimes worth considering based on wide band and narrow band signal.

\subsection{wideband signal photon counting}

The first case is when the signal bandwidth is wider than the acceptable output filter width $\gamma_{\vphi} > \gamma_q$. Here, take $\red{\SPEC}$ to be the representative signal power within the readout band.

Because we are freely changing the optical gain and the classical noises are currently defined to scale with the gain, they must be redefined into the parameter $\red{S}_{\cl{c}}$, using $\SNP + \SNA\rightarrow |g(\Omega)|^2\red{S}_{\cl{c}}$. This parameter appears in later expressions as $|\red{g}|^2\red{S}_{\cl{c}}(\Omega)$ for the classical noise of a broadband Michelson with a given power level.

From \cref{sec:psd_estimation}, specifically \cref{eq:classical_estimation}. If we have a noise budget of an interferometer decomposed into classical and quantum contributions, then
, and we know the optical gain for the instrument, then
\begin{align}
  \red{S}_{\cl{c}}(\Omega)
  &\approx
    \PSD{\text{Classical}(\Omega)}
    \label{eq:classical_noise_reduction}
\end{align}
Where the approximation indicates that the amplitude quadrature noise contribution is not included in this estimation. 
This expression is related to \cref{eq:classical_estimation} when the quantum noise contribution is given for a broadband Michelson design, where $g(\Omega)=\red{g}$. More generally, the relationship between the quantum and classical power spectra and the instrumental spectral density is:
\begin{align}
    |\red{g}|^2\red{S}_{\cl{c}}(\Omega)
    &\approx
      \frac{
      \PSD{\text{Classical}(\Omega)}
      }{
      \PSD{\text{Quantum}(\Omega)}
      }
      \frac{2|\red{g}|^2}{|g(\Omega) + \conj{g}(-\Omega)|^2}
      \label{eq:classical_noise_reduction}
  \end{align}

From these definitions, the photon counting estimators can be expressed using
\begin{align}
  \mathcal{N}_{q} &= \Delta T \int_{-\infty}^{\infty}\left|\frac{g'(\Omega)}{g(\Omega)}\mathfrak{q}(\Omega)\right|^2\frac{\rmd \Omega}{2\pi}
  &
    \red{G} &= \frac{\red{\SPEC} \mathcal{N}_{q}|\red{g}|^2}{2}
\end{align}
Which then enter the computation of the variance. Here an inverse 
\begin{align}
  \frac{1}{
  \VAR{\qo{G}_\param}
  }
  &=
    \frac{\mathcal{N}_{q}\red{\SPEC}^2|\red{g}|^4}{2}\left(
    |\red{g}|^2\red{S}_{\cl{c}}
    + \frac{2\Delta T \red{B} }{\mathcal{N}_{q}}
    \right)^{-1}
\end{align}
Here, whenever the bandwidth of the interferometer ${\gamma_g < \gamma_q}$, then the interferometer gain dominates the integrals. This gives:

\begin{align}
  \mathcal{N}_{q}
  &\approx
    \frac{\Delta T c }{2L}
    \frac{\eta_g \gamma_{q}}{\gamma_{q} + \gamma_{g} + \gamma_{\eta}}
  &&\text{with recycling}
  \label{eq:N_q}
  \\
  \mathcal{N}_{q} &= \Delta T \frac{\gamma_q}{2} &&\text{no recycling}
\end{align}
Note that the signal recycling case does not depend on the chosen interferometer bandwidth $\gamma_g$ as long as it is sufficiently smaller than the readout cavity bandwidth, $\gamma_{q}$ and above the loss-limited bandwidth, $\gamma_\eta$.

\subsubsection{Comparisons}

In the wideband signal case with no recycling, the sensitivities can be compared:
\begin{align}
  \frac{
  \VAR{\qo{G}_\param}_{\text{BB}}
  }{
  \VAR{\qo{\chi}'^2_{2N}}_{\text{BB}}
  }
  &\approx
    2
    \left(
    |\red{g}|^2\red{S}_{\cl{c}}
    + \frac{4 \red{B} }{\gamma_q}
    \right)
    \frac{2\gamma_{\vphi}}{\gamma_q}
    \text{ with } \gamma_{\vphi} > \gamma_{q}
\end{align}
This replicates the result from \cref{eq:readout_comparison}, using $\Delta F_{\HqM} = \gamma_{\vphi}/4$ and  $\Delta F_{\qo{E}} = \gamma_{q}/2$. These particular relationships between signal and cavity bandwidths and the frequency spans arise because  \cref{eq:readout_comparison} was computed assuming that the frequency response was constant. For homodyne readout, one can use a concept of an ``effective bandwidth'' to reason as if the response is constant. A non-constant response then gets converted into a bandwidth using the integral
\begin{align}
  \Delta F_{\HqM} \simeq \int_0^{\infty} \frac{\SPEC^2(\Omega)}{\red{\SPEC}^2} \frac{\rmd \Omega}{2\pi} \text{ with } \red{\SPEC} = \max_{0 \le \Omega < \infty} \SPEC(\Omega)
\end{align}

The comparison between photon counting and homodyne readout with signal recycling in the wideband case, $\gamma_{\vphi} < \gamma_{q}$, is
\begin{align}
  \frac{
  \VAR{\qo{G}_\param}_{\text{SR}}
  }{
  \VAR{\qo{\chi}'^2_{2N}}_{\text{SR}}
  }
  &\approx
    \left(
    |\red{g}|^2\red{S}_{\cl{c}}
    + \frac{\red{B}}{\mathcal{N}_{q}}
    \right)
    \left(\frac{c(\gamma_{q} + \gamma_{g} + \gamma_{\eta})}{4L\gamma_{q}(\gamma_{\vphi} + \gamma_{g} + \gamma_{\eta})}\right)
\end{align}
and in the typical wideband case $\gamma_{\eta} < \gamma_{g} < \gamma_{q} < \gamma_{\vphi}$, the Homodyne readout is optimal without signal recycling, while the photon counting is optimal with recycling. The ratio can be expressed as
\begin{align}
  \frac{
  \VAR{\qo{G}_\param}_{\text{SR}}
  }{
  \VAR{\qo{\chi}'^2_{2N}}_{\text{SR}/\text{BB}}
  }
  &\sim
    \left(
    |\red{g}|^2\red{S}_{\cl{c}}
    + \frac{\red{B}}{\mathcal{N}_{q}}
    \right)
    \left(\frac{c }{4L\gamma_{\vphi}}\right)^{\pm 1}
\end{align}
Where the exponents $+ 1$ and $- 1$ represent the respective broadband and signal recycled cases for the homodyne readout. For signals with $\gamma_{\vphi} \approx c/4L$, the cases are equivalent. For this particular signal bandwidth, signal recycling with photon counting avoids the degradation of $8\Delta F_{\HqM}/\Delta F_{\qo{E}}$ in its comparison to homodyne readout.

\subsection{Narrowband Signal photon counting}

The narrowband regime exists when the signal bandwidth is substantially more narrow than the readout cavities can be made with current technology, $\gamma_{\vphi} < \gamma_q$.  

Using the signal spectrum from \cref{eq:SPEC_narrow_ansatz}. The equations can be constructed
\begin{align}
  \mathcal{N}_{\SPEC}
  &=
    \Delta T \frac{\gamma_{\vphi}}{2}
    \int_{-\infty}^{\infty}
    \left|\frac{g'(\Omega)}{g(\Omega)}\mathfrak{q}(\Omega)\right|^2\SPEC(\Omega)
    \frac{\rmd \Omega}{2\pi}
  \\
    \red{G} &= \frac{\mathcal{N}_{\SPEC}|\red{g}|^2}{\gamma_{\vphi}}
\end{align}
And using $\mathcal{N}_{q}$ from \cref{eq:N_q}, gives the variance as:
\begin{align}
  \frac{1}{
  \VAR{\qo{G}_\param}
  }
    &=
      \frac{\omega^4}{c^4}\frac{F^2_{\text{BS}}}{\gamma_{\vphi}^2}
      \left(
      |\red{g}|^2\red{S}_{\cl{c}}\mathcal{N}_q
      + \Delta T \red{B}
      \right)^{-1}
      \mathcal{N}_{\SPEC}^2
\end{align}
Using the effective signal time-bandwidth as
\begin{align}
  \mathcal{N}_{\SPEC} &\approx \frac{\Delta T c}{2L}
                        \frac{\eta_g \gamma_{\vphi}}{\gamma_{\vphi} + \gamma_{g} + \gamma_{\eta}}
  &&\text{recycling }
  \\
  \mathcal{N}_{\SPEC} &\approx \Delta T \gamma_\vphi 
  &&\text{no recycling}
\end{align}

This parameterization collects the signal enhancement into the effective number of measurements, so that they can be directly compared. Using signal recycling with ${\gamma_g = \gamma_{\eta}}$ gives that recycling accelerates the search by $c/8L\gamma_{\eta}$ over no recycling.

\subsubsection{Comparisons}

The head to head comparison of photon counting to homodyne readout in the no recycling case with a narrowband search is given by the ratio:
\begin{align}
  \frac{
  \VAR{\qo{G}_\param}_{\text{BB}}
  }{
  \VAR{\qo{\chi}'^2_{2N}}_{\text{BB}}
  }
  &\approx
    2
    \left(
    |\red{g}|^2\red{S}_{\cl{c}}
    + \frac{2 \red{B} }{\gamma_q}
    \right)
    \frac{\gamma_{q}}{\gamma_{\vphi}}
    \text{ with } \gamma_{\vphi} < \gamma_{q}
\end{align}
and in narrowband signal recycling case $\gamma_{\vphi} < \gamma_{q}$
\begin{align}
  \frac{
  \VAR{\qo{G}_\param}_{\text{SR}}
  }{
  \VAR{\qo{\chi}'^2_{2N}}_{\text{SR}}
  }
  &\approx
    \left(
    |\red{g}|^2\red{S}_{\cl{c}}
    + \frac{\Delta T\red{B}}{\mathcal{N}_{q}}
    \right)
    \left(\frac{
    c\gamma_{q}(\gamma_{\vphi} + \gamma_{g} + \gamma_{\eta})
    }{
    4L\gamma_{\vphi}^2(\gamma_{q} + \gamma_{g} + \gamma_{\eta})
    }\right)
\end{align}
which reduces in the typical narrowband case ${\gamma_{\vphi} < \gamma_{\eta} < \gamma_{g} < \gamma_{q}}$ to
\begin{align}
  \frac{
  \VAR{\qo{G}_\param}_{\text{SR}}
  }{
  \VAR{\qo{\chi}'^2_{2N}}_{\text{SR}}
  }
  &\sim
    \left(
    |\red{g}|^2\red{S}_{\cl{c}}
    + \frac{\Delta T\red{B}}{\mathcal{N}_{q}}
    \right)
    \left(\frac{
    c\gamma_{g}
    }{
    4L\gamma_{\vphi}^2
    }\right)
\end{align}

Both the broadband and signal recycling cases show that photon counting is only advantageous in extreme cases, where the background noise rate, $|\red{g}|^2\red{S}_{\cl{c}}$, is substantially smaller than $\gamma_{\vphi}/2\gamma_{q}$ or smaller than $4L\gamma_{\vphi}^2/c\gamma_{g}$. When using a loss-limited recycling gain $\gamma_{g} \approx \gamma_{\eta}$, the classical noise should be lower than $2L\gamma_{\vphi}^2/c\gamma_\eta$.

\subsection{Scan search comparisons}
Narrowband searches for new physics are typically searches where the center frequency of the narrow signal spectrum is unknown. This changes the figure of merit of the searches. When the readout cavity is more wide than the signal, $\gamma_{q} > \gamma_{\vphi}$, multiple potential signal frequencies are searched simultaneously. This can ultimately save search time, offsetting the time lost from background noise with the time gained from search multiplicity. The improvement factor is $\gamma_{\vphi}(\gamma_{q}^{-1} + \gamma_{g}^{-1})$. This gives the ratio comparisons from the previous section as

\begin{align}
  \frac{
  \VAR{\qo{G}_\param}_{\text{BB,scan}}
  }{
  \VAR{\qo{\chi}'^2_{2N}}_{\text{BB,scan}}
  }
  &\approx
    2
    \left(
    |\red{g}|^2\red{S}_{\cl{c}}
    + \frac{2 \red{B} }{\gamma_q}
    \right)
\end{align}
For the broadband, no recycling case. There it is apparent that photon counting can provide a similar speedup as in the wideband search case.

For signal recycling. The comparison is more nuanced, as the homodyne and photon counting cases use different recycling cavity bandwidths. Photon counting should use $\gamma_{g} \approx \gamma_{q}$, while homodyne should use $\gamma_{g} \approx \gamma_{\eta}$. Using these values gives the ratio
\begin{align}
  \frac{
  \VAR{\qo{G}_\param}_{\text{SR,scan}}
  }{
  \VAR{\qo{\chi}'^2_{2N}}_{\text{SR,scan}}
  }
  &\approx
       \left(
       |\red{g}|^2\red{S}_{\cl{c}}
       + \frac{\Delta T \red{B}}{\mathcal{N}_q}
       \right)
       \frac{1}{4(1-\eta)}
       \frac{\gamma_{g}}{\gamma_{\vphi}}
    \label{eq:scan_ratio}
\end{align}
This shows that for scanning searches, the losses in the interferometer set the sensitivity limit. The homodyne search is limited by the ability to match $\gamma_{g}$ with $\gamma_{\eta}$ to the signal bandwidth $\gamma_{\vphi}$. The loss term in this ratio establishes this limit. Practically, interferometer losses in a high-power Michelson can be 1000ppm or smaller. This indicates that classical noises must be smaller than the quantum noise in the broadband interferometer by at least the factor of the losses, indicating that they must be smaller than 1000x in power or 30x in the amplitude spectral density. Under those conditions, scanning line searches using photon counting can outperform homodyne readout. Such ratios are possible in short high-frequency Michelson interferometers. 
\Cref{d:scan_comparisons} gives the full expressions without assuming loss limited bandwidths.

\section{Outlook}\label{sec:outlook}

This work establishes and analyzes photon counting as a promising methodology for using Michelson interferometers to search for new physics. The analysis focuses on incoherent stochastic-noise-like signals, yet utilizes an orthonormal Fourier-like temporal basis to decompose down the observables and statistics for both kinds of searches. The focus on incoherent signals avoids known quantum Fisher information limits in signal detection. This key aspect, and the surrounding analysis, motivates several additional avenues of research into photon counting interferometers:

\subsection{Experimental Demonstrations}

Photon counting is already widely applied for spectroscopy, quantum control of atoms, and most applications of light. It has yet to be developed in high-power Michelson interferometers for two major reasons. The first is that it has not yet been recognized to provide a substantial benefit to searches. Indeed, for phase-sensitive detection of displacement, quantum Fisher information arguments show that one method cannot be better than the other. This work shows that additional nuance must be considered, and that at least some classes of signals can greatly benefit with photon counting.

The second reason is the considerable experimental challenge. High power interferometers circulating kW of power have $10^{22}$ photons/s. Readout bandwidths for stochastic signals range from kHz to MHz, thus to surpass the standard quantum limit from shot noise requires isolating of the readout by $160$dB or more from the interferometer carrier light. Interferometer's can achieve >40dB of isolation in operating at a dark fringe, limited by wavefront error that creates contrast-defect light. High finesse optical cavities can isolate by >60dB each and can be operated in series. The requisite isolation appears experimentally feasible with today's optical cavity technology, but preventing any leakage of photons at that level of isolation is not assured. The experimental methodology to achieve this must be developed and demonstrated. The GQuEST experiment is being designed and developed to achieve this application. Its goal is to search for stochastic signals as signatures of quantum gravity.

\subsection{Generalized Temporal Basis}

The analysis of the photon counting performance in this section assumed a Fourier-like temporal basis that is passed through a readout filter cavity $\mathfrak{q}(\Omega)$.  Experimental realizations should be feasible which directly template-match to a chosen basis of the optical field, at frequencies $\omega \pm \Omega$. This acts as the converse of deterministic single-photon emitters, acting instead as a photon absorber into a quantum memory. Either single photons or N-photon state absorption may be needed, depending on the application. For a general basis, the absorber's coupling coefficient to the light must be modulated in phase and amplitude to implement the basis overlap integral. Raman transitions of atoms are a specific example of manipulating the coupling strength to a two-level system. 

Practically, this should be demonstrated both for single basis waveforms, as well as continuous, wideband signal processing. This extended goal is necessary to observe meaningful time-bandwidth product, while discriminating which specific basis or frequencies observe single photons. The event-base search section \cref{sec:events} relies on the assumption that this can be done. Having a customizable template basis may also help experimentally with the contrast/isolation issues indicated in ``experimental demonstrations''. For example, cavities naturally implement a Lorentzian temporal-mode basis, with limited spectral roll-off. Modulated-coupling quantum memories could instead implement Gaussian or other wavelet temporal-mode apodization with substantially lower spectral leakage, while providing an addressable readout.

\subsection{FIRSD Equivalent Representation Heuristic}

To speculate on the benefits of photon counting towards existing and proposed instruments, one can relate existing spectral densities to a ``Fisher-Information representative spectral density''(FIRSD) for chi-squared excess residual tests using photon counting. This can be done starting from \cref{eq:readout_comparison} applied to small bands of frequency. Assuming: the upper and lower sidebands are utilized, that there are no background detector counts counts, and that the classical noise follows \cref{eq:classical_noise_reduction}, together forms the relations
\begin{align}
  \SPEC^2(\Omega)\VAR{\qo{\chi}^2_{2}}
  &\equiv
    {\PSD{\text{Total}(\Omega)}}^2
  \\
  &=
    \left(
    {\PSD{\text{Cl}(\Omega)}} + {\PSD{\text{Qu}(\Omega)}}
    \right)^2
                             \\
  \SPEC^2(\Omega)\VAR{\qo{G}_{\param}}
  &\approx
    4\left(
    \frac{
    {\PSD{\text{Cl}(\Omega)}}
    }{
    2{\PSD{\text{Qu}(\Omega)}}
    }
    \right)
    \SPEC^2(\Omega)\VAR{\qo{\chi}^2_{2}}
\end{align}
The sensitivity of matched-template searches follows the relations
\begin{align}
  2\VAR{\qo{\chi}^2_{1}}
  &=
    \VAR{\qo{\chi}^2_{2}}
    \\
  \VAR{\qo{\chi}^2_{1}}
  &\propto
    {\PSD{\text{Total}(\Omega)}}
\end{align}
This leads to the FIRSD sensitivity of
\begin{align}
  {{\textsc{Firsd}[\Omega]}} &\equiv \sqrt{
  \SPEC^2(\Omega)\VAR{\qo{G}_{\param}}
  }
  \\
  &\approx
    \sqrt{
    \frac{
    4{\PSD{\text{Cl}(\Omega)}}
    }{
    {\PSD{\text{Qu}(\Omega)}}
    }
    }
    {\PSD{\text{Total}(\Omega)}}
    \label{eq:FIRSD}
\end{align}
This effective PSD representation of the photon counting Fisher Information can be used to heuristically estimate the significance of existing residual chi-square testing techniques, except where the excess power in the residuals is measured using counts. Note that this technique should be applied where the quantum noise spectrum does not use squeezing, or where the squeezing is removed from the quantum noise spectrum. Using filter cavities or a basis on only a single sideband will change the numerator factor 2 to a 4, but will pass the other sideband to the homodyne readout, providing sensitivity shared between both readout methods. 

At frequencies where the spectrum is dominated by quantum noise, the FIRSD effective sensitivity can be simplified to the form
\begin{align}
  \PSD{\text{Eff}(\Omega)}
  &
    \approx
    2\sqrt{
    \PSD{\text{Cl}(\Omega)}
    {\cdot}
    \PSD{\text{Qu}(\Omega)}
    10^{\text{dB}/10}
    }
    \label{eq:FIRSD_approx}
\end{align}
The ad-hoc factor $10^{\text{dB}/10}$ is included to account for squeezing in the quantum spectrum, and is $2$, $4$, or $10$ for 3dB, 6dB and 10dB of observed squeezing in the quantum noise spectrum, respectively. This allows the quantum noise spectrum to be directly used from gravitational wave liturature, with the assumed squeezing factor applied. 

The FIRSD spectrum representation is formulated here as a means of discussing the relative figures of merit between detection schemes. What it misses is a means to express any penalties, statistical or scientific, from using phase-insensitive observations. The event-based searches section \cref{sec:events} formulates arguments to bound this penalty. Any use of this spectral density representation must qualify that photon counting is not as universally accessible to signal analysis as the timeseries data from homodyne readout.

\subsection{Next Generation Gravitational Wave Detectors}
For photon counting, the temporal basis must be orthonormal and must be chosen apriori, before detection. Current technology also requires a basis of Fourier modes passed through readout cavities, though future technology can feasibly improve this restriction. All of this indicates major tradeoffs for matched template searches if switching to photon counting measurement, despite potential benefits in avoiding quantum noise.

However, homodyne readout and photon counting are not exclusive techniques and can be combined, when temporal-mode basis not directed to a counting readout, they will instead land on a standard readout. After the detection of an astronomical event, many additional science tasks follow to analyze the waveform. Many of these tasks can be framed as testing for excess residuals after subtracting the best-fit waveform or otherwise testing for deviations from a ``standard-model'' waveform. Thus, if the expected waveform can be established early during detection, then all deviations exist in orthogonal templates. This poses photon counting as a means to perform quantum hypothesis tests of excess residuals, with reduced quantum noise, for gravitational wave analysis.

Such potential applications for this are: tests of general relativity; searching for deviations from numerical relativity simulations; searching for weak, delayed gravitational wave ``echoes''; and testing for deviations between inspiral and merger portions of waveforms. Binary neutron star signals have similar potential applications: Searches for late-inspiral phase shifts in the signal from tidal damping in the equation of state; as well as searches for high-frequency resonances after the merger.
Applying photon counting techniques to gravitational wave interferometers will have profound experimental challenges, but further development of sub-shot-noise measurement techniques and data analysis may reveal new science possibilities. Future generations proposed gravitational wave detectors further motivate this study for two reasons.

First, they have a larger separation between their quantum and classical noise contributions due to the various scalings with the facility length \cite{abbottCQG17ExploringSensitivity, dwyerPRD15GravitationalWave}. The gain-bandwidth trade-offs\cite{mizunoPLA93ResonantSideband} cause quantum noise to scale less favorably than the next most dominant noises, primarily the coating thermal noise as well as the residual gas noise of molecules traversing the vacuum beam pipes.

\begin{figure}
  \centering
  \includegraphics[width=1\linewidth]{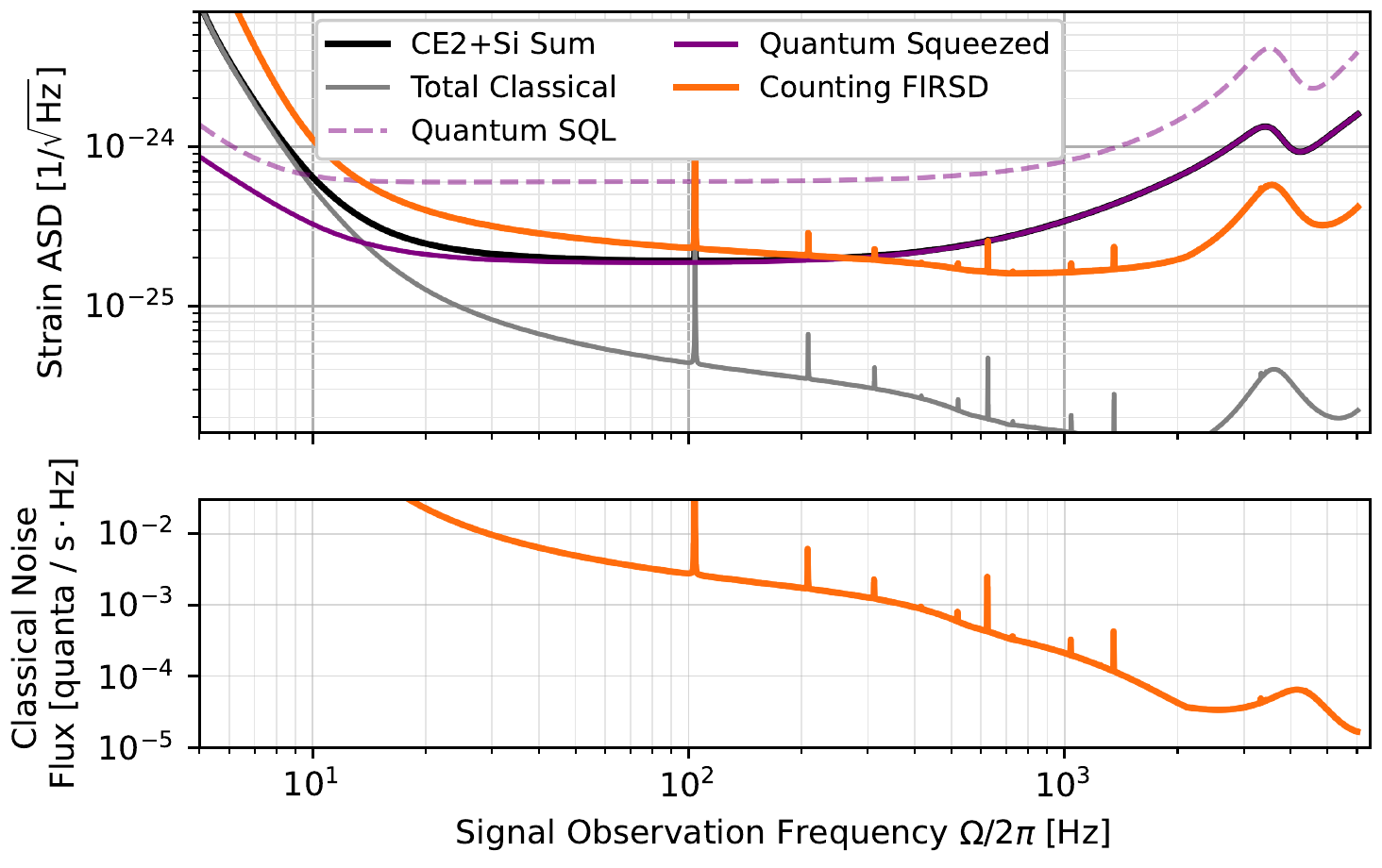}
  \caption{The classical and quantum noise of Cosmic Explorer\cite{evansAAP21HorizonStudy}, a proposed observatory design. \textbf{Top} shows the fundamental noises in a proposed upgraded configuration using cryogenic silicon technologies with 3 mega-Watts of arm power (per table IV of \cite{hallPRD21GravitationalwavePhysics}). The proposed design includes 10dB of squeezing, so the standard quantum limit noise contribution, without squeezing, is included. The FIRSD spectrum is computed from \cref{eq:FIRSD}. \textbf{Bottom} shows the background flux from classical noises in both upper and lower sidebands using \cref{eq:classical_estimation}. The noise curves are calculated using the scientific software \textit{pygwinc}.}
  \label{fig:CE2}
\end{figure}

\Cref{fig:CE2} shows the spectra of a configuration of the proposed Cosmic Explorer observatory design. The gap between the quantum and classical noise contribution can be orders of magnitude in the power spectral density. The FIRSD heuristic expression then indicates that photon counting -- even without further quantum enhancement -- can outperform squeezing in the science cases where it is found to be applicable. Furthermore, it can outperform narrow-band resonant tunings, where the optical cavities are optimized for specific high-frequency science objectives\cite{SrivastavaA22SciencedrivenTunable}. This is because the detector operates close to the limits of anticipated internal losses \cite{miaoPRX19QuantumLimit}.

Second, next generation observatories will detect substantially more events than the current generation \cite{evansAAP21HorizonStudy, maggioreJCAP20ScienceCase}. For the Cosmic Explorer design, $10^5$ binary neutron star coalescence events are predicted to be observed per year of operation. More events potentially mitigates some of the lost information of purely excess-residual searches, and indicates that the Fisher information as an applicable figure of merit in aggregating information from many events.

The bottom of \cref{fig:CE2} shows the low count rate expected from the fundamental classical noises. For each event detection, some number independent waveform tests might be defined. For binary neutron star postmerger physics, the number of independent tests can likely related to the maximum resonance Q-factor of oscillation features and the band over which they are expected to appear\cite{ganapathyPRD21TuningAdvanced}. The noise flux rate times the number of independent tests indicates the average number of background photons observed per event. Until a photon is observed, upper bounds on the postmerger signal power may be the primary statistical conclusion, but that thousands of events may allow postmerger-waveform morphology to be accessible even for the cosmologically-distant neutron star events comprising the majority of the detections.

It is worth reiterating the experimental challenges ahead for photon counting, but also reiterate that the technique may open new potential for future detectors. It shows promise for the observatories to even more fully utilize the advantages of their long baseline designs as well as the reduced classical noise from emerging physics instrumentation techniques \cite{adhikariCQG20CryogenicSilicon}.

\subsection{Quantum Enhancements}

The classical Fisher Information and variance calculations for photon counting (\cref{sec:counting}) show that the information scales proportionally to optical gain, as one would expect, yet inversely proportional to the noise background. An implication is that for arbitrarily low-background instruments with nonzero optical losses, the limits implied from \cite{tsangPRL11FundamentalQuantum} do not universally apply. As a result, the conclusions of \cite{demkowicz-dobrzanskiPRA13FundamentalQuantum} either, do not apply to photon counting, or must be interpreted differently.

Squeezed states are the principle method of incorporating a quantum enhancement into interferometers, to reduce quantum noise below the shot-noise SQL in homodyne readout. Soon, advanced frequency-dependent squeezed states will be utilized to also reduce measurement back-action. The technique of squeezing works by continuously preparing states over the measurement bandwidth using nonlinear optics and ``injecting'' them into the interferometer from the output port. They then reflect from the Michelson interferometer\cite{cavesPRD81QuantummechanicalNoise} towards the readout. Any signal modulations act as a displacement operation on those states, during the interferometer reflection.
Squeezed states carry energy and carry pairs of photons, entangled between upper and lower sideband frequencies \cite{schnabelPR17SqueezedStates, maNP17ProposalGravitationalwave}. This likely limits the utility of squeezed states for low-background photon counting, as losses decohere those pairs, leading to background counts. Photon counting is a nonlinear and non-Gaussian observation of the optical states in each temporal mode. This implies that, analogously to squeezing, non-Gaussian states should be prepared and injected into the dark port of the interferometer to enhance photon counting readouts.

Squeezed states are generated using relatively few optical elements, yet span all of the temporal basis read in the homodyne timeseries, changing the observed quantum contribution to the spectrum. Each and every temporal mode in any basis of the observation band then contains a squeezed state. Implementing this with non-Gaussian states to saturate a time-bandwidth basis will require considerable investigation and development. Such ideas are already being explored in microwave axion detectors \cite{dixitPRL21SearchingDark}, and may have analogous developments for optical interferometers -- if photon counting can be proven to be experimentally viable in high power instruments. Many experimental challenges of creating custom temporal mode quantum states remains, but may reveal new prospects for precision optical interferometer physics.

\subsection{Quantum Information}

Given known experimental limits, a principle question to ask is how well current observation and quantum enhancement techniques saturate provable theoretical measurement limits. To that end, interferometers have been extensively studied. When reading a timeseries, the optical gain and sensing noise are related through the ``energetic quantum limit\cite{braginskyACP00EnergeticQuantum}'', bounding the achievable precision when using light as a probe for displacement from the probe energy and its uncertainty. Extending these bounds and including the measurement back-action\footnote{Measurement back-action, which can imply other definitions for a ``standard quantum limit'', is mentioned here for completeness, but not analyzed in this work. } of ``too precisely'' probing the location of finite mass oscillators, continuously, leads to the Quantum Cramer-Rao Bound (QCRB)\cite{tsangPRL11FundamentalQuantum, miaoPRL17FundamentalQuantum}. Applied, this bound establishes fundamental limits of phase-sensitive time-series measurement in lossy, realizable detectors \cite{miaoPRA17GeneralQuantum, miaoPRX19QuantumLimit, tsangNJP13QuantumMetrology} with a realization that injecting Gaussian squeezed states approach saturating channel bounds.

The timeseries readout of interferometers is extremely convenient for all manner of scientific tests, as it linearly relates perturbations in the interferometer to modulations in the readout power. Standard, intuitive signal analysis is then applied to statistically define a ``detection test'' for: wide-bandwidth stochastic backgrounds, coherent or incoherent narrow-bandwidth fields, or finite-duration events. For events, a significant detection then leads to a different statistical problem, inferring the most likely waveform given measured data. Alternative to this waveform inference is the statistical test for significant deviations or ``features'' between the measured waveform and a set of expected event waveforms, either in a single instance or while aggregating many events, depending on the scientific goals. 

While convenient, the limitations of the timeseries readout has major implications \cite{demkowicz-dobrzanskiPRA13FundamentalQuantum}. In particular, the technique of using squeezed light approaches saturation of quantum waveform estimation bounds, implying that optical losses and decoherence establish the fundamental precision limit for a given interferometer circulating power.

Properly comparing theoretical and experimental quantum-measurement limits in a well-posed manner requires fully defining the statistical test necessary to achieve a science result. A specific example of this is the established knowledge that the timeseries readout and its associated QCRB does \emph{not} saturate more general quantum bounds. There are two notable examples of this. The first example is given by the Kennedy and Dolinar receivers\cite{ dolinar73ProcessingTransmission, kennedy73ProcessingTransmission, wittmannPRL08DemonstrationNearOptimal,tsujinoPRL11QuantumReceiver, becerraNP13ExperimentalDemonstration}, developed for hypothesis testing to discriminate between sets of candidate waveforms. These receivers are designed instead to saturate the Helstrom Bound\cite{helstrom76QuantumDetection, helstromJSP69QuantumDetection} in detecting a weakly-displaced coherent signal, are applicable to interferometer waveform estimation applications \cite{tsangPRA12FundamentalQuantum}. These \emph{do not} rely on non-classical states of light yet achieve the 3db improvement over homodyne readout from the Helstrom bound. Their statistical application generalizes as continuous hypothesis testing\cite{kiilerichPRA18HypothesisTesting}. The second example is the aforementioned quantum spectrum analysis \cite{ngPRA16SpectrumAnalysis}, which concludes that for the purposes of estimating a weak, stochastic signal over any bandwidth, the homodyne timeseries is inferior to using frequency-resolved photon counting. This work generalizes that result to signal-recycled interferometers with generalized a temporal-mode basis. In summary, optimizing for waveform estimation by implementing by a timeseries readout, even with squeezed light, can limit the output of instruments which either: search for incoherent signals, or frame their science as discriminating between finite sets of candidate waveform models carrying some total signal power.

The aforementioned quantum bounds have a common feature in that they are applied to detecting weak signals of some power density, rather than the specific sensitivity limit of an interferometer. This analysis does not include measurement back-action or quantum enhancements, in which case the Michelson interferometer can be considered a relatively simple transducer, with an associated optical gain, that emits weak signals from new physics. This leads to, for interferometers, the conclusion that the QCRB for waveform estimation\cite{tsangPRL11FundamentalQuantum} can be surpassed for certain physics applications. If this conclusion is connected to additional prospects for quantum enhancement using non-Gaussian states, then there is new potential for a wider variety pf application-specific quantum information bounds for interferometer signal science use-cases.

\newcommand{\ha}{h_0}
\newcommand{\hb}{h_1}
\newcommand{\rhb}{\red{h}_1}
\newcommand{\bparam}{\boldsymbol{\beta}}
\section{Event Based Searches}\label{sec:events}

Gravitational wave detection is a key application of modern high power Michelson interferometers. Some astrophysical science signals are linear time invariant stochastic backgrounds, but gravitational wave detections, so far, are from astronomical events, where a signal of finite duration is emitted and detected. The stochastic analysis developed in the earlier sections constructs chi-square power observations and their analogous implementation using photon counting. For event-based searches, this makes the formalism so far developed more similar to incoherent ``burst'' searches.
A work that extensively develops and describes similar time-bandwidth arguments to relate excess-power burst searches against matched-template is provided by Flanagan and Hughes (F\&H)\cite{flanaganPRD98MeasuringGravitational, flanaganPRD98MeasuringGravitationala}. In particular, the heuristic comparison for burst searches that appears to apply is the relation eqn (2.15) of \cite{flanaganPRD98MeasuringGravitational}.
\begin{align}
  \frac{\sigma_{\textsc{S/N}, \text{band-pass}}}{\sigma_{\textsc{S/N}, \text{matched}}}  &\approx \frac{1}{\sqrt{N_{\text{bins}}}} & N_{\text{bins}} \approx 2 \Delta T_{\text{bank}} \Delta F_{\text{bank}}
\end{align}
The $(S/N)$ in \cite{flanaganPRD98MeasuringGravitational} is analogous to the square-root of the Fisher information or, equivalently, the inverse of the standard deviation of an unbiased estimator. The factor $N_{\text{bins}}$ is, heuristically, the time and bandwidth spanned by the bank of templates. This section develops comparisons of photon counting with timeseries waveform inference, showing that the benefits of low classical noise remain, and that the statistical penalty is smaller than this heuristic expectation.  

\subsection{Template Search Analysis}

Event searches are implemented by convolving\footnote{Computationally, it is substantially more sophisticated than brute force convolution, using Fourier-domain waveform templates over an optimized grid of candidates.} a set of model waveform templates over the homodyne timeseries while checking a search statistic ``score'' against a threshold.  Sufficient score then signifies a significant detection. Over the set of candidate waveforms, this implements matched-template search. The temporal-mode basis used in this work, particularly for photon counting, is not compatible with convolution nor is it compatible with search over continuous space of parameterized waveforms. This is due to the ortho-normality property (cf. \cref{eq:orthogonality}) of the temporal-mode basis. Similarly, it cannot freely test a space of waveforms to perform parameter estimation. This might suggest that photon counting is not promising for event-based science; however, the event detection initiates follow-up science on the waveform. Photon counting can potentially improve deep, sub-shot noise tests for new features of waveforms.

To define such tests, start with the strain waveform broken into well-known and unknown components.
\begin{align}
  h(t) &= \ha(t) + \hb(t | \bparam)
\end{align}
$\hb(t | \bparam)$ represents the uncertain components of the waveform, and the uncertainty is parameterized with a vector $\bparam$ that has some apriori distribution $P(\bparam)$. The dimensionality of the parameter space is given by $N_{\text{param}} = \textsc{Dim}[\bparam]$.

One can assume that the majority of the signal energy for $\ha(t)$ is at times earlier than $\hb(t)$ and typically at frequencies that are lower than $\hb(t)$, which implies that $\ha(t)$ and $\hb(t | \bparam)$ are highly orthogonal. The measurement of $\ha$ is assumed to be performed using homodyne readout and estimated with high SNR. This assumption doesn't strongly affect formulas below, but is stated to indicate that waveform inference on $\ha(t)$ is needed to determine source properties and establish the prior distributions $P(\bparam)$ for the waveform models of $\hb(t | \bparam)$.

Assuming either the raw interferometer response $g(\Omega)$, or the response $\mathfrak{g}(\Omega)$ filtered through $\mathfrak{q}(\Omega)$, the field at the output port of the interferometer is:
\begin{align}
  \qo{d}(\Omega) 
  &= i \delta(\Omega) \red{l} + i \gq(\Omega) L h(\Omega)
    \nonumber\\ &\hspace{2em}+ \mathfrak{q}(\Omega)\big(i\VNP(\Omega) + \VNA(\Omega)\big) + \qo{a}(\Omega)
\end{align}
Where $L$ is the length of the interferometer arms that converts from gravitational metric strain to displacement.

To develop the timeseries search statistic, first define the unit-normalized template, weighted by the instrument response.
\begin{align}
  \rhb(\Omega | \bparam) &\equiv \frac{\red{\mathfrak{g}}(\Omega)h_1(\Omega | \bparam)}{\sqrt{\int_{-\infty}^{\infty}|\red{\mathfrak{g}}(\Omega)h_1(\Omega | \bparam)|^2 \frac{\rmd \Omega}{2\pi}}}
\end{align}

\subsubsection{Homodyne Detections}
First, we will derive known statistics for matched-template search using the framework and notation of this work.
To simplify, we will assume that the classical noise and optical gain is not varying with frequency $\mathfrak{q}(\Omega)\SNP(\Omega) \rightarrow \SNPr$, and $\red{\mathfrak{g}}(\Omega) \rightarrow \red{\mathfrak{g}}$. This then allows us to create a measurement operator, power operator, and then unbiased statistic on the single template $\K_{\bparam}(\Omega)=\rhb(\Omega | \bparam)$. For this, the statistic is built on the underlying Gaussian measurement, normalized by the homodyne measurement gain $\red{l}$, without creating the power measure chi-square of \cref{ssec:signal_power}.
\begin{align}
  \braket{\qo{D}_{\bparam}/\red{l}} &= 2|\red{\mathfrak{g}}|LH(\bparam)
  &
    \VAR{\qo{D}_{\bparam}/\red{l}} &= 2\SNPr + 1
\end{align}
These use $H(\bparam)$, for the RMS strain energy of the $\hb(\Omega | \bparam)$ waveform, expressed:
\begin{align}
  H^2(\bparam) &= \int_{-\infty}^{\infty} \left| \hb(\Omega | \bparam) \right|^2 \frac{\rmd \Omega}{2\pi}
\end{align}

Rescaling and shifting the measurement creates the unbiased estimators.
\begin{align}
  \qo{\mathcal{N}}_{\bparam}
  &=
    \frac{\qo{D}_{\bparam}}{2|\red{\mathfrak{g}}|LH(\bparam)\red{l}}
  &
  \VAR{\qo{\mathcal{N}}_{\bparam}} &= \frac{2\SNPr + 1}{\left( 2|\red{\mathfrak{g}}|LH(\bparam)\right)^2}
\end{align}

using \cref{eq:quantum_PSD} and \cref{eq:classical_PSD} to express the strain power-spectral-density in terms of the given noise density operators
\begin{align}
  L^2 S_h(\Omega) &= \PSD{\text{Qu}(\Omega)} + \PSD{\text{Cl}(\Omega)}
\end{align}

this then gives a relation between the variance and a score threshold
\begin{align}
  \red{\rho}_{\HqM}^2
  &\equiv
    \frac{1}{\VAR{\qo{\mathcal{N}}_{\bparam}}}
    = \frac{2H^2(\bparam)}{\red{S}_h}
    = \frac{2L^2H^2(\bparam)}{\PSD{\text{Qu}} + \PSD{\text{Cl}}}
\end{align}

Which agrees with the exact, optimally-weighted calculation from eqn 1.1 (or 2.7) of F\&H\cite{flanaganPRD98MeasuringGravitational}.
\begin{align}
  \rho_{\HqM}^2
  &\equiv
    4 \int_{0}^{\infty} \frac{|h_1(\Omega)|^2}{S_h(\Omega)} \frac{\rmd\Omega}{2\pi}
    \\
    &=
    2 \int_{-\infty}^{\infty} \frac{2|\red{g}(\Omega)|^2|L h_1(\Omega)|^2}{(1 + 2\SNP(\Omega))} \frac{\rmd\Omega}{2\pi}
      \approx
      \red{\rho}_{\HqM}^2
\end{align}
The difference is that the optimal template weighting moves the strain spectral density within the integral. The necessary weighting factors and their optimization were not included in this derivation for simplicity.

The search score $\rho^2$ indicates the expected significance for a given signal power. For a matched template search, there is some spacing of $\bparam$ over which the waveforms are orthogonal, and thus there is some total number, $N_{\text{shapes}}$ of independent tests of the search space (cf. 2.8 of \cite{flanaganPRD98MeasuringGravitational}). Additionally, there is some probability threshold, $\epsilon$ that is used to indicate a significant detection or event.

From eqn 2.8 of F\&H\cite{flanaganPRD98MeasuringGravitational}, the score threshold follows the relation
\begin{align}
  \text{erfc}\left(\frac{\rho_{\HqM}}{\sqrt{2}}\right) &= \frac{\epsilon}{N_{\text{shapes}}}
\end{align}

Leading to eqn 2.9 of F\&H\cite{flanaganPRD98MeasuringGravitational}
\begin{align}
  \red{\rho}_{\HqM} &\approx \sqrt{2\ln(N_{\text{shapes}}/\epsilon)}
\end{align}
The number of shapes is also related to the number of ``sufficiently independent'' templates in the computational search space, whereby the statistical penalty between $\ln(N_{\text{shapes}})$ and signal power loss from under-sampling are made similar.

For sub-shot noise detections, we assume that many events will need to be ``stacked'' or evaluated in a aggregate statistic. In some cases, it may be possible to re-scale and transform the timeseries across many events, such that waveforms can be coherently added, realizing the benefit of \cite{higginsN07EntanglementfreeHeisenberglimited}. In that case, the average-event search threshold becomes
\begin{align}
  \text{erfc}\left(\rho_{\HqM}\sqrt{\frac{N_{\text{events}}}{2 }}\right) &= \frac{\epsilon}{N_{\text{shapes}}}
\end{align}
This will improve the average threshold strain power by $\sqrt{N_{\text{events}}}$. Given that inspirals have many parameters, it is perhaps more reasonable to assume that only incoherent stacking may be possible for certain tests. these will have the scaling
\begin{align}
  \text{erfc}\left(\rho_{\HqM}^2\sqrt{\frac{N_{\text{events}}}{2 }}\right) &= \frac{\epsilon}{N_{\text{shapes}}}
\end{align}
which leads to the detection threshold for the signal power of
\begin{align}
  H^2_{\HqM}
  &\approx
    \red{S}_h\sqrt{\frac{\ln(N_{\text{shapes}}/\epsilon)}{2 N_{\text{events}}}}
  &
    \red{S}_h \equiv \frac{2\SNPr + 1}{2|\red{\mathfrak{g}}|^2L^2}
    \label{eq:ts_detection_threshold}
\end{align}
Before analyzing photon counting, it is worth noting the general scale of the number of shapes. From Eqn. 7.5 and eqn 7.12 F\&H\cite{flanaganPRD98MeasuringGravitationala} The number of independent waveforms scales can be expected to scale volume of the parameter space. In which case
\begin{align}
  \ln(N_{\text{shapes}}) \propto N_{\text{params}}
\end{align}
and thus ultimately
\begin{align}
  H^2_{\HqM}
  &\propto
    \red{S}_h\sqrt{\frac{N_{\text{params}}}{N_{\text{events}}}}
\end{align}
The crossover for when matched template search and excess-power burst search can then be considered as when $N_{\text{params}} \sim N_{\text{bins}}$, under the amended heuristic
\begin{align}
  \frac{\textsc{S/N}_{\text{band-pass}}}{\textsc{S/N}_{\text{matched}}} &\propto \sqrt{\frac{N_{\text{params}}}{N_{\text{bins}}}}
\end{align}

\subsubsection{Photon Counting Detections}
As mentioned, photon counting intrinsically appears more like a burst search, as it is forced to use excess-power measurements. This deleterious assumption is only partially true. Avoiding it requires us to assume that measuring the photon power in individual template measurements becomes possible. I.e. the ability to individually measure $\qo{E}_i$ rather than the summation $\qo{E}$. In the case of quantum spectrum analysis, individual line filters using cavities or gratings could achieve this \cite{ngPRA16SpectrumAnalysis}. For more general templates, or templates with optimized spectral leakage properties, this means using emerging technology such as quantum memories.

To start on the statistical analysis, we first must establish the required number of templates. The total number of templates, $N_{\text{bins}}$ is more exactly determined from the Von Neuman Entropy over a density matrix or phase-space representation, $K(t, t')$, of the templates, averaged over the parameter space. This representation of the waveform uncertainty has the construction.
\begin{align}
  K_{h_1}(t, t') &= \int_{\bparam}
      \rhb(t | \bparam)\rhb(t' | \bparam)
                      P(\bparam)\rmd \bparam
\end{align}
The number of bins is related to the information entropy on this phase-space density-matrix of the waveforms.
\begin{align}
  N_{\text{bins}}
  &= e^{S_{K}},
  &
    {S}_{K}
  &= -\int_{-\infty}^{\infty} K_{h_1}(t, t) \ln\big(K_{h_1}(t, t)\big) \rmd t
\end{align}
Where the number of bins also approximately corresponds to the time-bandwidth product spanned by the set of templates. This number of bins establishes the number of ortho-normal templates that will be needed to fully capture the event signal power for any given event.

Now assume a template basis set $\K_i(t) \in \mathcal{K}$ that are unit-normalized and ortho-normal. In the event that the local oscillator light is disabled or filtered away by $\mathfrak{q}(\Omega)$, emerging technology can template-match to the optical field, implementing the operators
\begin{align}
  \qo{d}_i
  &\equiv \int_{-\infty}^{\infty}\qo{d}(t)\K_{i}(t) \rmd t \equiv \int_{-\infty}^{\infty}\qo{d}(\Omega)\K_{i}(-\Omega) \rmd \Omega
\end{align}
A number operator can be built from these field operators, to measure how many photons have accumulated in each template.
\begin{align}
  \PqE_{i}
  &\equiv
    \qo{d}_i^\dagger 
    \qo{d}_i 
\end{align}

With the intent that each of these power measurements can be individually addressed during observation. Using an array of quantum memories, each with an orthogonal, modulated coupling rate then implements the template set $\mathcal{K}$. Each one can then be projectively-measured to determine the occupation.

Using the set of measurements of $\qo{E}_i$, an unbiased search statistic can be build using a weighted combination of them.
\begin{align}
  \qo{G}(\bparam)
  &\equiv
    \frac{1}{\red{G}(\bparam)}
    \sum_{i=0}^{N_{\text{bins}}} (\qo{E}_i - \red{S}_i) W_i(\bparam)
    \\
  \red{G}(\bparam) &\equiv
                     \sum_{i=0}^{N_{\text{bins}}} \braket{(\qo{E}_i - \red{S}_i) W_i(\bparam)}
\end{align}
A good choice of weights, determined in a manner similar to \cref{D:chi_weighted}, is based on the relative occupation of the templates by the true waveform.
\begin{align}
  W_i(\bparam) &= \left|
  \int_{-\infty}^{\infty}
  \rhb(\Omega | \bparam)\K_i(-\Omega)
  \frac{\rmd \Omega}{2\pi}
  \right|^2
\end{align}
This choice leads to the properties
\begin{align}
  \sum_{i=0}^{N_{\text{bins}}} W_i(\bparam) &= 1 & \sum_{i=0}^{N_{\text{bins}}} W_i^2(\bparam) &\equiv \frac{1}{N_{\text{split}}(\bparam)}
\end{align}
This $N_{\text{split}}(\bparam)$ is a measure of how many of the templates are covered by the waveform given by the parameter $\bparam$. For good choices of templates optimized for particularly high-probability parameters $N_{\text{split}} \sim 1$. For low probability parameters, or poorly optimized parameters $N_{\text{split}} \sim N_{\text{bins}}$.

The calculations on the normalization factors enter the computation for the gain-normalization factor, $\red{G}$, giving
\begin{align}
  \red{G}(\bparam)
  &=
  \frac{\int_{-\infty}^{\infty}|\mathfrak{g}(\Omega) L h_1(\Omega | \bparam)|^2 \frac{\rmd \Omega}{2\pi}}{N_{\text{split}}(\bparam)}
\end{align}

Now, for simplicity, we make the assumption that the classical noises are constant $\red{S}_i \rightarrow \red{S}$ and that the optical gain is constant $\mathfrak{g}(\Omega) \rightarrow \red{\mathfrak{g}}$. The variance for the unbiased estimator, $\braket{\qo{G}(\bparam)} = 1$, is then

\begin{align}
  \VAR{\qo{G}(\bparam)}
  &=
    \frac{
    N_{\text{split}}(\bparam)
    \red{S}
    }{
    \big(|\red{\mathfrak{g}}|^2 L^2 H^2(\bparam)\big)^2
    }
\end{align}
over the whole parameter space, the numerical efficiency factor can be averaged
\begin{align}
  \red{N}_{\text{split}} &= \int_{\bparam} N_{\text{split}}(\bparam) P(\bparam)\rmd \bparam
\end{align}
so the average penalty factor should be based on $\red{N}_{\text{split}}$ rather than $N_{\text{bins}}$.
One can expect that $N_{\text{split}}$ depends on the tiling of ortho-normal templates within the waveform parameter space, giving that, on average, one might assume $N_{\text{split}} \sim 2^{N_{\text{param}}}$ from a lattice geometry. More efficient ``sphere-packing'' of the templates within the waveform space might improve this bound, relating the statistical efficiency of photon counting to the Hamming Bound for waveform discrimination.

With the statistic constructed, we assume that many events are detected and incoherently stacked. This is the assumption behind Fisher Information arguments, and it allows us to assume a Gaussian error distribution in the limit of many events.
\begin{align}
\red{\rho}_{G}^2
  &\equiv
\frac{1}{\VAR{\qo{G}(\bparam)}}
    \approx
    \frac{\big(L^2H^2(\bparam)\big)^2}{2 \red{N}_{\text{split}} \PSD{\text{Cl}(\Omega)}\PSD{\text{Qu}(\Omega)}}
\end{align}
After many measurements, the score threshold relates to the statistical threshold by,
\begin{align}
  \text{erfc}\left(\rho_{G}\sqrt{\frac{N_{\text{events}}}{2 }}\right) &= \frac{\epsilon}{N_{\text{bins}}}
\end{align}
which leads to the detection threshold on the signal energy of
\begin{align}
  H^2_G
  &\approx
    \red{S}'_h\sqrt{\frac{2\red{N}_{\text{split}}\ln(N_{\text{bins}}/\epsilon)}{N_{\text{events}}}}
  &
    \red{S}'_h \equiv \frac{\sqrt{\SNPr + \SNAr}}{|\red{\mathfrak{g}}|^2L^2}
\end{align}

\subsubsection{Ratio Relation}

With the detection threshold estimates above, we can now estimate a modified heuristic for the FIRST of \cref{eq:FIRSD} to evaluate the scaling with parameter and time-bandwidth bin dimensionality.

\begin{align}
  \red{S}'_h &\approx \frac{
  \sqrt{2\PSD{\text{Qu}}\PSD{\text{Cl}}}
  }{
  L^2
  }
  \\
  \red{S}_h &\approx \frac{
  \PSD{\text{Qu}}+\PSD{\text{Cl}}
  }{
  L^2
  }
\end{align}

For a system that is primarily dominated by quantum noise, this gives

\begin{align}
  H^2_G \propto \sqrt{8\frac{\red{N}_{\text{split}}}{\red{N}_{\text{params}}}\frac{\PSD{\text{Cl}}}{\PSD{\text{Qu}}}} H^2_{\HqM}
\end{align}
With a constant of proportionality that will depend on the specifics of the template search and will require further study. The key point of this analysis is to show that there is some level of classical noise at which photon counting over multiple events is always superior over a finite set of templates. That level does depend on information arguments related to the size of the search parameter space, but that the new definition of $\red{N}_{\text{split}}$ allows us to analyze the continuum between fully continuous ``waveform inference'' and fully ``discrete'' hypothesis testing. In actuality, the difference in performance depends critically on the relative dimensionality and distribution of the waveforms between the dimensionality relations $1 \le \red{N}_{\text{params}}  \le \red{N}_{\text{split}} \le N_{\text{bins}}$.

\section{Conclusions}

This work methodically describes the measurement process and readout of a Michelson interferometer, starting with a description of the measurement process in \cref{sec:descriptions}. The standard homodyne readout, typically implemented using the interferometer destructive fringe light, provides a continuous timeseries. This gives a linear, classical recording of the passing signals. At the quantum sensitivity limit of the interferometer, there is a background noise on the timeseries, from the shotnoise of independently arriving photons at the detector. In this case, the majority of the photons are provided by the local oscillator field, rather than the signal itself. The derivation of \cref{sec:homodyne} shows how the local oscillator provides a static field for the signal field modulations to with. Calculating an overlap integral with any single temporal mode, the quantum noise statistics will be Gaussian, where any signal in the chosen temporal mode will offset the mean value. Squaring the overlap integral then constructs a chi-square test which can indicate whether a signal is present. By Fisher-information, this chi-square test is as efficient as any other test for the presence of a signal, when using the timeseries data.

In short, searching for signals in a timeseries uses the sequence: define a temporal-mode, perform an overlap-integral, then square it to test for signal power in excess of shot-noise power. This led to an analogous derivation for photon counting, whereby temporal modes are defined, and the overlap integral is performed as part of the derivation, rather than as a calculation. The act of squaring is then performed using single-photon detectors, which directly observe signal power. The conclusion of this derivation is that now only the signal and background classical noises introduce shot-noise, so the Fisher Information can be arbitrarily high, and the associated measurement variance low.

This is a curious conclusion, as weak signals may only rarely result in observed photon emissions, and thus limit the information available. This is true and reflects that quantum limits are still at play, but the so called ``standard quantum limit'' from ``vacuum fluctuations'' are not a necessary limit to Michelson interferometers. When searching for weak stochastic signals from new physics, the search acceleration of this new technique can be dramatic, when interferometers are designed to minimize their classical noise contributions to be well-below the shot noise. For such searches, the derivation provided used the temporal-mode basis only as a formal tool to build the search statistics. Physically, the temporal-modes can be selected using optical cavities, and this is an immediate prospect for implementing photon counting searches for interferometers.

Cavities have many drawbacks, physically and statistically, and in time may be surpassed by a physical implementation of matched-template-search, directly in the optical domain with high-contrast and low spectral-leakage. This suggests interesting new experimental avenues using quantum memory devices for fundamental physics measurements.

After deriving the statistics and benefits of photon counting, \cref{sec:sensitivity_comparison} considered wider classes of Michelson interferometers - those with signal recycling cavity enhancements. There, the measurement bandwidth is traded for optical gain. Both wideband and narrow band searches were considered. For wide-band searches, where the signal has a bandwidth or coherence-time comparable to light-crossing time of the interferometer, signal recycling does not provide substantial benefit and photon counting maintains its advantages. A special case for this statement is when the filter cavities used to implement photon counting pass less bandwidth than the signal provides. In this case, signal recycling can added to recover the full statistical benefit of photon counting over homodyne readout. For narrow-band signals,  particularly those with signal power at an unknown frequency - such as in searches for dark-matter, photon counting is beneficial only when the interferometer has extremely low classical noise, such that even the resonantly-enhanced classical noise is below the shot noise limit. 

The formal sections outline specific areas where photon counting can benefit interferometer searches. The outlook, \cref{sec:outlook}, then speculates on further application areas. In particular, photon counting shows that known limits to squeezing from losses do not imply as universal limits to quantum measurement in interferometers, or at least must be adapted or reinterpreted for photon counting.  For applications that can be posed as binary or N-ary tests, rather than continuous waveform inference, a new statistical measure, the ``Fisher-Information representative spectral density'', is introduced to used to reason about potential future benefits of photon counting for domain-specific applications. Applications, such as gravitational-wave astronomy and instrumentation, that typically use power-spectral density as the figure of merit can then project more complex statistical measures, under some bold assumptions that eventual technology can provide similar spectral resolution as current classical timeseries computations.

In total, the technical derivations in this work bridge recent development in quantum measurement theory to physics applications best suited for optical interferometers. This motivates the substantial work ahead to demonstrate the experimental feasibility to realize ``Counting Quantum Limited'' rather than ``Standard Quantum Limited'' performance.

\textbf{Acknowledgements}

LM gratefully acknowledges helpful comments and discussions from Chris Stoughton, Jocelyn Read, Rana Adhikari, and Yanbei Chen. This research towards the GQuEST experiment is supported by the Heising-Simons Foundation through grant 2022-3341.

\nocite{apsrev42Control}
\bibliographystyle{apsrev4-2-trunc}
\bibliography{control, GQStats}

\appendix    
\clearpage
\begin{widetext}

\section{Derivations and Notes}

This section carries snippets of technical qualifications and derivations that are useful but combersome to include  inline in the main text

\subsection{Band Limited Power Observations} \label{D:band_limit}

The form \cref{eq:P_d_form} is ill-defined due to the singular nature of the quantum density operators $\qo{a}(t)$ contained in $\qo{d}(t)$. More realistically, it requires the convolution with some kernel $K(t)$ that encapsulates the finite bandwidth of the photo detector opto-electronic properties.
\begin{align}
  \qo{P}_d(t) = \int_{\infty}^{\infty}\qo{d}^\dagger(t)\qo{d}(t - t')K(t')\rmd t'
\end{align}
The power operator in the frequency domain for the output port of the interferometer is then expressed as
\begin{align}
  \qo{P}_d(\Omega) = K(\Omega)
  \Big(
  \qo{d}^\dagger(\omega + \Omega)
  \circledast
  \qo{d}(\omega + \Omega)
  \Big)
\end{align}
The $K(t)$ is introduced here for correctness, but is cumbersome to carry. The main text assumes the limit $K(t') \rightarrow \delta(t')$,  $K(\Omega) \rightarrow 1$. 

\subsection{Homodyne Field Observable Gaussian Characteristic Function, Equation~\ref{eq:PCF_Gaussian}}\label{D:PCF_Gaussian}

This is a short derivation of the Characteristic function of a Gaussian using the Homodyne observable, related to \cref{eq:PCF_Gaussian}.
\begin{align}
  \PCF{\Hpm, \HqM_{ci}}
  &=
    \Braket{e^{-i \HqM_{ci} \Hpm}}
  =
    \Braket{e^{-i (2\red{g}\cl{\vphi}_{ci} + 2\cl{\theta}_{ci} + i\qo{a}_{c i} - i\qo{a}_{c i}^\dagger)\red{l}\Hpm}}
  \\
  &=
    e^{i\red{l}\Hpm(2\red{g}\cl{\vphi}_{ci} + 2\cl{\theta}_{ci})} \Braket{e^{ \red{l}\Hpm \qo{a}_{c i}^\dagger - \red{l}\Hpm \qo{a}_{c i}}} \label{eq:deriv_PCF_Gaussian}
       =
        e^{i\red{l}\Hpm(2\red{g}\cl{\vphi}_{ci} + 2\cl{\theta}_{ci})}  \Braket{e^{(\red{l} \Hpm)^2\frac{[\qo{a}_{c i}^\dagger, \qo{a}_{c i}]}{2}}e^{\red{l}\Hpm \qo{a}_{c i}^\dagger} e^{-\red{l}\Hpm \qo{a}_{c i}}}
  \\ &=
       e^{i\red{l}\Hpm(2\red{g}\cl{\vphi}_{ci} + 2\cl{\theta}_{ci}) - \frac{(\red{l}\Hpm)^2}{2}}
\end{align}
Where the lines \cref{eq:deriv_PCF_Gaussian} utilizes the Campbell-Baker-Hausdorff formula to normally-order the raising and lowering operators.

\subsection{Full Variance of Homodyne Chi-Square}\label{D:chi_weighted}

This section utilizes the weighting parameters of \cref{eq:Homodyne_statistic_full} to create the optimal search statistic. The weighting is an ad-hoc factor that is included in a manner that preserves the unbiased mean of the estimator. It represents taking a weighted average when aggregating many independent chi-square measurements.
We start from \cref{eq:Homodyne_statistic_full}  and write the full variance for the weighted expression.
\begin{align}
  \qo{\chi}^2_{2N}
  &\equiv
    \frac{1}{(\sum_{i=0}^{N} W_i) }
    \sum_{i=0}^{N} \frac{W_{i}}{2|\red{g}(\Omega_i)|^2\SPEC(\Omega_i)}\left(\qo{V}_{+i} - \braket{\qo{V}_{+i}}\Big|_{S_{\varphi} =0}\right)
    \\
  \VAR{\qo{\chi}^2_{2N}}
  &\approx
    \frac{1}{(\sum_{i=0}^{N} W_i)^2}
    \sum_{i=1}^{N}
    \left( \frac{W_{i}}{2|\red{g}(\Omega_i)|^2\SPEC(\Omega_i)} \right)^2
    \left(
    2|\red{g}(\Omega)|^2S_{\cl{\vphi}}(\Omega_{i}) +
    2\SNP(\Omega_{i}) + 1
    \right)
\end{align}
The derivative of the variance with respect to $W_i$ shows that the variance is minimized, derivative zero, when
\begin{align}
  W_i
  &= 
    \frac{
    \left( 
    2|\red{g}(\Omega_i)|^2\SPEC(\Omega_i)
    \right)^2
    }{
    2|\red{g}(\Omega)|^2S_{\cl{\vphi}}(\Omega_{i}) +
    2\SNP(\Omega_{i}) + 1
    }
    \approx
    \frac{
    \left( 
    2|\red{g}(\Omega_i)|^2\SPEC(\Omega_i)
    \right)^2
    }{
    2\SNP(\Omega_{i}) + 1
    }
\end{align}
Where the approximation is in the weak signal limit, but includes classical noise contributions. Plugging in this optimal weighting recovers the usual formulas for optimal stochastic searches such as eqn. 7.293 of Creighton and Anderson \cite{creighton11GravitationalWave}. The summation can also be converted, using \cref{eq:S_idx_expr_full}, into an integral form.
\begin{align}
  \frac{1}{
  \VAR{\qo{\chi}^2_{2N}}
  }
  &\approx
    \sum_{i=1}^{N}
    \frac{
    \left(
    2|\red{g}(\Omega_i)|^2\SPEC(\Omega_i)
    \right)^2
    }{
    2\SNP(\Omega_{i}) + 1
    }
    \approx
    \Delta T
    \int_0^{\infty}
    \frac{
    \left(
    2|\red{g}(\Omega)|^2\SPEC(\Omega)
    \right)^2
    }{
    2\SNP(\Omega) + 1
    }
    \frac{
    \rmd \Omega
    }{2\pi}
  \approx
    \Delta T
    \int_0^{\infty}
    \FISH{{\HqM_{+i}} | \param}
    \frac{
    \rmd \Omega_{+i}
    }{2\pi}
    \label{eq:chi_sq_fully_weighted}
\end{align}
The last integral represents a summation over all possible orthogonal templates within the detection time $\Delta T$. The integral forms indicate that the optimal weighting saturates the Fisher information over the whole frequency band.

\subsection{Properties of Gaussian distributions}\label{D:Gaussian_Props}
When first introduced, the signal and noise processes were described as random noise processes with two-point correlation functions of \cref{eq:phi_expectations} that include delta-function factors. This makes them singular, and difficult to reason about their distribution. Once those signals are integrated with templates, they become substantially better behaved and higher-order correlation functions and moments can be defined.

First, the characteristic function for the Gaussian processes is given by:
\begin{align}
  \braket{e^{i \Ppe \REof{\cl{\vphi}_{\pm i}}}}
  &= e^{-S_{\vphi}(\Omega_i)\Ppe^2/4}
  &
    \braket{e^{i \Ppe \IMof{\cl{\vphi}_{\pm i}}}}
  &= e^{-S_{\vphi}(\Omega_i)\Ppe^2/4}
\end{align}
Another useful higher order moment is the characteristic function of the square of an expectation value. Those are computed using the Gaussian PDF over the variable $\vphi_R$
\begin{align}
  \braket{e^{i \Ppe \REof{\cl{\vphi}_{\pm i}}^2}}
  &=
    \sqrt{\frac{2}{\pi S_{\vphi}}}
    \int_{-\infty}^{\infty}
    e^{i e \vphi_R^2}
    e^{-\frac{2 \vphi_R^2}{S_{\vphi}}}
    \rmd \vphi_{R}
  =
    \sqrt{\frac{2}{\pi S_{\vphi}}}
    \int_{-\infty}^{\infty}
    e^{-\vphi_R^2(\frac{2}{S_{\vphi}} + ie)}
    \rmd \vphi_{R}
  =
    \sqrt{\frac{2}{2 + ieS_{\vphi}}}
    \\
  \braket{e^{i \Ppe |\cl{\vphi}_{\pm i}^2|^2}}
  &=
    \braket{e^{i \Ppe( \REof{\cl{\vphi}_{\pm i}}^2 + \IMof{\cl{\vphi}_{\pm i}}^2)}}
  =
    \frac{2}{2 + ieS_{\vphi}}
    \label{eq:ChiSq_character_by_Gaussian}
\end{align}
These characteristic functions are equivalent to those of a scaled chi-square distribution.

\subsection{Single-mode Energy Characteristic Function, Equation~\ref{eq:CF_Energy}}\label{D:CF_Energy}

This section derives the number operator for a Michelson interferometer in the Heisenberg picture.
This derivation starts from \cref{eq:E_template_observable} or \cref{eq:E_power_d} and will result in the expressions used for \cref{eq:CF_Energy}. First expressions for $\qo{E}_{\pm i}$ are factorized with the classical-fluctuation operator $\cl{Y}_{\pm i}$
\begin{align}
  \PqE_{\pm i}
  &=
    \big(\cl{Y}_{\pm i} - i\qo{a}_{\pm i}\big)^\dagger \big(\cl{Y}_{\pm i} - i\qo{a}_{\pm i}\big)
  &
    \cl{Y}_{\pm i}
  &=
    \gq(\Omega_{\pm i}) \cl{\vphi}_{\pm i} + q(\Omega_{\pm i})\cl{n}_{\pm i}
\end{align}
Similarly to the homodyne observation derivation of \cref{D:PCF_Gaussian}, the expectations can be reduced have applying a normal ordering to the expressions. For this case, the Campbell-Baker-Hausdorff formula is not sufficient, and a generalized normal ordering relation \cite{blasiakAJP07CombinatoricsBoson} is used:
\begin{align}
  e^{X \qo{a}_{\pm i}^\dagger \qo{a}_{\pm i}} &= :e^{\qo{a}_{\pm i}^\dagger \qo{a}_{\pm i}(e^X-1)}:
\end{align}
Here, the $::$ symbols indicate that the operators of the expression are normally-ordered.

To calculate the PMF, the normal ordering is immediately applied to eliminate the quantum variables. The PDF of the classical signal and noise is then used to calculate the expectations from the factor $\red{S}_{\pm i}$ of \cref{eq:constituent_S}. The derivation can be simplified by applying \cref{eq:ChiSq_character_by_Gaussian} at \cref{eq:CF_Geom_mid} to skip to the end.
\begin{align}
  \PCF{\Ppe, \PqE_{+ i}}
  &\equiv
    \Braket{e^{i \PqE \Ppe}}
           =
    \braket{
    e^{i \Ppe(\cl{Y}_{\pm i} - i\qo{a}_{\pm i})^\dagger(\cl{Y}_{\pm i} - i\qo{a}_{\pm i})}
    }
       =
    \braket{
    :e^{(\cl{Y}_{\pm i} - i\qo{a}_{\pm i})^\dagger(\cl{Y}_{\pm i} - i\qo{a}_{\pm i})(e^{i\Ppe }-1)}:
    }
       =
    \braket{
    e^{|\cl{Y}_{\pm i}|^2(e^{i\Ppe }-1)}
    }
    \label{eq:CF_Geom_mid}
  \\
  &=
    \frac{1}{\pi \red{S}_{\pm i}}
  \int_{-\infty}^{\infty}
  \int_{-\infty}^{\infty}
    e^{-(\vphi_R^2+\vphi_I^2)(e^{i\Ppe }-1)}
  e^{-\frac{\vphi_R^2 + \vphi_I^2}{\red{S}_{\pm i}}}
  \rmd \vphi_{R}
  \rmd \vphi_{I}
       =
    \frac{1}{\red{S}_{\pm i}}
    \int_{0}^{\infty}
    e^{-\vphi(e^{i\Ppe }-1+\frac{1}{\red{S}_{\pm i}})}
    \vphi
    \rmd \vphi
  \\ &
       =
    \frac{1}{\red{S}_{\pm i}(e^{i\Ppe }-1)+1}
\end{align}
Which is the characteristic function of a Geometric distribution with a Bernoulli trial probability $p=1/(1 + \red{S}_{\pm i})$. The Geometric distribution implements Bose statistics on a single temporal mode of the light, with a thermal occupation. This implies a mean occupation number and resulting temperature:
\begin{align}
  \braket{\qo{E}_{\pm i}} = (1-p)/p = \red{S}_{\pm i} = e^{\frac{\hbar \omega}{k_{\text{B}} T_\omega}} = e^{\frac{\hbar \Omega_{+ i}}{k_{\text{B}} T_\Omega}}
\end{align}
The two definitions of temperature are for the absolute temperature, $T_\omega$, of a system that would create photons at frequency $\omega$ vs. the sideband temperature, $T_\Omega$, for creating acoustic modulations.

\subsection{Single-mode Energy Probability Mass, Equation~\ref{eq:PMF_Energy}}\label{D:PMF_Energy}

This derivation extends the previous one to generate the probability mass function for photon counting a single signal mode \cref{eq:PMF_Energy}. It starts form of the characteristic equation which has not been entirely simplified

\begin{align}
  \PDF{\PpE, \PqE_{+ i}}
  &=
    \frac{1}{2\pi}\int_{-\pi}^{\pi}
    e^{-i \PpE \Ppe}
    \braket{e^{|\cl{Y}_{\pm i}|^2(e^{i\Ppe }-1)}}
    \rmd \Ppe
       =
    \frac{1}{2\pi}
    \Braket{
    e^{-\cl{Y}_{\pm i}^2}
    \int_{-\pi}^{\pi}
    \sum_{n=0}^{\infty}\frac{\cl{Y}_{\pm i}^{2n}}{n!}
    e^{-i \PpE \Ppe}
    e^{i n \Ppe }
    \rmd \Ppe
    }
       =
    \Braket{
    e^{-\cl{Y}_{\pm i}^2}
    \frac{\cl{Y}_{\pm i}^{2E}}{\PpE!}
    }
  \\ &
       =
    \label{eq:E_PDF_intint}
    \frac{1}{\PpE!}
    \frac{1}{\pi \red{S}_{\pm i}}
    \int_{-\infty}^{\infty}
    \int_{-\infty}^{\infty}
    (\vphi_R^2 + \vphi_I^2)^{\PpE}
    e^{-\vphi_R^2-\vphi_I^2}
    e^{-\frac{\vphi_R^2 + \vphi_I^2}{\red{S}_{\pm i}}}
    \rmd \vphi_{R}
    \rmd \vphi_{I}
  \\ &
       =
    \frac{1}{\PpE!}
    \frac{2\pi}{\pi \red{S}_{\pm i}}
    \int_{-\infty}^{\infty}
    \vphi^{2E}
    e^{-\vphi^2}
    e^{-\frac{\vphi^2}{\red{S}_{\pm i}}}
    \vphi \rmd \vphi
       =
    \frac{1}{\PpE!}
    \frac{2}{\red{S}_{\pm i}}
    \int_{-\infty}^{\infty}
    \vphi^{2E}
    e^{-\vphi^2(1 + \frac{1}{\red{S}_{\pm i}})}
    \vphi \rmd \vphi
  \\ &
       =
    \frac{1}{\PpE!}
    \frac{1}{\red{S}_{\pm i}}
    \int_{0}^{\infty}
    x^{\PpE}
    e^{-x(1 + \frac{1}{\red{S}_{\pm i}})}
    \rmd x
       =
    \frac{1}{\PpE!}
    \frac{1}{\red{S}_{\pm i}}
    \left(1 + \frac{1}{\red{S}_{\pm i}}\right)^{-1 - \PpE}
    \int_{0}^{\infty}
    y^{\PpE}
    e^{-y}
    \rmd y
  \\
  &=
    \frac{1}{\red{S}_{\pm i}}
    {\left(1 + \frac{1}{\red{S}_{\pm i}}\right)}^{-1 - \PpE}
\end{align}

This relates the distribution fully back to the geometric distribution.
\begin{align}
    p &\equiv \frac{1}{\red{S}_{\pm i} + 1}
  &
  \left(1 + \frac{1}{\red{S}_{\pm i}}\right)^{-1 - \PpE} &\equiv p{(1-p)}^{\PpE}
  &
  \frac{1}{\red{S}_{\pm i} + 1}
  \left(1 - \frac{1}{\red{S}_{\pm i} + 1}\right)^{\PpE} &\equiv p{(1-p)}^{\PpE}
\end{align}

\label{D:PCF_full_variance}
From above, the variance of the Geometric distribution can be utilized to calculate the full expressions of \cref{eq:var_Energy_sm} and \cref{eq:photon_counting_G_VAR}.
\begin{align}
  \VAR{\qo{E}}
  &=
    \frac{1-p}{p^2} = \red{S}_{\pm i}(\red{S}_{\pm i} + 1)
            &
  \VAR{\qo{G}_\param}
  &\approx
    \frac{2}{N_{\qo{E}}}\left(\frac{1}{|\red{g}|^2\red{\SPEC}}\right)^2
    \red{S}_{\pm i}(1 + \red{S}_{\pm i})
    \label{eq:photon_counting_G_VAR_full}
    &
      \red{S}_{\pm i}
  &=
    |\red{g}|^2\param \red{\SPEC} + \red{S}_{\VNP} + \red{S}_{\VNA} + \frac{2\red{B}\Delta T}{N_{\qo{E}}} 
\end{align}

\subsection{Full Narrowband Search Scan}\label{d:scan_comparisons}

These expressions are for the full relationship for \cref{eq:scan_ratio} without assuming optimal bandwidths.

\begin{align}
  \frac{
  \VAR{\qo{G}}_{\text{SR,scan}}
  }{
  \VAR{\qo{\chi}'^2_{2N}}_{\text{SR,scan}}
  }
  &=
    \left(
    |\red{g}|^2\red{S}_{\cl{c}}
    + \frac{\Delta T \red{B}}{\mathcal{N}_q}
    \right)
    \left(\frac{c}{8L\gamma_{\eta}}\right)
    \left( \frac{\Delta T c}{2L}\right)
    \frac{\mathcal{N}_q\gamma_{\vphi}(\gamma_{q}^{-1} + \gamma_{g}^{-1})}{\mathcal{N}_{\SPEC}^{2}}
  \\ &=
       \left(
       |\red{g}|^2\red{S}_{\cl{c}}
       + \frac{\Delta T \red{B}}{\mathcal{N}_q}
       \right)
       \left(\frac{c}{8L\gamma_{\eta}}\right)
       \frac{\left(\gamma_{\vphi} + \gamma_{g} + \gamma_{\eta} \right)^2}{\eta_g \gamma_{\vphi}}
       \frac{\gamma_{q}(\gamma_{q}^{-1} + \gamma_{g}^{-1})}{\gamma_{q} + \gamma_{g} + \gamma_{\eta}}
\end{align}

\subsection{Application Example from Statistical Equations}\label{D:application_relations}

Here, the heuristic argument developed to demonstrate the application of photon counting towards a stochastic signal search from quantum gravity of \cref{sec:application} is derived from the equations of \cref{sec:homodyne} and \cref{sec:counting}. Both start from the relationship between the shot-noise limited displacement sensitivity and the optical gain of an interferometer:
\begin{align}
  S_{\HqM} &\equiv \frac{1}{2|\red{g}|^{2}} = \PSD{\text{Quantum}}
\end{align}

Using this definition and the optimal search from \cref{eq:chi_sq_fully_weighted}, the expressions for the search time can be derived. The $1\sigma$ observing time is established from the condition $\VAR{\qo{\chi}^2_{2N}} = 1$. In the case where the dominant noise is quantum noise, and classical noise is a negligible contribution, \cref{eq:chi_sq_fully_weighted} can be given.

\begin{align}
  1 = 
  \frac{1}{
  \VAR{\qo{\chi}^2_{2N}}
  }
  &
    \approx
    \Delta T_{\HqM}^{1\sigma}
    \int_0^{\infty}
    \left(
    2|\red{g}(\Omega)|^2\SPEC(\Omega)
    \right)^2
    \frac{
    \rmd \Omega
    }{2\pi}
    \approx
    \Delta T_{\HqM}^{1\sigma}
    \left( 2|\red{g}|^2\red{\SPEC} \right)^2
    \int_0^{\infty}
    \frac{
    \SPEC(\Omega) ^2
    }{\red{\SPEC}^2}
    \frac{
    \rmd \Omega
    }{2\pi}
\end{align}
And then reduced to the following equations.
\begin{align}
  \Delta T_{\HqM}^{1\sigma} &= \frac{N_{\HqM}}{\Delta F_{\HqM}}
  &
  N_{\HqM}
  &\equiv
    \Delta F_{\HqM} \Delta T_{\HqM}^{1\sigma}
    =
    \frac{S_{\HqM}^2}{(\param \red{\SPEC})^2}
  &
    \Delta F_{\HqM}
  &=
    \int_0^{\infty} \left( \frac{\SPEC(\Omega)}{\red{\SPEC}} \right)^2 \frac{\rmd \Omega}{2\pi}
   \label{eq:homodyne_time_scaling_2}
\end{align}

The application section \cref{sec:application} then uses a heuristic set of arguments to determine the search time from photon counting, resulting in the following equations
\begin{align}
  E_{\vphi} &= \frac{\param\red{\SPEC}}{4 S_{\HqM}} \Delta \red{T}_E \Delta F_E
  &
    \Delta F_{E} &= \int_{-\infty}^{\infty} \frac{\SPEC(\Omega)}{\red{\SPEC}} \frac{\rmd \Omega}{2\pi}
                   &
  E_{\vphi} &= \Delta T_E^{1\sigma} \int_{-\infty}^{\infty} P_{\vphi}(\Omega) \frac{\rmd \Omega}{2\pi}
              &
              P_{\vphi}(\Omega) &= \frac{ S_{\vphi}(\Omega)}{4 S_{\HqM}(\Omega)}
\end{align}
using $E_{\vphi}=1$ to imply the $1\sigma$ statistical time is done first under the assumption of negligible classical noise. In the equations above, negligible classical noise is assumed by $P_{\vphi}(\Omega)$ including only the power contributions from the signal. The reduced time $\red{T}_E^{1\sigma}$ denotes the time to measure one signal photon, while the full time, $T_E^{1\sigma}$ includes the classical noise contributions. In the application section, these appeared ad-hoc, but can be related to the statistics of the photon-counting section.

\begin{align}
  \Delta \red{T}_E^{1\sigma}
  &\equiv \frac{4 S_{\HqM}}{\param\red{\SPEC}\Delta F_E}
    = \frac{2 }{|\red{g}|^{2}\param\red{\SPEC}\Delta F_E}
    \label{eq:counting_SNR_example_3}
  & 
    \Delta T_E^{1\sigma}
  &=
    \Delta \red{T}_E^{1\sigma}
    \left(1 + \frac{\PSD{\text{Cl}}}{\red{\SPEC}} + \red{B}\Delta \red{T}_E^{1\sigma} \right)
\end{align}

The full, classical noise included, integration time can be derived from \cref{eq:photon_counting_G_VAR} (including the signal contribution term) where $1 = \VAR{\qo{G}_\param}$ indicates the $1\sigma$ observing time. Expanding that condition leads to the expressions
\begin{align}
  \Delta T_E^{1\sigma} \Delta F_E
  &\approx
    2\left(\frac{1}{|\red{g}|^2\red{\SPEC}}\right)^2
    \left(|\red{g}|^2\red{\SPEC} + \red{S}_{\VNP} + \red{S}_{\VNA} + \frac{2\red{B}\Delta T}{N_{\qo{E}}} \right)
    = 
    \frac{4 S_{\HqM}}{\red{\SPEC}}
    \left(1
    + \frac{\red{S}_{\VNP} + \red{S}_{\VNA}}{|\red{g}|^2\red{\SPEC}}
    + \frac{2\red{B}}{|\red{g}|^2\red{\SPEC}\Delta F_E} \right)
    \label{eq:photon_counting_G_VAR_2}
\end{align}
The right hand side is directly convertible into the expressions of \cref{eq:counting_SNR_example_3}.

\end{widetext}

\end{document}